\documentclass[fleqn,usenatbib]{mnras}
\usepackage[T1]{fontenc}
\usepackage{ae,aecompl}
\usepackage{color}
\usepackage{graphicx}
\usepackage{natbib}
\usepackage[switch,columnwise]{lineno}
\newcommand{\result}[1]{{\color{black} {{\it \color{black} Result:}} 
\color{black} #1}}
\newcommand{\hattuuslippa}{{\color{black} reproducible}}

\newcommand{\Data}[2]{{{\sc data}$_{\mathrm{{#1},{#2}}}$}}

\newcommand{\flip}{flip--flop}
\newcommand{\Inc}{incompatibility}
\newcommand{\cpsmethod}{CPS-method}

\newcommand{\sdr}{SDR}
\newcommand{\dimethod}{DI-method}
\newcommand{\lcmethod}{LC-method}
\newcommand{\lsmethod}{LS-method}

\newcommand{\Testone}{{\color{black}   \it 1st test}}
\newcommand{\Testtwo}{{\color{black}   \it 2nd test}}
\newcommand{\Testthree}{{\color{black} \it 3rd test}}
\newcommand{\Testfour}{{\color{black}  \it 4th test}}
\newcommand{\Testfive}{{\color{black}  \it 5th test}}
\newcommand{\Testsix}{{\color{black}   \it 6th test}}
\newcommand{\Testseven}{{\color{black} \it 7th test}}
\newcommand{\Testeight}{{\color{black} \it 8th test}}
\newcommand{\JHLhyp}{{JHL-hy\-poth\-e\-sis}}
\newcommand{\JHLtext}{{``The {\it observed} 
light curves of chromospherically active binary
and single stars are interference of two {\it real} 
constant period light curves of long-lived starspots.
These periods are the non-stationary active longitude
period $P_{\mathrm{act}}$ and the stationary rotation period
$P_{\mathrm{rot}} \approx P_{\mathrm{orb}}$.''}}
\newcommand{\weakJHLtext}{{``The {\it observed} 
light curves of chromospherically active binary
and single stars are interference of two {\it real} 
constant period light curves of long-lived starspots.''}}
\newcommand{\Testonetext}{\Testone: 
Why have these two constant 
$P_{\mathrm{act}}$ 
and 
$P_{\mathrm{rot}} \approx P_{\mathrm{orb}}$ 
periods  {\it ~not} 
been detected in the light curves?
Why are so {\it many different} 
$P_{\mathrm{phot}}$ 
periods detected in the {\it same} star?
}
\newcommand{\Testtwotext}{ \Testtwo:
\citet{Hac13} applied  the Kuiper method to
the % $t_{\mathrm{CPS,min,1}}$
light curve minimum epochs of FK Com.
Did this analysis give 
an {\it unambiguous} 
$P_{\mathrm{act}}$ period estimate?
}
\newcommand{\Testthreetext}{\Testthree: 
The Kuiper method detects the $P_{\mathrm{act}}$
period from 
the % $t_{\mathrm{CPS,min,1}}$
light curve minimum epochs
of binary  \citep[][]{Jet17}
and single stars \citep[][]{Hac13}.
Why does this method {\it not} detect 
the $P_{\mathrm{rot}}\approx P_{\mathrm{orb}}$
period?}
\newcommand{\Testfourtext}{\Testfour: 
The long-term mean light curves of 
binaries
followed the 
$P_{\mathrm{rot}} \approx P_{\mathrm{orb}}$ period
in \citet{Jet17}.
Why did these light curves {\it not} follow
the $P_{\mathrm{act}}$ period?
}
\newcommand{\Testfivetext}{\Testfive:
  What explains the {\it abrupt}
  \flip ~events,
  if the long-lived
  starspots rotate with the two
  {\it regular 
constant} 
$P_{\mathrm{act}}$ 
and 
$P_{\mathrm{rot}} \approx P_{\mathrm{orb}}$ periods?
}
\newcommand{\Testsixtext}{\Testsix: 
Why have these two constant 
$P_{\mathrm{act}}$ 
and 
$P_{\mathrm{rot}} \approx P_{\mathrm{orb}}$
periods 
{\it ~not} been detected in the 
spot models?
}
\newcommand{\Testseventext}{\Testseven: 
Why have these two constant
$P_{\mathrm{act}}$ 
and 
$P_{\mathrm{rot}} \approx P_{\mathrm{orb}}$
periods
{\it ~not} been detected in the 
Doppler images?
}
\newcommand{\Testeighttext}{\Testeight:
Why do the light curves and the Doppler images
give {\it different} periods 
even for the {\it same} 
individual star?
}
\newcommand{\simuone}{{\sc{sim\-u\-la\-tion}}$_{a_1=a_2}$}
\newcommand{\simutwo}{{\sc{sim\-u\-la\-tion}}$_{a_1<a_2}$}
\newcommand{\firstcase}{{{\sc \color{black} case}$_1$}}
\newcommand{\secondcase}{{{\sc \color{black} case}$_2$}}
\newcommand{\thirdcase}{{{\sc \color{black} case}$_3$}}

\newcommand{\migrone}{{{\sc migration}$_1$}}
\newcommand{\migrtwo}{{{\sc migration}$_2$}}
\newcommand{\migrthree}{{{\sc migration}$_3$}}

\title[Real light curves of FK Com]
{Real light curves of FK Comae Berenices:  
Farewell flip-flop}
\author[L. Jetsu]{
L. Jetsu\thanks{E-mail: lauri.jetsu@helsinki.fi} \\
$$Department of Physics, 
P.O. Box 64, FI-00014 University of Helsinki, 
Finland}
\date{Accepted XXX. Received YYY; in original form ZZZ}
\pubyear{2018}
\begin{document}
\label{firstpage}
\pagerange{\pageref{firstpage}--\pageref{lastpage}}
\maketitle
\begin{abstract}
%\revised{81/100}
For seven decades,
the widely held view has been 
that the formation, the migration and the decay
of short-lived starspots explain
the constantly changing
light curves of chromospherically active stars.  
Our hypothesis is that these deceptive observed 
light curves are interference 
of two real constant period light curves
of long-lived starspots. 
The slow motion of these long-lived 
starspots with respect to each other
causes the observed light curve changes.
This hypothesis contradicts
  the current views of starspots.
Therefore, we subject it to eight \hattuuslippa ~tests.
  Our new period finding  method
  detects the two real light curves of FK Com. 
  Our hypothesis is a total success:
  all real light curve parameters 
  are directly connected to the long-lived starspots
  which are also seen in the Doppler images of FK Com.
  These parameters are spatially and temporally
  correlated just like in the Sun, including
  weak solar-like surface differential rotation. 
  As for other chromospherically active stars, all eight
  \hattuuslippa ~tests also support our hypothesis. 
It explains many 
spurious phenomena:
the rapid light curve changes,
the short starspot life-times,
the rapid rotation period changes,
the active longitudes,
the starspot migration,
the period cycles, 
the amplitude cycles 
and 
the minimum epoch cycles. 
It also explains why 
the light curves and the Doppler images give
contradicting
surface differential rotation estimates 
even for the same individual star,
as well as the abrupt 180 degrees shifts of activity 
(the flip-flop events) and the long-term mean light curves.
We argue that the current views of starspots 
need to be revised.
\end{abstract}
%%%%%%%%%%%%%%%%%% BODY OF PAPER %%%%%%%%%%%%%%%%%% 

\section{Introduction}
\label{Intro}

The ancient Egyptian papyrus Cairo 86637 
%%%% which contains the calendar of lucky and unlucky days,
is the oldest preserved 
document of the discovery of a variable star,
Algol
\citep[][]{Por08,Jet13,Jet15,Por18}.
Algol's changes can be observed with
naked eye, but the solar luminosity changes
only with satellites
\citep{Wil91,Rad18}.
John of Worcester made
the oldest preserved drawing of a sunspot
 in the year 1128 \citep[][]{van96}.
\citet{Sch44} discovered the eleven
years cycle in the number 
of sunspots.
\citet{Hal08} discovered
the Zeeman effect of solar magnetic field,
and that this magnetic field 
is stronger in the sunspots.

\citet{Kro47} discovered the 
starspots in the light curves of 
the eclipsing binary AR Lac.
He observed short-term % light curve
changes  ``within a few weeks to a few months''.
FK Comae Berenices (HD 117555, FK Com)
was among the first late--type stars where 
the starspots were also discovered 
\citep{Chu66}.
This chromosperically active single G4 giant \citep{Str09}  
is the prototype of a class of variable stars,
the FK~Com class, defined by \citet{Bop81}
as rapidly rotating single G--K giants.
Only a few stars belonging to this class have been found
\citep[][]{Puz14,How16,Puz17}. 
These stars may represent
recently coalesced W~UMa binaries \citep{Web76,Egg89,Wel94}. % Bop81

The starspots of FK Com seem to
concentrate
on two  long--lived
active longitudes separated by 180 degrees, 
and undergo abrupt shifts between these longitudes.
These shifts are called the \flip ~events
\citep{Jet91,Jet93}.
They occur also 
in binaries
\citep[e.g.][$\sigma$ Gem]{Jet96}.
Kuiper method analysis of the light curve
minimum epochs gave
the 
$P_{\mathrm{act}}=2.^{\mathrm{d}}401155 \pm 0.^{\mathrm{d}}000092$
active longitude period of FK Com
\citep{Hac13}.
There are numerous photometric 
\citep[e.g.][]{Kor02,Ola06,Pan07}
and Doppler imaging
\citep[e.g.][]{Kor00,Kor04,Kor07,Hac13,Vid15}
 studies of FK Com. % Kor09, Kor99

\citet{Jet17} presented a general light curve
model for the Chromospherically Active Binary Stars\footnote{Table \ref{abbre}
gives the abbreviations used in this paper.}
(hereafter CABS).
They studied the long-term 
Mean Light Curve (hereafter MLC)
of fourteen CABSs.
%and also applied their model to FK Com.
\citet[][JHL = Jet\-su, Henry, Leh\-ti\-nen]{Jet17} presented 
the  hypothesis

\begin{description}

\item[-] \JHLhyp: \JHLtext

\end{description}
At first sight, this hypothesis would
seem to contradict the current views of starspots and
stellar Surface Differential Rotation (hereafter \sdr).
Two methods are widely used 
to measure SDR \citep[][Sect. 7]{Str09}.
In the \lcmethod,
the $P_{\mathrm{phot}} \! \approx \! P_{\mathrm{rot}}$ estimates 
are obtained from period analysis
\citep[e.g.][]{Rei13,Leh16}
or spot modelling
\citep[e.g.][]{Kip12,Lan14} of light curves.
In the \dimethod, Doppler imaging gives
$P_{\mathrm{rot}}$ periods
of starspots at different latitudes
\citep[e.g.][]{Pet04,Col07,Kov17}.
These two methods can give 
{\it different} \sdr ~estimates
for the {\it same} star, e.g.
the \dimethod ~study by \citet{Kor00} 
indicated ``solid body rotation'' in FK Com,
while \citet{Hac13} measured  photometric
period changes of 3.1\% with the \lcmethod.

If the \JHLhyp ~is true, it should predict the
results for the following eight undermining tests.

\begin{description}
\item[-] \Testonetext
\item[-] \Testtwotext
\item[-] \Testthreetext
\item[-] \Testfourtext 
\item[-] \Testfivetext 
\item[-] \Testsixtext 
\item[-] \Testseventext 
\item[-] \Testeighttext
\end{description}

\noindent
  We summarize the results for all tests in 
  Sect. \ref{Conclusions}.

\section{Data}
\label{Data}

We analyse the standard Johnson
$V$ photometry of FK Com
from \citet[][]{Hac13}.
These observations were made with 
the ``T7'' (TEL=1) and ``Ph10'' (TEL=2) telescopes.
\cite{Hac13} stored them as two separate files 
into the CDS database.
We analyse these files separately.
%as \citet{Hac13} also did.
We discard $n\!=\!47$ temporally isolated observations 
and $n\!=\!21$ outliers, and publish
the remaining observations
in
Table\footnote{Full Tables \ref{fkdata} and \ref{electric}
  are available only in electronic form
  via anonymous ftp  to
cdsarc.u-strasbg.fr (130.79.128.5) 
or via http://cdsarc.u-strasbg.fr/viz-
bin/qcat?J/...
}
\ref{fkdata}.
We divide the data into segments
in Table \ref{tabledata} (SEG).
Our notation for TEL=x data
in segment SEG=y 
is \Data{x}{y}.
The accuracy 
is between $0.^{\mathrm{m}}004$
and $0.^{\mathrm{m}}008$ in good photometric nights
\citep{Hen95A,Str97}.
%All observations where
%the standard deviation of three
%measurements exceeds $0.^{\mathrm{m}}020$
%are automatically discarded.
\citet{Hac13} applied the Continuous Period Search 
method (hereafter the \cpsmethod)
to these data, but
did not publish the results.
We publish those results in 
Table \ref{electric}.

\section{Models}
\label{Models}

\subsection{General model}
\label{General}

The $n$ observations at times
$t_i$ are $y_i\!=\!y(t_i)$. 
Their mean and standard deviation
are $m_{y}$ and $\sigma_{y}$.
%Our zero point in time is $t_1$. 
%which is subtracted from all observing times $t_i$. 
The time span % of $n$
$\Delta T\!=\!t_n\!-t_1$
gives 
the distance between independent frequencies
\begin{eqnarray}
f_0=1/\Delta T.
\label{indepedentfreq}
\end{eqnarray}
\noindent
The first part of our general model is a $K_0$ order polynomial
\begin{eqnarray}
m_0(t)=  m_0(t,\bar{\beta_0})=\sum_{k=0}^{K_0} M_k (f_0 t)^k,
\label{gzero}
\end{eqnarray}
where the free parameters are
$\bar{\beta}_0=$ $[M_0, ..., M_{K_0}]$ measured in magnitudes.
The second part is a $K_1$ order harmonic
\begin{eqnarray}
g_1(t)\!\!=\!\!
g_1(t,\bar{\beta}_1) \!\!=\!\!
\sum_{k=1}^{K_1} \! B_k \cos{(k 2 \pi f_1 t)} \! + \!  C_k \sin{(k 2 \pi f_1 t)},
\label{gone}
\end{eqnarray}
where
$\bar{\beta}_1=$ $[B_1, ..., B_{K_1}, C_1, ..., C_{K_1},f_1]$.
The amplitudes are
are measured in magnitudes
and the frequency $f_1$ in ${\mathrm{d}}^{-1}$.
We determine the following  $g_1(t)$ parameters
\begin{description}
\item $P_1=$ period $=f_1^{-1}$
\item $A_1=$ peak to peak amplitude
\item $t_{\mathrm{g1,min,1}}=$ epoch of primary (i.e. deeper) minimum
\item $t_{\mathrm{g1,min,2}}=$ epoch of secondary minimum (if present)
\end{description}
The units are
$[P_1]\!=\!{\mathrm{d}}$,
$[A_1]\!=\!{\mathrm{mag}}$ and
$[t_{\mathrm{g1,min,1}}]\!=\!$
$[t_{\mathrm{g1,min,2}}]\!=\!$ ${\mathrm{HJD}}-2~400~000$.
%We use the epoch of the first primary and secondary minimum
%within each individual segment.
The phases of $g_1(t)$ are 
\begin{eqnarray}
\phi_1={\mathrm{FRAC}}[(t-t_1)/P_1],
\label{phaseone}
\end{eqnarray}
\noindent
where ${\mathrm{FRAC}}[x]$ removes the integer part of its argument $x$.
The third part of our model is a $K_2$ order harmonic
\begin{eqnarray}
g_2(t)\!\!=\!\!
g_2(t,\bar{\beta}_2) \!\!=\!\!
\sum_{k=1}^{K_2} \! D_k \cos{(k 2 \pi f_2 t)} \!+ \! E_k \sin{(k 2 \pi f_2 t)},
\label{gtwo}
\end{eqnarray}
where $\bar{\beta}_2=$ $[D_1, ..., D_{K_2}, E_1, ..., E_{K_2},f_2]$.
The units are the same as in $g_1(t)$.
% "It fulfills $f_2 < f_1$." 
% This will be mentioned when periodogram
% is calculated in two dimensions, i.e. the limits in tested frequencies.
The following parameters 
\begin{description}
\item $P_2=$ period $=f_2^{-1}$
\item $A_2=$ peak to peak amplitude
\item $t_{\mathrm{g2,min,1}}=$ epoch of primary minimum
\item $t_{\mathrm{g2,min,2}}=$ epoch of secondary minimum (if present)
\end{description}
of the $g_2(t)$ function
are determined.
The units are the same as 
those used for the $g_1(t)$ function.
The $g_2(t)$ phases are
\begin{eqnarray}
\phi_2={\mathrm{FRAC}}[(t-t_1)/P_2].
\label{phasetwo}
\end{eqnarray}
The $g_1(t)$ and $g_2(t)$
functions are the ``real light curves''.
The duplication of their parameter explanations is required,
because we can show that they are directly connected
to the starspots on the surface of FK Com
(Sect. \ref{FKsdr}).

Our general model is the sum of the above three parts
\begin{eqnarray}
g(t)=g(t,\bar{\beta})=m_0(t)+g_1(t)+g_2(t).
\label{fullmodel}
\end{eqnarray}
This nonlinear model has
\begin{eqnarray}
p=(K_0+1)+(2K_1+1)+(2K_2+1)
\label{freeparam}
\end{eqnarray}
free parameters $\bar{\beta}=[\bar{\beta}_0,\bar{\beta}_1,\bar{\beta}_2]$.
The $g_1(t)$ and $g_2(t)$ functions have unique phases
(Eqs. \ref{phaseone} and \ref{phasetwo}),
but no such phases exist for the constantly changing
$g(t)$ function.

We use $(f_0 t)^k$ argument in Eq. \ref{gzero}
to ensure that every $k \ge 1$ term
of $m_0(t)$ undergoes a change of $M_k$ during $\Delta T$.
This $M_1$, ... $M_{K_0}$ scale
is the same as the scale for all $2(K_1+K_2)$
amplitudes of trigonometric functions
of $g_1(t)$ and $g_2(t)$.

The residuals of the model
\begin{eqnarray}
\epsilon_i=y_i-g(t_i)
\label{modelepsilon} 
\end{eqnarray}
give the sum of squared residuals
\begin{eqnarray}
R=\sum_{i=1}^n\epsilon_i^2.
\label{squaredresiduals}
\end{eqnarray}
We will later normalize $R$ with $1/n$ in 
Eqs. \ref{twoz} and \ref{onez}.
The mean for the absolute values of all residuals is
\begin{eqnarray}
|\epsilon|={1 \over n} \sum_{i=1}^n |\epsilon_i|.
\label{meanresiduals}
\end{eqnarray}
\citet{Jet93} detected flares in the light curves of FK Com.
These flares could be identified from the standard Johnson
$U$ and $B$ magnitudes of FK Com, but not from the $V$ magnitudes
studied here \citep[][see Figs. 1-3]{Jet93}.
Even a single flare can significantly increase
the $R$ estimate because its residual is 
squared. Its effect on $|\epsilon|$ is
smaller, and therefore we also use this parameter
to estimate the light curve models for FK Com.

\subsection{Complex and simple models}
\label{Particular}

We study two special cases of the
general model (Eq. \ref{fullmodel}),
the complex and the simple model.
Our complex model 
\begin{eqnarray}
g_{\mathrm{C}}(t)=m_0(t)+g_1(t)+g_2(t)
\end{eqnarray}
has $K_0=2$, $K_1=2$, $K_2=2$ and
$p_{\mathrm{C}}=13$ free parameters
(Eq. \ref{freeparam}).
The complex model parameters $R$ 
(Eq. \ref{squaredresiduals})
and $|\epsilon|$
(Eq. \ref{meanresiduals})
 are denoted with 
$R_{\mathrm{C}}$ and
$|\epsilon|_{\mathrm{C}}$.
Our simple model 
\begin{eqnarray}
g_{\mathrm{S}}(t)=m_0(t)+g_1(t)
\end{eqnarray}
has $K_0=2$, $K_1=2$ and 
$p_{\mathrm{S}}=8$ free parameters (Eq. \ref{freeparam}).
There is no third part $g_2(t)$.
We use the notations $R_{\mathrm{S}}$ and
$|\epsilon|_{\mathrm{S}}$ for the parameters
$R$ 
(Eq. \ref{squaredresiduals})
and $|\epsilon|$ 
(Eq. \ref{meanresiduals}).
The $g_{\mathrm{C}}(t)$ and 
$g_{\mathrm{S}}(t)$ models are nested.
They are
the same model,  if
\begin{description}

\item \firstcase: $f_1=f_2$ in Eqs. \ref{gone} and \ref{gtwo}
\item \secondcase: $B_1=B_2=C_1=C_2=0$ in Eq. \ref{gone}
\item \thirdcase: $D_1=D_2=E_1=E_2=0$ in Eq. \ref{gtwo}.

\end{description}

\noindent
These ``models approach'' each other,
$g_{\mathrm{C}}(t) \leftrightarrow g_{\mathrm{S}}(t)$,
when
\begin{eqnarray}
 f_1-f_2  \rightarrow  0 & \Rightarrow & g_1(t) \rightarrow g_2(t) \label{1case} \\ 
 A_1/A_2  \rightarrow  0 & \Rightarrow & g_1(t) \rightarrow 0      \label{2case} \\
 A_2/A_1  \rightarrow  0 & \Rightarrow & g_2(t) \rightarrow 0. \label{3case} 
\end{eqnarray}
\noindent

\noindent
If the \JHLhyp ~is true, the observed light curve
contains one or two periodic signals.
If {\it both} signals can be detected,
we use the complex model. 
The simple model is used,
if only {\it one} signal can be detected.

\section{Period finding method}
\label{Method}

\subsection{Finding two periods 
with $g_{\mathrm{C}}(t)$-method}
\label{Cmethod}

For the
$g_{\mathrm{C}}(t)$ model periods $P_1$ and $P_2$,
the mid points for the tested frequency ranges are 
$f_{\mathrm{mid}}\!=\!1/P_1$ for $f_1$ 
and
$f_{\mathrm{mid}}\!=\!1/P_2$ for $f_2$.
The lowest and highest tested frequencies are
\begin{eqnarray}
f_{\mathrm{min}} =  (1-a)f_{\mathrm{mid}}, & &  
f_{\mathrm{max}} =  (1+a)f_{\mathrm{mid}},
\label{fsimraja}
\end{eqnarray}
where $a=0.03 \!\equiv \!\pm 3$\%.
Our step in tested frequencies is
\begin{eqnarray}
f_{\mathrm{step}}=f_0/{\mathrm{OFAC}},
\label{fstep}
\end{eqnarray}
\noindent
where ${\mathrm{OFAC}}=60$ 
is the over-filling factor. 
For both $f_1$ and $f_2$,
we test all integer multiples of $f_{\mathrm{step}}$ 
between
$f_{\mathrm{min}}$ and $f_{\mathrm{max}}$.
We have performed numerous tests 
to confirm that this high ${\mathrm{OFAC}}$ value
gives reasonably accurate values 
for the correct periods $P_1$ and $P_2$,
as well as for their errors.
For a tested $f_1$ and $f_2$ pair,
the $g_{\mathrm{C}}(t)$-method periodogram is
\begin{eqnarray}
z_{\mathrm{C}}(f_1,f_2)=
%\sqrt{
%\left(
%{ {1} \over {n} }
%\right)
%\sum_{i=1}^{n} 
%\epsilon_i^2
%}=
\sqrt{
R_{\mathrm{C}}/n.
}
\label{twoz} 
\end{eqnarray}
The units are
$[z_{\mathrm{C}}]={\mathrm{mag}}$.
When the tested $f_1$ and $f_2$ frequencies
are fixed,
the $g_{\mathrm{C}}(t)$ model has only eleven free parameters.
These are the $M_0, M_1, M_2$ coefficients of Eq. \ref{gzero},
the $B_1,B_2,C_1, C_2$ amplitudes of Eq. \ref{gone},
and
$D_1,D_2,E_1, E_2$ amplitudes of Eq. \ref{gtwo}.
This model is linear, and the least squares fit solutions for
these eleven free parameters are unambiguous.
The frequencies 
$1/P_1$ and $1/P_2$ 
of the best periods 
are at the {\it global} 
minimum of the $z_{\mathrm{C}}(f_1,f_2)$
periodogram.

We determine the errors of the  $g_{\mathrm{C}}(t)$ model parameters 
 with the bootstrap procedure
\citep{Efr86,Jet99}.
The tested $f_1$ and $f_2$ frequency pairs connected 
to the global $z_{\mathrm{C}}(f_1,f_2)$ minimum
are within 
\begin{eqnarray}
[(f_1-P_1^{-1})/(f_0/2)]^2
+
[(f_2-P_2^{-1})/(f_0/2)]^2
=1,
\label{redcircle} 
\end{eqnarray}
where $P_1$ and $P_2$ are the pair of best periods for the 
{\it original data.}
During each bootstrap round,
we select a random sample $\bar{\epsilon}^*$ from
the residuals $\bar{\epsilon}$ of the best model $g_i=g_{\mathrm{C}}(t_i)$
for the {\it original} data $\bar{y}$ 
(Eq. \ref{modelepsilon}).
Any particular $\epsilon_i$ value of all $n$ values can enter into this
sample $\bar{\epsilon}^*$ as many times as the random selection 
happens to favour it.
This random sample of residuals gives the sample of {\it artificial} data
\begin{eqnarray}
y_i^*=g_i+\epsilon_i^*
\label{bootsample}
\end{eqnarray} 
during each bootstrap round.
The best $g_{\mathrm{C}}(t)$ model for this artificial $\bar{y}^*$ 
data sample gives one estimate for the model parameters
$P_1$, $P_2$, $A_1$, $A_2$, 
$t_{\mathrm{g1,min,1}}$, $t_{\mathrm{g1,min,2}}$, $t_{\mathrm{g2,min,1}}$
and $t_{\mathrm{g2,min,2}}$.
Our error estimate for each particular 
model parameter is the standard deviation
of all estimates obtained for this parameter 
in all bootstrap rounds.

\subsection{Finding one period with $g_{\mathrm{S}}(t)$-method}
\label{Smethod} 

The tested $f_1$ range 
of the $g_{\mathrm{S}}(t)$-method
is the same (Eq. \ref{fsimraja}).
For any fixed tested $f_1$ value,
the $g_{\mathrm{S}}(t)$ model becomes linear,
and the solutions for the remaining seven
free parameters are unambiguous.
The $g_{\mathrm{S}}(t)$-method periodogram is
\begin{eqnarray}
z_{\mathrm{S}}(f_1)=
%\sqrt{
%\left(
%{ {1} \over {n} }
%\right)
%\sum_{i=1}^{n} 
%\epsilon_i^2
%}
%=
\sqrt{
R_{\mathrm{S}}/n
}.
\label{onez} 
\end{eqnarray}
The units are $[z_{\mathrm{S}}]={\mathrm{mag}}$.
A similar bootstrap
procedure as in Sect. \ref{Cmethod}.
gives 
the $P_1$, $A_1$, $t_{\mathrm{g1,min,1}}$
and $t_{\mathrm{g1,min,2}}$ errors.

\section{Results for FK Com photometry}
\label{sectallresults}

\subsection{Results for three particular segments}

\subsubsection{\Data{1}{9}}
\label{SectTel1Seg9}

% 22.03.2019
%/method/progs$ cp Tel1Seg9a.eps /home/jetsu/method/texts/
%/method/progs$ cp Tel1Seg9a.eps /home/jetsu/method/progs/valmis2/
\begin{figure} 
\begin{center}
\resizebox{7.2cm}{!}{
\includegraphics{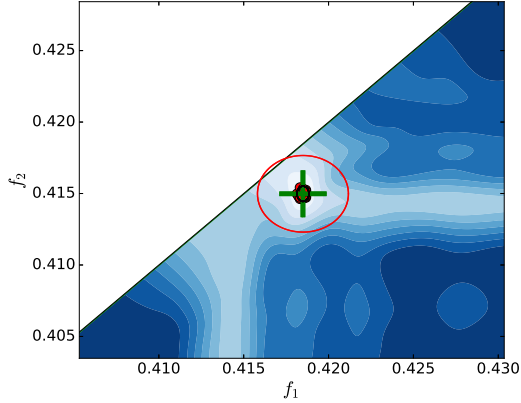}}
\end{center}
\caption{$z_{\mathrm{C}}(f_1,f_2)$ periodogram for \Data{1}{9}.
Ten different shades are 
between white minimum and dark blue maximum levels. 
Diagonal black line is $f_1=f_2$.
Due to $z_{\mathrm{C}}(f_1,f_2)=z_{\mathrm{C}}(f_2,f_1)$ symmetry,
periodogram is shown only  
for triangular area $f_2 <(1-b)f_1$, where $b=0.01$
eliminates instability of Eq. \ref{1case}.
Large green cross denotes 
$1/P_1$ and $1/P_2$ of the best
$g_{\mathrm{C}}(t)$ model.
Red circle (Eq. \ref{redcircle}) 
and black ellipse (Eq. \ref{blackellipse})
surround this location.
Small  red circles show the
best models in 50 bootstrap samples.
}
\label{zcfig1}
\end{figure}

% 22.03.2019
%/method/progs$ cp Tel1Seg9b.eps /home/jetsu/method/texts/
%/method/progs$ cp Tel1Seg9b.eps /home/jetsu/method/progs/valmis2/
\begin{figure} 
\begin{center}
\resizebox{7.2cm}{!}{
\includegraphics{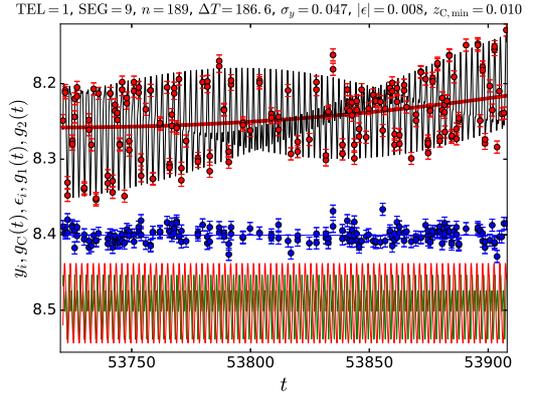}}
\end{center}
\caption{Best $g_{\mathrm{C}}(t)$ model for \Data{1}{9}.
Red  circles are $n=189$ observations made during
$\Delta T=186.^{\mathrm{d}}7$ and having $\sigma_y=0.^{\mathrm{m}}047$.
Thin black and thick red continuous lines denote 
$g_{\mathrm{C}}(t)$ and $m_0(t)$, respectively.  
Blue  circles are residuals
offset to $V=8.4$ level.
They give $|\epsilon|_{\mathrm{C}}=0.^{\mathrm{m}}008$
and $z_{\mathrm{C,min}}=0.^{\mathrm{m}}010$.
Green and orange continuous lines are $g_1(t)$ and
$g_2(t)$ curves offset to $V=8.5$ level.}
\label{gc1}
\end{figure}

Our tested frequency range is
  $a=0.03$, $f_{\mathrm{mid}}\!=\!1/P_1$ for $f_1$
  and       $f_{\mathrm{mid}}\!=\!1/P_2$ for $f_2$
  in Eq. \ref{fsimraja}, where
  $P_1\!=\!2.^{\mathrm{d}}39321$
  and
  $P_2\!=\!2.^{\mathrm{d}}40413$
  (Eqs. \ref{epheone} and \ref{ephetwo}).
The $z_{\mathrm{C}}(f_1,f_2)$ periodogram for \Data{1}{9}
is shown in Fig. \ref{zcfig1}.
We give the $g_{\mathrm{C}}(t)$ model parameters 
in Table \ref{Cresults}.
The periodogram minimum  
$z_{\mathrm{C,min}}=0.^{\mathrm{m}}010$
and
$|\epsilon|_{\mathrm{C}}=0.^{\mathrm{m}}008$ % OK: 15.04.2019
(Eq. \ref{meanresiduals})
are close to the $0.^{\mathrm{m}}008$ accuracy of the data.
This is an excellent result for $n=189$ observations
during six months.
The $g_1(t)$ and $g_2(t)$ signals for the periods
$P_1=2.^{\mathrm{d}}3895\pm 0.^{\mathrm{d}}0005$
and
$P_2=2.^{\mathrm{d}}4098\pm 0.^{\mathrm{d}}0011$
have high amplitudes,
$A_1=0.^{\mathrm{m}}085 \pm 0.^{\mathrm{m}}005$ 
and
$A_2=0.^{\mathrm{m}}106 \pm 0.^{\mathrm{m}}004$. % OK 15.04.2019
The small black ellipse 
in Fig. \ref{zcfig1} is
\begin{eqnarray}
[(f_1-P_1^{-1})/(3 \sigma_{f_1})]^2
+
[(f_2-P_2^{-1})/(3 \sigma_{f_2}])^2
=1,
\label{blackellipse}
\end{eqnarray}
\noindent
where $\sigma_{f_1}$ and $\sigma_{f_2}$ are the errors 
for $1/P_1$ and $1/P_2$.
For \Data{1}{9}, 
this black ellipse does not
intersect the $f_1=f_2$ line.
The best $g_{\mathrm{C}}(t)$  model is shown in Fig. \ref{gc1}.
The amplitude of this thin black curve decreases 
first, and then increases.
The thick red $m_0(t)$ curve shows a minor increase
in the mean brightness. 
The residuals are stable 
(Fig. \ref{gc1}: blue  circles).
The green lower amplitude $g_1(t)$ curve has two
minima
(Table \ref{Cresults}: 
$t_{\mathrm{g1,min,1}}$ and $t_{\mathrm{g1,min,2}}$).
The orange higher amplitude $g_2(t)$ curve has one
minimum
(Table \ref{Cresults}: $t_{\mathrm{g2,min,1}}$).

% 22.03.2019
%/method/progs$ cp Tel2Seg1a.eps /home/jetsu/method/texts/
%/method/progs$ cp Tel2Seg1a.eps /home/jetsu/method/progs/valmis2/
\begin{figure} 
\begin{center}
\resizebox{7.2cm}{!}{
\includegraphics{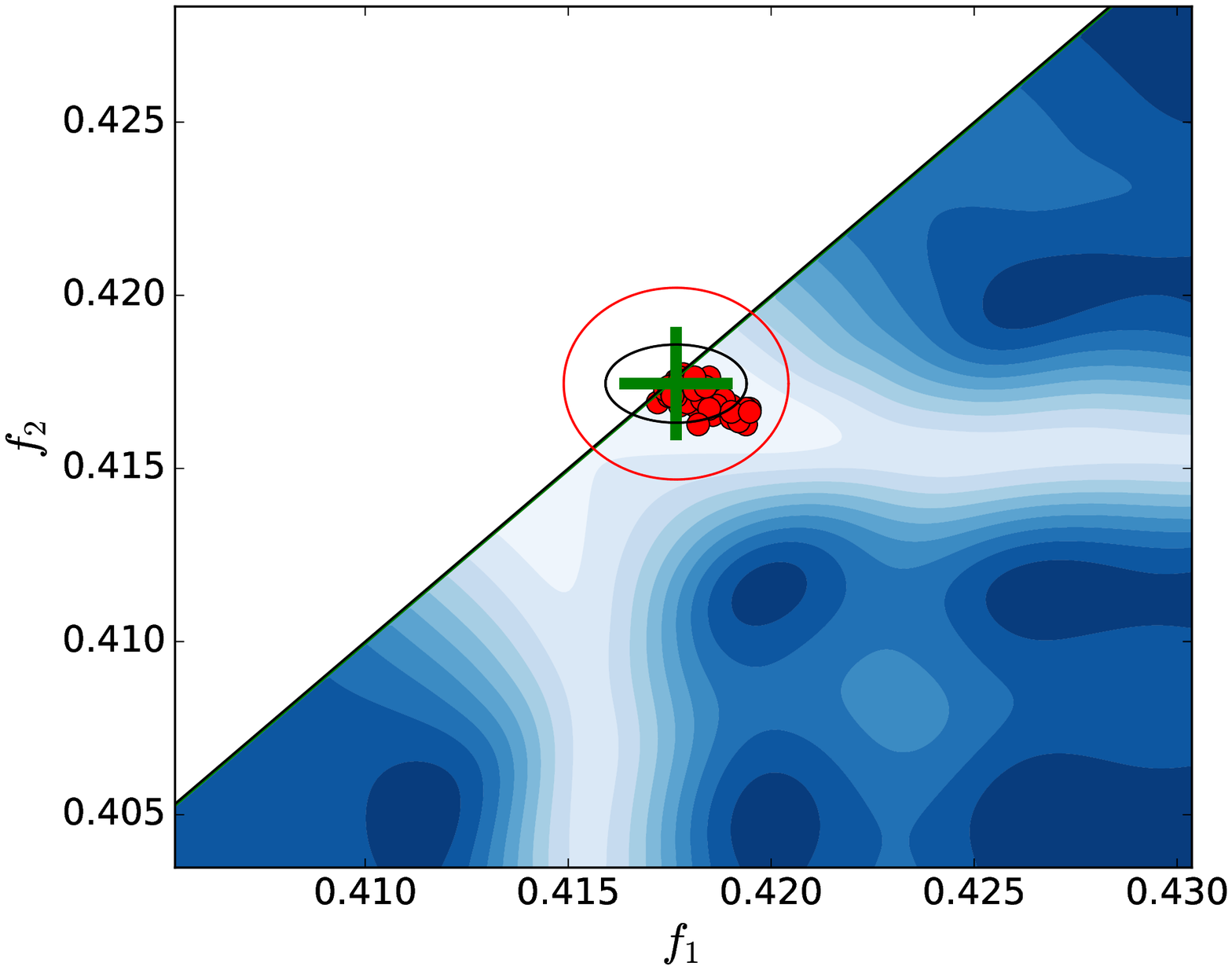}}
\end{center}
\caption{$z_{\mathrm{C}}(f_1,f_2)$ for \Data{2}{1}.
Otherwise as in Fig. \ref{zcfig1}}
\label{zcfig2}
\end{figure}

% 22.03.2019
%/method/progs$ cp Tel2Seg1b.eps /home/jetsu/method/texts/
%/method/progs$ cp Tel2Seg1b.eps /home/jetsu/method/progs/valmis2/
\begin{figure} 
\begin{center}
\resizebox{7.2cm}{!}{
\includegraphics{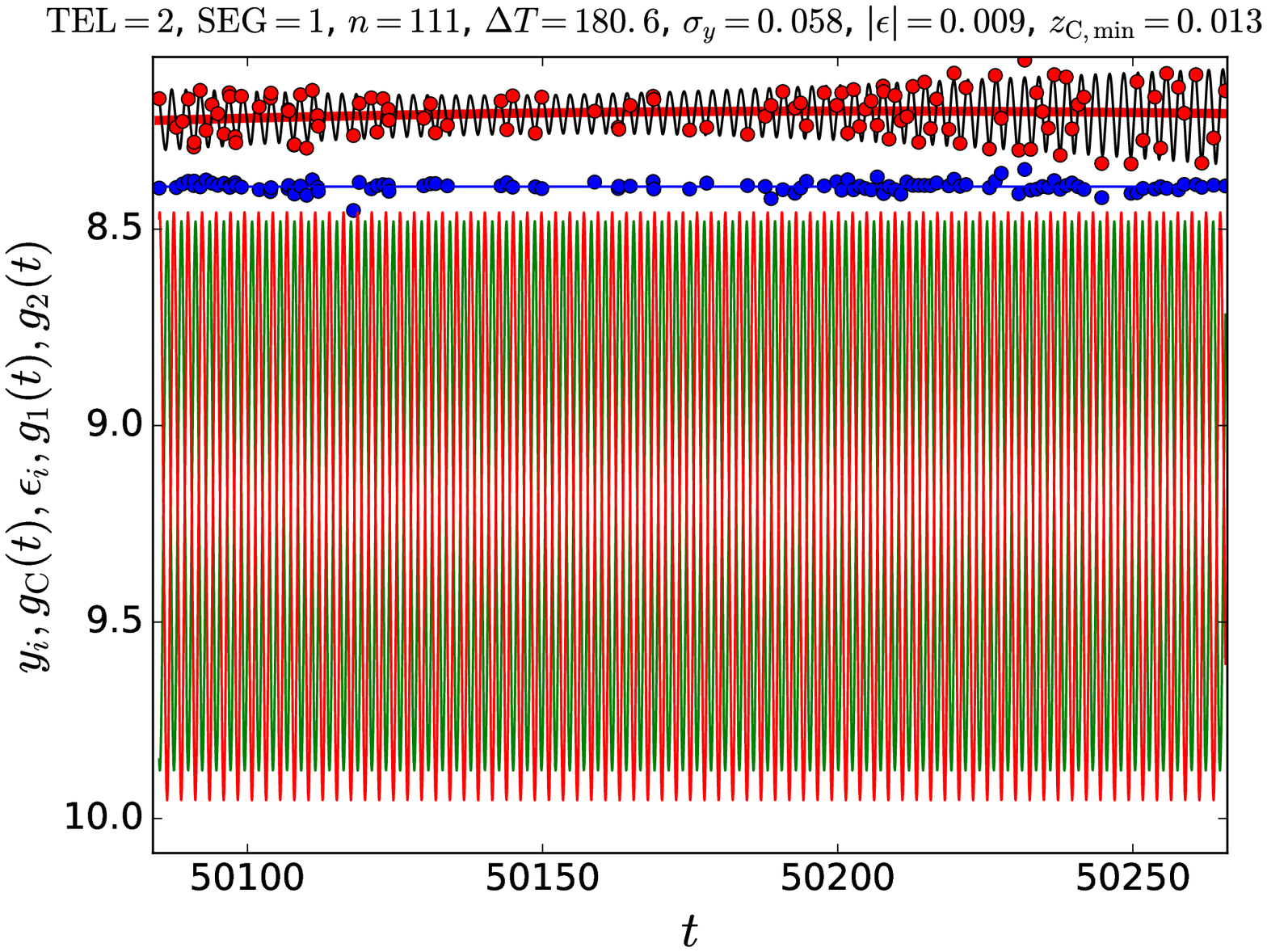}}
\end{center}
\caption{Best $g_{\mathrm{C}}(t)$ model for \Data{2}{1}.
Otherwise as in Fig. \ref{gc1}}
\label{gc2}
\end{figure}

\subsubsection{\Data{2}{1}}
\label{SectTel2Seg1}

The best  $g_{\mathrm{C}}(t)$ model for \Data{2}{1} is so close to
$f_1=f_2$ that
the black ellipse of Eq. \ref{blackellipse} intersects 
this diagonal
(Fig. \ref{zcfig2}).
We refer to this event as ``intersects $f_1=f_2$''.
The dramatic consequences are displayed in
Fig. \ref{gc2}, where the amplitudes of the $g_1(t)$ and $g_2(t)$
curves of the complex model are totally unrealistic.  
``Amplitude dispersion''
can occur in two different ways:
the $A_1$ and $A_2$ estimates and/or
the $\sigma_{A_1}$ and  $\sigma_{A_2}$ estimates are large.
Due to $f_1-f_2 \rightarrow 0$ (Eq. \ref{1case}),
the complex and the simple models approach,
$g_{\mathrm{C}}(t) \leftrightarrow g_{\mathrm{S}}(t)$.
The former becomes unstable close to $f_1=f_2$,
because the linear least squares routine
tries to fit the same function twice.
It solves this problem by summing up
the difference of two large
nearly opposite-phase oscillations.
This does not mean
that our model is flawed. 
This confirms that 
it is impossible to model one constant period signal
with two constant period signals.
The Doppler images confirm that only one signal dominated
when we correctly dismiss the complex model 
(Sect. \ref{FKsdr}:
\Data{2}{1}, \Data{2}{2}, \Data{2}{3} and \Data{1}{13}).  

Table \ref{Cresults} gives
the results for all segments,
where ``intersects $f_1=f_2$'' and ``amplitude dispersion'' do not occur.
We use the complex $g_{\mathrm{C}}(t)$ model 
in these segments.
If ``intersects $f_1=f_2$'' or ``amplitude dispersion''
does occur,
we use the simple $g_{\mathrm{S}}(t)$ model.
Table \ref{Sresults} gives those results.
``Intersects $f_1=f_2$''
always causes ``amplitude dispersion'',
but the latter occurs once without the former 
(Table \ref{Sresults}: \Data{1}{10}).

For \Data{2}{1},
the $z_{\mathrm{C}}(f_1,f_2)$ periodogram shows
a horizontal and a vertical white arm
(Fig. \ref{zcfig2}).
This means that if 
$f_1$ has a suitable value, many $f_2$ values give 
a reasonable model (vertical arm),
or vice versa (horizontal arm).
These arms indicate
that the simple $g_{\mathrm{S}}(t)$ model 
may be the best model for \Data{2}{1}.
Note that the $z_{\mathrm{C}}(f_1,f_2)$ periodogram
for \Data{1}{9} has similar, but weaker,
white arms (Fig. \ref{zcfig1}).

\begin{figure} 
\begin{center}
\resizebox{7.2cm}{!}{
\includegraphics{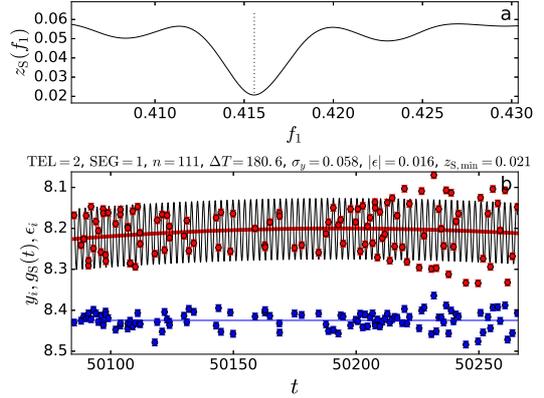}}
\end{center}
\caption{(a) $z_{\mathrm{S}}(f_1)$ periodogram (continuous line) and
best period $1/P_1$ (dotted line) for \Data{2}{1}.
(b) Best $g_{\mathrm{S}}(t)$ model. 
Red circles are $n=111$ observations
having $\sigma_y=0.^{\mathrm{m}}058$ and 
$\Delta T=180.^{\mathrm{d}}6$.
We denote $g_{\mathrm{S}}(t)$ and $m_0(t)$ curves
with thin black and thick red continuous lines.
Residuals are
offset to $V=8.43$ level (blue  circles).
They give $|\epsilon|_{\mathrm{S}}=0.^{\mathrm{m}}016$
and $z_{\mathrm{S,min}}=0.^{\mathrm{m}}021$.}
\label{gs1}
\end{figure}

The $g_{\mathrm{S}}(t)$-method periodogram $z_{\mathrm{S}}(f_1)$ 
(Eq. \ref{onez}) for \Data{2}{1} is shown in
Fig. \ref{gs1}a. The vertical dotted
line denotes the 
best period 
$P_1=2.^{\mathrm{d}}4063 \pm 0.^{\mathrm{d}}0013$. % OK 15.04.2019
The simple $g_{\mathrm{S}}(t)$ model for \Data{2}{1} is
shown in Fig. \ref{gs1}b. 
The  $|\epsilon|_{\mathrm{S}}=0.^{\mathrm{m}}016$ 
and $z_{\mathrm{S,min}}=0.^{\mathrm{m}}021$ % OK 15.04.2019
estimates
are about two
times larger than the $0.^{\mathrm{m}}008$ accuracy of data.
The level of residuals is stable,
because the $m_0(t)$ part of the $g_{\mathrm{S}}(t)$ model
can follow the light curve mean changes.
The scatter of the residuals is
large in the end of the segment,
because the $g_1(t)$ part of
the simple $g_{\mathrm{S}}(t)$
model can not reproduce amplitude changes.
Hence,
the complex model $|\epsilon|_{\mathrm{C}}$ and $z_{\mathrm{C,min}}$
values in Table \ref{Cresults} are 
systematically smaller than the simple model
$|\epsilon|_{\mathrm{S}}$ and $z_{\mathrm{S,min}}$
values in Table \ref{Sresults}.
Regardless of these limitations, 
this simple model must be used for \Data{2}{1},
because the simple and the complex model
approach each other, $g_{\mathrm{C}}(t) \leftrightarrow g_{\mathrm{S}}(t)$.

% 22.03.2019
%/method/progs$ cp Tel2Seg1c.eps /home/jetsu/method/texts/
%/method/progs$ cp Tel2Seg1c.eps /home/jetsu/method/progs/valmis2/

% 22.03.2019
%/method/progs$ cp Tel1Seg12a.eps /home/jetsu/method/texts/
%/method/progs$ cp Tel1Seg12a.eps /home/jetsu/method/progs/valmis2/
\begin{figure}
\begin{center} 
\resizebox{7.2cm}{!}{\includegraphics{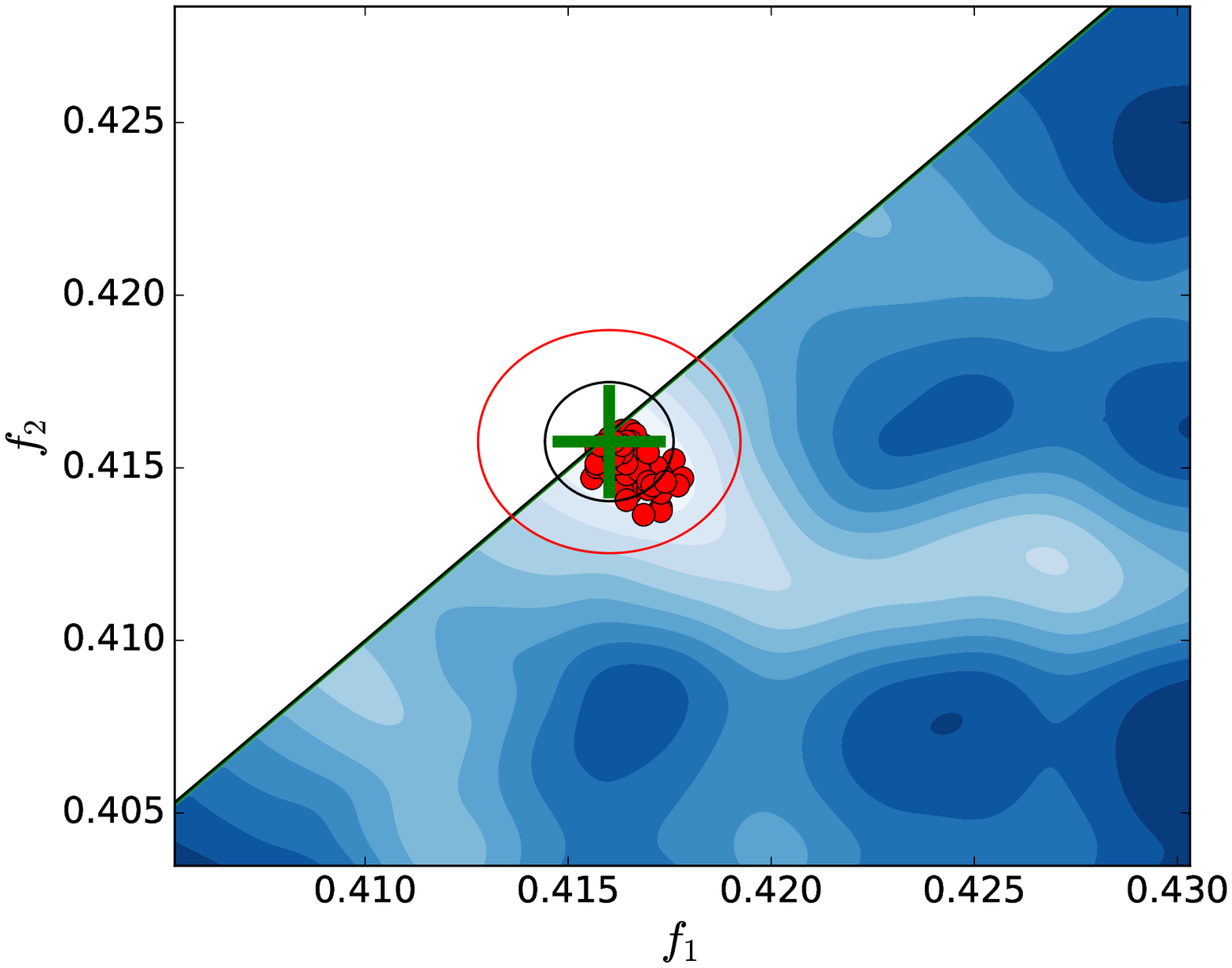}}
\end{center}
\caption{$z_{\mathrm{C}}(f_1,f_2)$ for \Data{1}{12}.
Otherwise as in Fig. \ref{zcfig1}}
\label{zcfig3}
\end{figure}

\subsubsection{\Data{1}{12}}
\label{SectTel1Seg12}

The $z_{\mathrm{C}}(f_1,f_2)$ periodogram for \Data{1}{12}
is ``pathological''
(Fig. \ref{zcfig3}).
The simple and the complex model approach each other,
because the black ellipse of Eq. \ref{blackellipse}
``intersects $f_1=f_2$''.
Due to ``amplitude dispersion'', 
the unrealistic $A_1$ and $A_2$ amplitudes reach 
$1.^{\mathrm{m}}5$.
However, the $z_{\mathrm{C}}(f_1,f_2)$ periodogram
does not show white arms similar
to those seen in Figs. \ref{zcfig1} and \ref{zcfig2}.
Except for this particular segment \Data{1}{12}, 
the $z_{\mathrm{C}}(f_1,f_2)$ periodograms
of {\it all other} segments show such white arms.
The absence of these arms indicates that the
simple $g_{\mathrm{S}}(t)$ model,
having
$|\epsilon|_{\mathrm{S}}=0.^{\mathrm{m}}020$ 
and $z_{\mathrm{S,min}}=0.^{\mathrm{m}}025$,
is neither a good model for \Data{1}{12}.
In this segment,
the $y_i$ variations first decrease
and approach zero 
just before there is a gap in the observations.
After this gap, these variations increase rapidly.
If the photometry before and after
this gap is plotted with 
the period
$P_1=2.^{\mathrm{d}}4049$, there is a phase shift
of half a rotation.
Such a shift is predicted by Eq. \ref{abruptphase}
(see Sect. \ref{explanation}).
Our modelling probably fails, because there are
some outliers close to the gap
when the $y_i$ variation approaches zero.
Since the complex and
the simple models fail for \Data{1}{12},
we use only the mean brightness $M$ (Table \ref{Sresults}).

% 22.03.2019
% figlatex.py produces figlatex.eps 
% ~/method/progs$ cp figlatex.eps /home/jetsu/method/texts/
% %%BoundingBox: 130 280 480 612
\begin{figure*}
\begin{center} 
\resizebox{9.0cm}{!}{\includegraphics{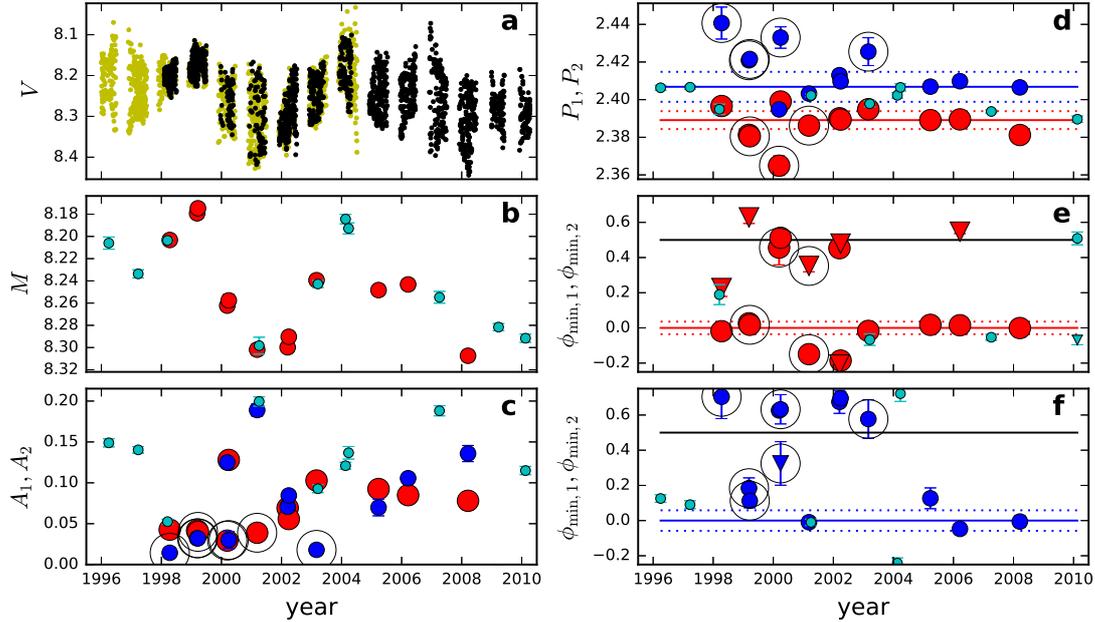}} 
\end{center}
\caption{Results for FK Com.
(a) Data: 
TEL=1 (black dots) and 
TEL=2 (yellow dots).
(b) Mean: 
$M$ 
(Table \ref{Cresults}: red  circles, 
 Table \ref{Sresults}: cyan  circles).
(c) Amplitude: 
$A_1$ 
(Table \ref{Cresults}: red  circles, 
 Table \ref{Sresults}: cyan  circles)
and
$A_2$ 
(Table \ref{Cresults}: blue  circles).
Large black circles highlight amplitudes lower than 
$0.^{\mathrm{m}}041$. % OK 15.04.2019 figlatex.py, raja=0.041
Same segments 
are highlighted in "def".
(d) Periods: 
$P_1$ 
(Table \ref{Cresults}: red  circles, 
 Table \ref{Sresults}: cyan  circles)
and
$P_2$ 
(Table \ref{Cresults}: blue  circles).
Horizontal lines show
levels of 
Eqs. \ref{Pwone} ($P_{\mathrm{w,1}}$: red continuous and dotted lines)
and 
\ref{Pwtwo} ($P_{\mathrm{w,2}}$: blue continuous and dotted lines).
(e) Minimum phases with Eq. \ref{epheone} ephemeris:
$t_{\mathrm{g1,min,1}}$
(Table \ref{Cresults}: red  circles, 
 Table \ref{Sresults}: cyan  circles)
and
$t_{\mathrm{g1,min,2}}$
(Table \ref{Cresults}: red  triangles, 
 Table \ref{Sresults}: cyan  triangles).
Levels are
% figlatex.py prints: mid phase error   =  0.035237285283
$\phi=0.000 \pm 0.035$
(red continuous and dotted lines) 
and $\phi=0.5$
(black continuous line).
(f) Minimum phases with Eq. \ref{ephetwo} ephemeris:
$t_{\mathrm{g1,min,1}}$
(Table \ref{Cresults}: blue  circles, 
 Table \ref{Sresults}: cyan  circles)
and
$t_{\mathrm{g1,min,2}}$
(Table \ref{Cresults}: blue  triangles). 
Levels are
% figlatex.py prints: mid phase error   =  0.0498000603994
$\phi=0.000 \pm 0.050$
(blue continuous and dotted lines) 
and $\phi=0.5$
(black continuous line).}
\label{fkresults}
\end{figure*}

\subsection{Results for all segments}
\label{sectallsegments}

The variations in Fig. \ref{fkresults}a
agree for overlapping
the TEL=1 (black dots) and 
TEL=2 data (yellow dots).
We use the mean ($M \pm \sigma_M$)
for all 
observations in each segment,
not the $M_0$ of $m_0(t)$ curve (Eq. \ref{gzero}).
This requires no model,
which is missing for \Data{1}{12}
(Sect. \ref{SectTel1Seg12}).
The $M$ changes 
are quasi-periodic (Fig. \ref{fkresults}b). 
The maxima are five (1999 and 2004)
and the minima seven (2001 and 2008) years apart.
The $A_1$ and $A_2$ amplitudes of overlapping
segments between 
1998 and 2004 agree, and it is difficult 
to separate some symbols from each other (Fig. \ref{fkresults}c).
We solve this ``problem''
by denoting the complex model amplitudes 
with larger red and blue  circles,
and the simple model amplitudes 
with smaller cyan  circles.
We use
these sizes and colours also
in Figs. \ref{fkresults}d-e.
Large circles
highlight amplitudes below $0.^{\mathrm{m}}041$. % OK 15.04.2019
These segments 
are also highlighted in
Figs. \ref{fkresults}d-f.
%\comment{{{\color{blue} X?}}}{
%The largest amplitudes in 2001 and 2007
%nearly coincide with the $M$ minima,
%but there is no clear
%linear correlation between these parameters.}

The complex models give $P_i\pm \sigma_{P_i}$
$(n=12)$ estimates for $P_1$ and $P_2$ 
(Table \ref{Cresults}). 
We compute their weighted mean
\begin{eqnarray}
P_{\mathrm{w}}= 
{
             {\sum_{i=1}^n w_i P_i}
             \over
             {\sum_{i=1}^n w_i}
}
\label{Pw}
\end{eqnarray}
and their weighted standard deviation
\begin{eqnarray}
\Delta P_{\mathrm{w}}= 
\sqrt{
             {
             {\sum_{i=1}^n w_i (P_i -P_{\mathrm{w}})^2}
             \over
             {\sum_{i=1}^n w_i}
             }
      },
\label{ePw}
\end{eqnarray}
\noindent
where $w_i=\sigma_{P_i}^{-2}$.  
% /home7jetsu/method/progs/figlatex.py prints
% p1w = 2.38919134673 +/- 0.00576601885598 n= 12
% p2w = 2.40659131516 +/- 0.00752397569987 n= 12
The complex model $P_1$ and $P_2$ periods 
give
\begin{eqnarray}
P_{\mathrm{w,1}} \pm \Delta P_{\mathrm{w,1}} & = 
& 2.^{\mathrm{d}}3892 \pm 0.^{\mathrm{d}}0057 ~(n=12) 
\label{Pwone} \\
P_{\mathrm{w,2}} \pm \Delta P_{\mathrm{w,2}} & = & 
2.^{\mathrm{d}}4066 \pm 0.^{\mathrm{d}}0075 ~(n=12).
\label{Pwtwo} 
\end{eqnarray}
\noindent
One rotation difference during
$\Delta T=5213^{\mathrm{d}}$ 
requires
$\delta_P=(P \Delta T )/(\Delta T \pm P)=\pm0.^{\mathrm{d}}0011$
period change. Thus,
$|P_{\mathrm{w,1}}-P_{\mathrm{w,2}}|=15.8  \delta_P$ indicates 
that these levels are separated.  % OK 15.04.2019

The red continuous and dotted lines
show the $P_{\mathrm{w,1}} \pm \Delta P_{\mathrm{w,1}}$
level in Fig. \ref{fkresults}d.
The respective $P_{\mathrm{w,2}}$ levels are blue.
Three cyan circle symbols for the
simple model $P_1$ periods are close to
$P_{\mathrm{w,1}}$ 
(Table \ref{Sresults}: $P_{\mathrm{w}}=P_{\mathrm{w,1}}$),
and the remaining six cyan circles are
close to $P_{\mathrm{w,2}}$
(Table \ref{Sresults}: $P_{\mathrm{w}}=P_{\mathrm{w,2}}$).
Although these nine simple model $P_1$ periods 
{\it are not} used 
to compute the two complex model levels of
Eqs. \ref{Pwone} and \ref{Pwtwo},
they {\it all} coincide with these levels.
This strongly indicates that the periods of
FK Com have remained at two stable constant levels
during the full fourteen years time span of our data.
These separate $P_{\mathrm{w,1}}$ and $P_{\mathrm{w,2}}$
levels would stand out even more clearly,
if all low amplitude signals
were ignored (Fig. \ref{fkresults}d:
highlighted large open circles).

We use {\it only} the
complex model $P_1$ and $P_2$ periods
to compute $P_{\mathrm{w,1}}$ and $P_{\mathrm{w,2}}$ levels.
If these levels represent real physical structures,
the distributions of light curve minima
connected to these $P_{\mathrm{w,1}}$ and $P_{\mathrm{w,2}}$ 
levels should also be regular.
In other words,
the time points of these light curve
minimum epochs should be approximate
multiples of the above two periods,
which means that
their phase distributions should be regular.
These complex model minima in
Table \ref{Cresults} are

\begin{itemize}

\item[] $P_{\mathrm{w,1}}$: $g_1(t)$-epochs
$t_{\mathrm{g1,min,1}}$ and $t_{\mathrm{g1,min,2}}$
$(n=18)$ 
\item[] $P_{\mathrm{w,2}}$:
$g_2(t)$-epochs 
$t_{\mathrm{g2,min,1}}$ and  $t_{\mathrm{g2,min,2}}$ 
$(n=13)$.

\end{itemize}

\noindent
Both samples are small,
considering 
the fourteen years time span of data.
The $t_i$ time points
are a year apart.
Pairs of $t_i$ 
from overlapping segments
between 1998 and 2004 
are nearly simultaneous.
This regular $t_i$ spacing 
induces spurious periods.
Our period search must
rely on the most accurate $t_i \pm \sigma_{t_i}$ values.
The 
ordinary non-weighted Kuiper method test statistic $V_n$ 
can not utilize the information provided by the accuracy
$\sigma_{t_i}$ of time points \citep[][Eq. 20]{Jet96A}.
Hence, we apply the weighted Kuiper method
separately to the $g_1(t)$- and  $g_2(t)$-epoch samples
\citep[see][their Eq. 28]{Jet96A},
because this method can also utilize the 
$\sigma_{t_i}$ information.
The test statistic $V_n$ of 
this weighted Kuiper method utilizes the weights
\begin{eqnarray}
w_i=
{ 
{n ~ \sigma_{t_i}^{-2}} 
\over 
{\sum_{i=1}^n  \sigma_{t_i}^{-2}} 
}.
\label{wkuiper}
\end{eqnarray}
\noindent
As for earlier application examples,
\citet{Jet96} applied this 
weighted Kuiper method
to the light curve minima of four CABS. His
samples were at least three
times larger than our
$g_1(t)$- and  $g_2(t)$-epoch samples.
He obtained no significance estimates
for the detected periods,
because the scatter of
weights $w_i$ was large.
This scatter is larger in our samples,
e.g.  the two largest $w_i$ 
% figlatex.py prints these 
% FKCom_minima1.dat : weights  1 = 6.70722272467  cumulative  6.70722272467
% FKCom_minima1.dat : weights  2 = 3.37411594242  cumulative  10.0813386671
% FKCom_minima1.dat : weights  3 = 1.69201472514  cumulative  11.7733533922
contribute $56\% \equiv 6.7+3.4=10.1$ out of $n=18$ time points
in our $g_1(t)$-epoch sample.
This exceeds the $\equiv 44\%$ contribution
of the other 16 less accurate time points.
Our samples are small, sparse and regularly
spaced, and the scatter of weights is large.
Therefore,
we use the weighted Kuiper method only to
find the $V_n$ periodogram peaks 
closest to the $P_{\mathrm{w,1}}$ and $P_{\mathrm{w,2}}$ periods.
Those peaks are at
\begin{eqnarray}
P_{\mathrm{WK,1}} & = & 
2.^{\mathrm{d}}39321 \pm 0.^{\mathrm{d}}00036 ~(n=18) \label{Kone}\\
P_{\mathrm{WK,2}} & = & 
2.^{\mathrm{d}}40413 \pm 0.^{\mathrm{d}}00048 ~(n=13). \label{Ktwo}
\end{eqnarray}
\noindent
We determine the zero epochs for
these periodicities from
the most accurate $t_i$ values in 
our  samples.
Their weights are 
$37\% \equiv w_i=6.7$ out of $n=18$ 
and
$38 \% \equiv w_i=4.9$ out of $n=13$,
respectively.
%They give the largest amount of information.
Of all other epochs, we select those that are
closer than $\pm 0.20$ in phase to these most 
accurate $t_i$ values.
The weighted mean 
phase of all these epochs, and their weighted error,
gives us the
ephemerides 
\begin{eqnarray} 
{\mathrm{HJD~}}2453722.898\pm 0.084 & \!\!+\!\! 
& 2.39321\pm0.00036{\mathrm{E}} 
\label{epheone} \\
{\mathrm{HJD~}}2451874.437\pm 0.120 & \!\!+\!\! 
& 2.40413\pm0.00048{\mathrm{E}}
\label{ephetwo}
\end{eqnarray}
for the phases $\phi_1=0$
and $\phi_2=0$ of Eqs. \ref{phaseone} and \ref{phasetwo}
(Figs. \ref{fkresults}ef: continuous 
and dotted horizontal red and blue lines).
We also display % OK 15.04.2019
a continuous black horizontal line,
which is half a period
above the ephemeris zero epoch in Figs. \ref{fkresults}ef.
The majority of the time points $t_i \pm \sigma_{t_i}$
are close to the phases 0.0 or 0.5 
in Figs. \ref{fkresults}ef,
especially the more accurate ones.

% 21.03.2019
% produced by /home/jetsu/method/progs/twomlc.py
% ~/method/progs$ cp mlc.eps /home/jetsu/method/texts/
% ~/method/progs$ cp mlc.eps /home/jetsu/method/progs/valmis2/
\begin{figure}
\begin{center}
\resizebox{7.0cm}{!}{
\includegraphics{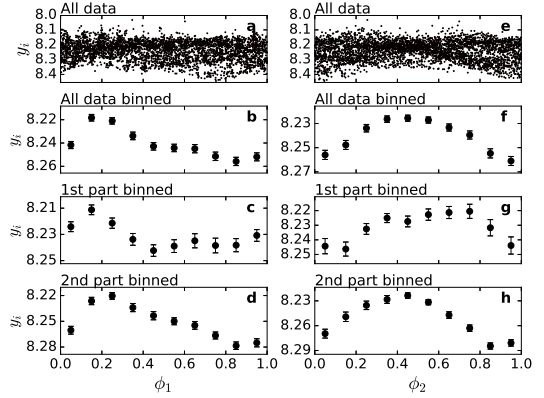}}
\end{center}
\caption{MLC of FK Com. (a) All data as function 
of phase $\phi_1$ computed from the
ephemeris of Eq. \ref{epheone}.
(b) All data in 10 bins. (c) 1st part binned data.
(d)  2nd part binned data.
(e-h) Phases $\phi_2$ computed from the
ephemeris of Eq. \ref{ephetwo}, otherwise as in ``a-d''.}
\label{figmlc}
\end{figure}

We use the complex model epochs to determine
the ephemerides of Eqs. \ref{epheone} and \ref{ephetwo},
but {\it not} the simple model epochs.
Regardless of this, these simple model
epochs follow these ephemerides.
Three $t_{\mathrm{g1,min,1}}$ (small cyan circles) 
and one $t_{\mathrm{g1,min,1}}$ (small cyan triangle) 
simple model epochs are connected to 
the complex model $g_1(t)$-epochs
(Table \ref{Sresults}).
Their phases are close to $\phi_1=0.0$ and 0.5
(Fig. \ref{fkresults}e). 
Six simple model $t_{\mathrm{g1,min,1}}$ (small cyan circles) 
epochs are connected 
to the complex model $g_2(t)$-epochs
(Table \ref{Sresults}). They are close to $\phi_2=0.0$ 
(Fig. \ref{fkresults}f). 
The $\phi_1$ distribution is bi-modal, 
because the epochs are close
to $\phi_1=0.0$ and 0.5. It
contains {\it seven} secondary minima 
(Fig. \ref{fkresults}d: six red and one cyan triangles).
The uni-modal $\phi_2$ distribution
is close to $\phi_2=0.0$,
and contains only {\it one} secondary minimum
(Fig. \ref{fkresults}e: blue triangle).
Both of these $\phi_1$ and $\phi_2$
phase distributions are indeed regular. This regularity
would become even more pronounced,
if the epochs of low amplitude signals
were removed 
(Fig. \ref{fkresults}ef: highlighted large circles).

We detect the two separate period levels
and the two separate minimum
epoch levels from {\it all} complex models.
{\it All} simple models fit to these levels.
{\it Both} models confirm the expected
period and epoch connection of
two structures rotating with 
their own constant angular velocities.

These two constant angular velocity structures
should be seen in the long-term
$g_1(t)$ and $g_2(t)$ curves,
the MLC of FK Com.
The ephemerides 
of Eq. \ref{epheone} and \ref{ephetwo}
give these two MLC.
We compute their mean for ten phase bins. 
The first MLC
in Figs. \ref{figmlc}a-d has a
minimum  at $\phi_1 \approx 0.0$.
Its shape and amplitude are similar
in the 1st and the 2nd part of data.
The ephemeris of Eq. \ref{ephetwo} gives the second
MLC (Figs. \ref{figmlc}e-h).
Its minimum is close to $\phi_2=0.0$,
and its shape and amplitude are 
similar in the 1st and the 2nd part.
The low amplitudes of these two MLC signals
are of the same order as those of some
CABSs \citep[][e.g. EL Eri and FG UMa]{Jet17}.
Both MLC signals
confirm that
the two structures rotating with constant
angular velocities are real.

We will show that the
one-dimensional period finding methods
fail to detect the above two structures
rotating with different angular velocities.
Therefore, it is crucial to test,
which one is the better model for the light curves
of FK Com, the complex or the simple model?
These two models can not be compared
in the segments of Table \ref{Sresults},
because the complex and the
simple model approach each other, 
$g_{\mathrm{C}}(t) \leftrightarrow g_{\mathrm{S}}(t)$,
and only the latter can be used.
However, they can be compared 
in the segments of Table \ref{Cresults}.
The test statistic 
\begin{equation}
F=
\left(
{
{R_{\mathrm{S}}}
\over
{R_{\mathrm{C}}}
}
-1
\right)
\left(
{
{n-p_{\mathrm{C}}-1}
\over
{p_{\mathrm{C}}-p_{\mathrm{S}}}
}
\right).
\label{fvalue} 
\end{equation}
reveals
the better model for the data.
The null hypothesis is
\begin{description}

\item $H_{\mathrm{0}}$: {\it ``The complex model $g_{\mathrm{C}}(t)$ 
does not provide
a significantly better fit than the simple model $g_{\mathrm{S}}(t)$.'' }

\end{description}

\noindent
Under $H_0$, the test statistic of Eq. \ref{fvalue}
has an $F$ distribution with 
$(\nu_1,\nu_2)$ degrees of freedom,
where $\nu_1=p_{\mathrm{C}}-p_{\mathrm{S}}$ and $\nu_2=n-p_{\mathrm{C}}$
\citep{Dra98}. % Chapter 9: 217-221{Dra98} 
The probability that $F$ reaches or exceeds a fixed level $F_0$
is called the critical level $Q_{F} = P(F \ge F_0)$. 
We will reject $H_0$, if and only if
\begin{eqnarray}
Q_F < \gamma_F=0.001,
\label{oneortwo}
\end{eqnarray}
where $\gamma_F$ is a pre-assigned significance level.
It is the probability for falsely
rejecting $H_0$ when it is in fact true.

The results for this $F$-test are given in the last
column of Table \ref{Cresults}.
In seven out of twelve segments, the critical level $Q_{\mathrm{F}}$
is below the computational accuracy $10^{-16}$.
The smallest segment, \Data{1}{3}, gives 
the highest $Q_{\mathrm{F}}=0.00025$ value, 
but even this value fulfills the rejection
criterion of Eq. \ref{oneortwo}.
Hence, the $H_0$ hypothesis must be
rejected in {\it all} twelve segments of Table \ref{Cresults}.
For the light curves of FK Com,
the complex model is certainly 
better than the simple model. 
Although all this does not mean that
the complex model itself is correct,
the extreme  $Q_{\mathrm{F}}<10^{-16}$
values indicate that it
is certainly a major improvement to the correct direction.
% All results for FK Com support the \JHLhyp.

% jetsu@dx5-flspa-02:~/ilana/progs$ python FKpreltest.py 
% DOC 28.05.2019: Double-check
% DOC 27.06.2019: Triple-check
\begin{figure}
\begin{center} 
\resizebox{8.5cm}{!}{\includegraphics{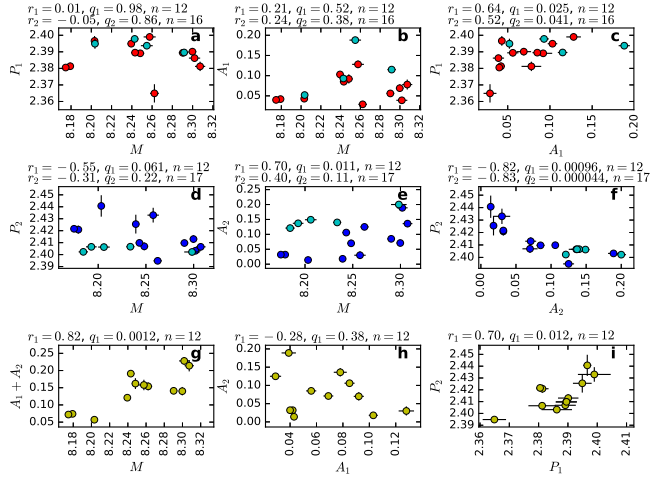}} 
\end{center}
\caption{{\color{black} Correlations.
  (a) Linear correlation coefficient $r_1=0.01$ and
       its significance $q_1=0.98$ for complex model $ M$ and $P_1$
       from Table \ref{Cresults} ($n=12$ red circles).
       Respective $r_2$ and $q_2$ values when corresponding
       simple model $M$ and $P_1$ are included
       from Table \ref{Sresults}
       ($P_{\mathrm{w,1}}$-group: $n=4$ cyan circles)
       (b) $M$ and $A_1$
       from Table \ref{Cresults},
       and corresponding
       $M$ and $A_1$ 
       from Table \ref{Sresults}.      
       (c) $A_1$ and $P_1$
       from Table \ref{Cresults},
       and corresponding
       $A_1$ and $P_1$ 
       from Table \ref{Sresults}.
       (d) $M$ and $P_2$
       from Table \ref{Cresults},
       and corresponding
       $M$ and $P_1$ 
       from
       Table \ref{Sresults} ($P_{\mathrm{w,2}}$-group: $n=5$).      
       (e) $M$ and $A_2$
       from Table \ref{Cresults},
       and corresponding
       $M$ and $A_1$ 
       from
       Table \ref{Sresults}.      
       (f) $A_2$ and $P_2$
       from Table \ref{Cresults},
       and corresponding
       $A_1$ and $P_1$ 
       from
       Table \ref{Sresults}.      
     (g) Complex model $M$
        and $A_1+A_2$ 
       (Table \ref{Cresults}).
       (h) Complex model $A_1$ and $A_2$
       (Table \ref{Cresults}).
       (i) Complex model $P_1$ and $P_2$
       (Table \ref{Cresults}).
       }
}
\label{FKpreltest}
\end{figure}

% jetsu@dx5-flspa-02:~/ilana/progs$ python FKspots.py 
% 6.0cm & BoundingBox: 75 180 400 610
\begin{figure}
\begin{center} 
\resizebox{6.0cm}{!}{\includegraphics{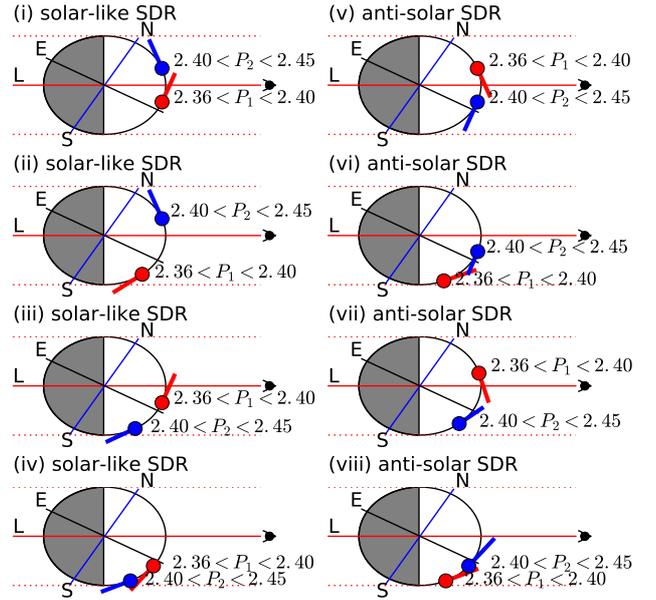}} 
\end{center}
\caption{{\color{black} Alternatives \ref{sdrone}-\ref{sdreight}
of Table \ref{eightalternatives}.
Red horizontal continuous and dotted lines denote
line of sight (L).
%Shaded area can not be observed.
Blue continuous line shows rotation axis
between north (N) and south (S). 
Black continuous line is equator (E).
Red and blue circles denote
starspots $S_1$ and $S_2$, respectively.
Lines extending from these circles show
directions, where 
periods $P_1$ and $P_2$ increase.  }}
\label{FKspots}
\end{figure}

\subsection{Solar-like versus anti-solar \sdr}
\label{FKsdr}

There are interesting correlations between
the parameters of the real $g_1(t)$ and $g_2(t)$
light curves of FK Com.
If the %two-tailed 
probability for some correlation 
is below 0.05,
we call it
significant.

The $g_1(t)$ light curve parameters
are shown in Figs. \ref{FKpreltest}a-c.
If the lowest inaccurate
period value were ignored in Fig. \ref{FKpreltest}a,
$P_1$ would first increase and then decrease
when $M$ increases.
The $M$ 
and $A_1$ correlations are not significant (Fig. \ref{FKpreltest}b),
but the positive $A_1$ and 
$P_1$ correlations are 
(Fig. \ref{FKpreltest}c: $q_1=0.025$ and $q_2=0.041$).

The $g_2(t)$ light curves show some
correlation between $M$ and $P_2$ of complex models
(Fig. \ref{FKpreltest}d: $q_1=0.061$).
However,
it becomes weaker
when the corresponding 
five simple model values are included
(Fig. \ref{FKpreltest}d: $q_2=0.22$).
The complex model $M$ and $A_2$ correlation
is significant (Fig. \ref{FKpreltest}e: $q_1=0.011$),
but again the five corresponding
simple model values weaken
this correlation (Fig. \ref{FKpreltest}e: $q_2=0.11$).
The extremely significant 
negative $A_2$
and $P_2$ correlations confirm
that
here we encounter a real physical connection
(Fig. \ref{FKpreltest}f: $q_1\!=\!0.00096$ and
$q_2\!=\!0.000044$).
However, these strong
amplitude-period correlations are negative in Fig. \ref{FKpreltest}f,
while the weaker amplitude-period correlations
in Fig. \ref{FKpreltest}c are positive!

We also study separately the correlations of
segments,
where both starspots are detected simultaneously, 
and the complex model can be used
(Figs. \ref{FKpreltest}g-i).
The $M$
and $A_1+A_2$ correlation is significant,
and supports dark starspots 
(Fig. \ref{FKpreltest}g: $q_1=0.0012$).
The $A_1$ and $A_2$ correlation is not significant
(Fig. \ref{FKpreltest}h).
The significant positive $P_1$ and
$P_2$ correlation indicates that both periods increase and decrease
simultaneously (Fig. \ref{FKpreltest}i. $q_1=0.012$).

Let us assume that the starspots $S_1$ and $S_2$
at the latitudes $b_1$ and $b_2$ cause
the real $g_1(t)$ and $g_2(t)$ light curves.
Their periods are $P_1$ and $P_2$,
their amplitudes are $A_1$ and $A_2$, 
and $S_1$ rotates faster than $S_2$,
because $P_1 < P_2$.
We compare the eight alternatives 
summarized in Table \ref{eightalternatives}.
These alternatives are illustrated in
Fig. \ref{FKspots}, 
where the starspots $S_1$ and $S_2$
are on the same hemisphere
or on different hemispheres,
and \sdr ~is solar-like or anti-solar.
We fix the positions of the north pole (N),
the south pole (S), 
the equator (E) 
and the line of sight (L)
in Fig. \ref{FKspots}.
The inclination of FK Com is fixed to
$i=60 \degr$,
as in 
\citet[][Table 2]{Kor99}      % Correct 24.06.2019 in jetsufk.bib
or \citet[][Table 4]{Kor07}.  % Correct 24.06.2019 in jetsufk.bi
Our results would be qualitatively
the same, if the north pole (N) or
the south pole (S) were above the line of sight (L).
Therefore, we discuss only the alternative,
where the full northern hemisphere is seen,
but parts of the southern hemisphere are never seen.
Our results are valid, if the following
simple assumptions are true.

% Hand-written
\begin{table}
  \caption{{\color{black}
      Alternatives \ref{sdrone}-\ref{sdreight}. Next two columns give
hemispheres of starspots $S_1$ and $S_2$
(N = North or S = South).
Last three columns give \sdr ~(solar-like or anti-solar),
starspot closer to equator E ($S_1$ or $S_2$),
and directions where periods $P_1$ of $S_1$ and $P_2$ of $S_2$ 
increase in Fig. \ref{FKspots}
(same or opposite).}}
\begin{center}
\begin{tabular}{ccclcl}
\hline
Alternative           & $S_1$  & $S_2$ & \sdr      & Closer      &  Directions  \\
\hline
\ref{sdrone}    & N      & N     & solar-like     & $S_1$       & same   \\
\ref{sdrtwo}    & S      & N     & solar-like     & $S_1$       & opposite  \\
\ref{sdrthree}  & N      & S     & solar-like     & $S_1$       & opposite  \\
\ref{sdrfour}   & S      & S     & solar-like     & $S_1$       & same    \\
\ref{sdrfive}   & N      & N     & anti-solar& $S_2$       & same \\
\ref{sdrsix}    & S      & N     & anti-solar& $S_2$       & opposite \\
\ref{sdrseven}  & N      & S     & anti-solar& $S_2$       & opposite \\
\ref{sdreight}  & S      & S     & anti-solar& $S_2$       & same \\
\hline
\end{tabular}
\end{center}
\label{eightalternatives}
\end{table}

\begin{itemize}

\item[1.] \sdr ~is latitudinally monotonic and symmetric, like 
  $P(+b)\!=\!P(-b)$ in Eq. \ref{rotationsun}.

\item[2.] Starspots  $S_1$ and $S_2$ never cross the equator E.
  
\item[3.]
  Starspot $S_1$ and $S_2$ latitude ranges do not overlap,
  because $P_1$ and $P_2$ ranges
  do not overlap.

\item[4.] Amplitudes $A_1$ and $A_2$ depend on
  the projected areas of starspots $S_1$ and $S_2$.
  This area increases from S to E,
  and from E to L.
  It decreases from L to N.
  
\item[5.] Starspots  $S_1$ and $S_2$ are dark
  (Fig. \ref{FKpreltest}g).

\end{itemize}

\noindent
Periods $P_1$ and $P_2$ increase or decrease simultaneously
(Fig. \ref{FKpreltest}i).
For {\it both} solar-like and anti-solar \sdr, this indicates
that

\begin{itemize}

 \item[6.] For starspots $S_1$ and $S_2$ on {\it different}
  hemispheres, their periods must increase
  to the {\it opposite} directions in Fig. \ref{FKspots}.

\item[7.] For starspots $S_1$ and $S_2$ on the {\it same}
  hemisphere, their periods must increase
  to the {\it same} direction in Fig. \ref{FKspots}.

\end{itemize}

\noindent
At all latitudes,
  both starspots of FK Com
   migrate
to the same direction with respect to the equator,
as in the Sun.

Here, we evaluate each alternative \ref{sdrone} - \ref{sdreight}
with
the elimination criteria given in Table \ref{sdrresults}.

\begin{table}
  \caption{{\color{black}
    Alternative \ref{sdrone}-\ref{sdreight} elimination criteria.
    Next
two columns indicate if this alternative can explain correlations of 
Figs.  \ref{FKpreltest}c and  \ref{FKpreltest}f (Yes or No).
Last column states, if this alternative has
starspots that
induce rotational modulation of
brightness, 
and are located above line of sight L in Fig. \ref{FKspots}
(Yes or No).
}}
\begin{center}
\begin{tabular}{ccccc}
\hline
               & & \multicolumn{3}{c}{Criterion} \\
\cline{3-5}
               & & Fig. \ref{FKpreltest}c &  Fig. \ref{FKpreltest}f  & Starspots \\
Alternative    & & $A_1 \propto + P_1$       &  $A_2 \propto - P_2$       & above L   \\  
\hline
\ref{sdrone}   & & Yes & Yes & Yes \\
\ref{sdrtwo}   & & No & Yes & Yes \\
\ref{sdrthree} & & Yes & Yes & No \\
\ref{sdrfour}  & & No & Yes & No \\
\ref{sdrfive}  & & Yes & Yes & Yes \\
\ref{sdrsix}   & & Yes & Yes & No \\
\ref{sdrseven} & & Yes & No & Yes \\
\ref{sdreight} & &Yes & No & No \\
\hline
\end{tabular}
\end{center}
\label{sdrresults}
\end{table}

\begin{enumerate}

 % Case I
% =======
\item
Both starspots are on the northern hemisphere.
For solar-like \sdr, starspot $S_1$ must be closer to E
than starspot $S_2$ ($P_1 < P_2$).
If the latitude range of $S_1$
were between E and L,
solar-like \sdr ~would explain
the positive correlation of
Fig. \ref{FKpreltest}c.
It would also explain the negative
correlation of Fig. \ref{FKpreltest}f,
if the latitude range of $S_2$ was
between L and N.
This negative correlation would be
stronger than the positive correlation,
if starspot $S_2$ stayed above L,
but starspot $S_1$ occasionally
migrated above L.
The negative correlation may also be stronger,
because the suitable latitude range for $S_2$
$(+30\degr\! < \! b_2 \! < \!+90\degr)$
is two times larger than that for $S_1$
($+0\degr \! < \! b_1 \! < \! +30\degr$). \label{sdrone}

% Case II
% =======
\item For solar-like \sdr,
the southern hemisphere
starspot $S_1$ must be closer to E
than the northern hemisphere
starspot $S_2$, because $P_1 < P_2$.
There is a contradiction in this alternative,
because solar-like \sdr ~can not explain the
positive correlation between $A_1$ and $P_1$ 
(Fig. \ref{FKpreltest}c). 
It could explain the
negative correlation between $A_2$ and $P_2$
(Fig. \ref{FKpreltest}f),
if the latitude range of $S_2$ were
between L and N. \label{sdrtwo}

% Case III
% ========
\item
For solar-like \sdr,
the southern hemisphere starspot $S_2$ must
be further away from E than the northern
hemisphere starspot $S_1$.
Solar-Like \sdr ~would explain the
correlation of Fig. \ref{FKpreltest}c, 
if the latitude range
of $S_1$ were between E and L.
It would also explain the negative
$A_2$ and $P_2$  
correlation for starspot $S_2$
(Fig. \ref{FKpreltest}i).
However, if the $S_1$ maximum latitude were L $\equiv b_1=+30\degr$,
then the $S_2$ latitudes would have to be close to the limb of visible
stellar disk at
$-60\degr < b_2 < -30\degr$, and it would be difficult to
explain the strong negative correlation in
Fig. \ref{FKpreltest}f.
If the $S_1$ maximum latitude were below $b_1=+30\degr$,
the $S_2$ starspot latitudes could be closer to E.
In any case, 
all area above L, or equivalently half of the visible stellar disk,
would be free of starspots causing rotational
modulation of brightness. \label{sdrthree}

% Case IV
% =======
\item Over half the visible stellar disk,
the northern hemisphere, is free 
of starspots causing rotational modulation
of brightness.
Starpots $S_1$ and $S_2$ are on the southern hemisphere,
and the former is closer to E for solar-like \sdr.
They are in the narrow latitude range
between $0\degr$ and  $-60\degr$.
The contradiction in this alternative is that
solar-like \sdr ~can
not explain the positive correlation between $A_1$ and $P_1$
(Fig. \ref{FKpreltest}c).
Furthermore, starspot $S_2$ must be so
close to the limb of the visible stellar disk
that due to the limb darkening, the strong negative
correlation between $A_2$ and $P_2$ becomes
difficult to explain (Fig. \ref{FKpreltest}f). \label{sdrfour}

% Case V
% ======
\item Both starspots are on the northern
hemisphere.
For anti-solar \sdr,
starspot $S_2$ must be closer to E.
If the latitude range of starspot $S_1$ is between N and L,
then anti-solar \sdr ~explains the
positive $A_1$ and $P_1$ correlation
(Fig. \ref{FKpreltest}c).
Anti-solar \sdr ~would also explain
why $A_2$ decreases when $P_2$ increases
(Fig. \ref{FKpreltest}f),
if the latitude range of starspot $S_2$ is
between L and E.
However, is difficult to explain
why the negative correlation for $S_2$ (Fig. \ref{FKpreltest}f)
is so much stronger than the positive correlation for $S_1$
(Fig. \ref{FKpreltest}c),
because the suitable latitude range for $S_2$
$(+0\degr < b_1 < +30\degr)$
is two times smaller than that for $S_1$
($+30\degr < b_2 <+90\degr$).
\label{sdrfive}

% Case VI
% =======
\item For anti-solar \sdr,
the northern hemisphere starspot $S_2$
must be closer to E than the southern
hemisphere starspot $S_1$.
Anti-solar \sdr ~would explain
the positive correlation between $A_1$ and  $P_1$
(Fig. \ref{FKpreltest}c),
as well as the negative correlation
between $A_2$ and  $P_2$ if starspot $S_2$ stays below L
(Fig. \ref{FKpreltest}f).
For this combination,
there is no contradiction.
However,
the strong negative correlation in Fig. \ref{FKpreltest}f
requires that starspot $S_2$ migrates over
most of the latitudes between E and L,
but its highest latitude
should not exceed L $\equiv b_2=+30\degr$. 
Hence, the latitude range of
starspot $S_1$ is approximately
$-60\degr < b_1 < -30\degr$.
In this combination,
there are
no starspots above L,
on half of the visible stellar disk, causing rotational modulation
of brightness. \label{sdrsix}

% Case VII
% ========
\item For anti-solar \sdr,
the southern hemisphere starspot $S_2$ must be closer to
E than the northern hemisphere starspot $S_1$.
Anti-solar \sdr ~would explain the positive
correlation between $A_1$ and $P_1$ (Fig. \ref{FKpreltest}c),
if the latitudes of starspot $S_1$ were above L.
This combination contradicts the strong negative
correlation in Fig. \ref{FKpreltest}f,
because $A_2$ does not decrease when $P_2$ increases. \label{sdrseven}

% Case VIII
% =========
\item Both starspots are below E in the southern
hemisphere, and more than half of the visible
disk is free of starspots causing rotational modulation of brightness.
Anti-solar \sdr ~would explain increasing $A_1$
when $P_1$ increases for starspot $S_1$ further away from
E (Fig. \ref{FKpreltest}c).
For starspot $S_2$, it would predict
that $A_2$ increases when $P_2$ increases.
This contradicts Fig. \ref{FKpreltest}f. \label{sdreight}

\end{enumerate}

We can reject the contradictory 
alternatives \ref{sdrtwo}, \ref{sdrfour},
\ref{sdrseven} and \ref{sdreight}
(Table \ref{sdrresults}).

Alternatives \ref{sdrthree} and \ref{sdrsix}
have no starspots
above the line of sight at $b=30\degr$
(Table \ref{sdrresults}: Starspots above L = ``No'').
We can also reject these two alternatives, because
many Doppler images of FK Com have shown
high latitude starspots causing rotational modulation
of brightness \citep[e.g.][]{Kor00,Vid15}.

Only alternatives \ref{sdrone} and \ref{sdrfive} remain.
The available Doppler images of FK Com reveal 
the correct alternative.

\citet{Kor00} published six Doppler
images of FK Com between
the years 1994 and 1997. 
%They used 
%$P_{\mathrm{phot}}=2.^{\mathrm{d}}4002466
%\pm0.^{\mathrm{d}}0000056$
%in their inversions.
These images
showed long-lived starspots rotating with
a constant period
$P_{\mathrm{DI}}=
2.^{\mathrm{d}}4037 \pm 0.^{\mathrm{d}}0005$.
Their images overlap our photometry
     between 1995 and 1997.
    Their  $P_{\mathrm{DI}}$ period  
    agrees perfectly with
    our ephemeris of Eq. \ref{ephetwo}.
  Because we detect
  only one signal, we use
  the simple model.
  It
  reproduces
  this $P_{\mathrm{DI}}$  signal of Doppler images
  (Fig. \ref{fkresults}d: two cyan circles)
  and its minima
  (Fig. \ref{fkresults}f: two cyan circles).
  This signal of the longer $P_2$ periods (i.e. $S_2$ spots),
  which is centered at
  $P_{\mathrm{WK,2}}=2.^{\mathrm{d}}40413 \pm 0.^{\mathrm{d}}00048$,
  dominated
  the light curves of FK Com
  between 1996
  and 1997.
  Because only one signal dominated, 
the Kuiper method can also detect
$P_{\mathrm{act}}=
2.^{\mathrm{d}}4035 \pm 0.^{\mathrm{d}}0005$
from the overlapping
thirteen first statistically independent
$t_{\mathrm{CPS,min,1}}$ epochs 
(Table \ref{electric}: IND=1).
% Checked with /newprogs/thirteen.py ... 19.06.2018
% documented to analysis.tex 
This $P_{\mathrm{act}}$ value also agrees with  $P_{\mathrm{DI}}$.
The above
  six Doppler images indicate that our simple harmonics
  $g(t)$ model (Eq. \ref{fullmodel}) may not suffer
  from the ambiguity problems described by
  \citet{Rus06} and \citet{Jef09}.
  We will also discuss these ambiguity problems
  later in Sect. \ref{sectfrontback}.
 Perhaps the simple \JHLhyp ~model really is
  the correct solution
  for this ill-posed problem.
The starspots in the Doppler images
followed 
``solid body rotation''
and their latitudes
were between $45\degr$ and $70\degr$ 
\citep{Kor00}.
In this {\it particular} case,
the \dimethod ~undoubtedly detects
long-lived starspots
of FK Com rotating with a
constant period $P_{\mathrm{WK,2}}=P_{\mathrm{act}}$
(see \Testseven).

These high latitude
starspots in the Doppler
images of \citet{Kor00} were connected to 
the slower rotating 
$S_2$ starspots $(P_2>P_1)$ which dominated
during that time interval
(Fig. \ref{fkresults}d two cyan circles in 1996 and 1997).
This result supports alternative \ref{sdrone} and
contradicts alternative \ref{sdrfive}, 
where the slower rotating $S_2$ starspots
should remain at lower latitudes
below L $\equiv b_2 = 30\degr$, and 
the faster rotating $S_1$ starspots should
remain at higher latitudes.

The above Doppler imaging study by \citet{Kor00}
supports the rejection of alternative \ref{sdrfive}.
We can also show that the $S_1$ and $S_2$ starspot latitudes
in other Doppler images support this rejection.
Our simple model $P_1$ periods show that the $S_1$ starspots
dominated in the beginning of the years 1998, 2003, 2007
and 2010 (Fig.  \ref{fkresults}d, 
$P_{\mathrm{w,1}}$-group: four cyan circles).
This predicts that
these low latitude
faster rotating $S_1$ starspots should dominate in
the Doppler images during those time intervals. 
No simultaneous images are available for the beginning
of 2003 and 2007, but the
two available simultaneous images confirm our prediction:

\begin{itemize}

\item[] January 1998: Dominating $S_1$ starspot at
        $b_1 \approx 45\degr$ and no high latitude $S_2$ starspots
        \citep[][Fig. 1]{Kor07}          % jetsufk.bib
\item[] January 2010: Dominating $S_1$ starspots at
         $b_1 \approx 20\degr$ and no high latitude $S_2$ starspots
        \citep[][Fig. 5]{Vid15}          % jetsufk.bib

\end{itemize}

\noindent
This result also contradicts alternative \ref{sdrfive}
and supports alternative \ref{sdrone}. 
We conclude that
the correlations in Figs. \ref{FKpreltest}a-i,
together with the published Doppler images, definitely 
support only alternative \ref{sdrone}:
weak solar-like \sdr ~in FK Com. 

We detect significant correlations 
between the parameters
of the real $g_1(t)$ and $g_2(t)$ light curves of FK Com
(Figs. \ref{FKpreltest}c, f, g and i).
As far as we know,
no one else has detected this many 
significant and clear correlations
between any light curve parameters of any chromospherically
active star. % , nor between the light curves and the Doppler images.
They connect our period analysis parameters 
directly to the starspots of FK Com.
This supports the \JHLhyp,
because these correlations would not exist unless
the $S_1$ and $S_2$ starspot life-times
were of the same order as the length of the
segments, from a few months to over half a year.
Furthermore, the Doppler images confirm that
the $S_1$ and $S_2$ starspot periods
can be directly interpreted as stellar surface structures.
We call the following relations
the {\it starspot activity mode} of FK Com
(Fig. \ref{FKspots}: Alternative \ref{sdrone}).
  \begin{equation}
     \left\{
     \begin{array}{l}
       M \propto c_1 (A_1+A_2) \\
       P_1 \propto c_2 (A_1)  \\
       P_2 \propto c_3 (A_2)  \\ 
       P_1 \propto c_4 (P_2),
     \end{array}
   \right.
   \label{FKmode}
   \end{equation}
   \noindent
where  $c_1=+1$, $c_2=+1$, $c_3=-1$ and $c_4=1$.
It states that the mean brightness $(M)$ is proportional to
the total projected area  $(A_1+A_2)$ of dark starspots $(c_1=1)$.
The starspot rotation periods ($P_1$ and $P_2$) are
proportional to these projected areas below $(c_2=1)$ and
above $(c_3=-1)$ the line of sight $L$ in Fig. \ref{FKspots}.
The periods $P_1$ and $P_2$ increase and decrease
simultaneously $(c_4=1)$.
Therefore, the mean ($\propto$ both starspot areas),
the amplitudes      ($\propto$ separate starspot areas)
and the periods     ($\propto$ separate starspot latitudes)
follow the {\it same activity cycle}.
Everything is just like in the Sun, except that the dark starpots
change the mean, not the bright structures
like faculae or plages.
This confirms the results by
\citet{Rad94}:
dark and bright structures
change the luminosity of more and less active stars,
respectively.

Scientific arguments must be \hattuuslippa.
Our period analysis of FK Com is \hattuuslippa,
because we publish 
the data (Sect. \ref{Data}),
the model (Sect. \ref{Models}),
the method (Sect. \ref{Method})
and the results (Sect. \ref{sectallresults}).
Especially in this last Sect. \ref{FKsdr},
we show beyond any reasonable doubt
that the parameters of the real 
$g_1(t)$ and $g_2(t)$ light curves are
directly connected
to real starspots on the surface of FK Com. 
Our results support the \JHLhyp: the presence of
two periodic signals in the observed light curves
of FK Com. We can show that the
activity of FK Com resembles the activity of the Sun.
Hence, our \JHLhyp ~should also
predict the results for the \Testone ~- \Testeight,
if the light curves of other chromospherically active
stars also contain such signals.
We will perform
two easily \hattuuslippa ~simulations which show
that it simply
makes no sense to analyse such light curves
with one-dimensional period analysis methods 
~(Sect. \ref{LCconnectionsect}: \Testone ~- \Testfive). 
For the past seven decades,
this approach has given incompatible results (Sect. \ref{incomp}),
like the former \cpsmethod ~analysis of our 
data (Sect. \ref{cpsexplained}).
The following argument is very often \hattuuslippa
\begin{itemize}
  \item[] {\it
  ``This works in practice, but does this work in theory?''}
\end{itemize}
Since our period analysis reveals a clear connection to
the starspots of FK Com, the argument
\begin{itemize}
  \item[] {\it
  ``This works in practice for 
the period analysis of real starspots,
  but does this work for 
the modelling of theoretical starspots?''}
\end{itemize}
should be \hattuuslippa.
While we do not try  
to reproduce this argument too,
it might be reasonable to use our period analysis 
parameters for the real starspots as input for the
theoretical starspot models.  
Although our paper 
concentrates only on period analysis, 
we will also address the above argument, 
but only for the modelling of real starspots
(Sect. \ref{sectfrontback}: \Testsix). 
The Doppler images of FK Com show
the long-lived starpots predicted by the \JHLhyp 
~(Sect. \ref{FKsdr}).
We will show that such starspots
have also been detected in the Doppler
images of other stars (Sect. \ref{SIsect}: \Testseven).
The results for the above two spot modelling and Doppler imaging tests 
are certainly \hattuuslippa, because they have already been published.
Our results for the \Testone ~and the \Testseven ~will 
provide the answer for the \Testeight ~(Sect. \ref{Conclusions}).

\section{LC-method connection}
\label{LCconnectionsect}

In the next Sects. \ref{LCconnectionsect}
and \ref{incomp}, we show that simple
mathematical relations reproduce 
practically all phenomena that have been
detected with the one-dimensional period 
analysis methods from
the light curves of chromospherically active 
stars: 
rapid light curve changes, 
short starspot life-times, 
rapid period changes,
active longitudes, 
starspot migration patterns, 
amplitude and period cycles, 
and \flip ~events.

\subsection{\lcmethod ~parameters and earlier results} 
\label{lcmethodsect}

The observed $P_{\mathrm{phot}}$ is connected to
the stellar rotation period $P_{\mathrm{rot}}$.
In the \sdr ~context, 
the observed $P_{\mathrm{phot}}$ may tell
something about the starspot latitude.
If $P_{\mathrm{phot}}$ has changed,
it is logical to assume that the starspot
latitude has changed.
Thus, the range of  $P_{\mathrm{phot}}$ changes
could measure \sdr.

The one-dimensional period
finding methods determine only
one $P_{\mathrm{phot}}$ period
value.
We compare
three such methods.
Their models are summarized
in Table \ref{methodstable}.
The first one is our $g_{\mathrm{S}}(t)$-method
(Sect. \ref{Smethod}).
The second one is the most widely used one-dimensional
period finding method, 
the Lomb-Scargle periodogram
(Table \ref{methodstable}: \lsmethod).
It finds the best least squares
sinusoidal model for the data.
The third one, the \cpsmethod, can determine
the best modelling order for the data
(Table \ref{methodstable}: $K_1=0, 1$ or 2).

The 
$P_{\mathrm{phot}}$ are usually compared to 
the law of solar \sdr
\begin{eqnarray}
P(b)={
      {P_{\mathrm{eq}}} 
       \over
      {1- k \sin^2 b}
      },
\label{rotationsun}
\end{eqnarray}
\noindent
where $b$ is the latitude,
$P_{\mathrm{eq}}$ is the equatorial rotation period 
and $k$ is the \sdr ~coefficient.
A useful parameter is 
\begin{eqnarray}
h= \sin^2 (b_{\mathrm{max}}) -  \sin^2 (b_{\mathrm{min}}),
\label{latrange}
\end{eqnarray}
\noindent
where $b_{\mathrm{min}}$ and $b_{\mathrm{max}}$ are the
lowest and the highest starspot latitudes.
The 
$b_{\mathrm{max}}\!-b_{\mathrm{min}}\!= \!90\degr$ range
has $h\!=\!1$, other
ranges $h<1$.
This gives
$k= \Delta P / (h P_{\mathrm{eq}})$,
where $\Delta P=
P_{\mathrm{phot,max}}-P_{\mathrm{phot,min}}$
is the largest 
observed range.
Since $h\le1$, 
the stellar \sdr ~coefficient fulfills
\begin{eqnarray}
  |k| \ge {{P_{\mathrm{phot,max}}- P_{\mathrm{phot,min}}}
  \over {P_{\mathrm{phot,mean}}}}=
  {\Delta P \over  P_{\mathrm{phot,mean}}},
\label{k_two}
\end{eqnarray}
\noindent
where $P_{\mathrm{phot,mean}}$ 
is the mean of observed 
$P_{\mathrm{phot}}$.
There are four uncertainties.
Firstly, 
$P_{\mathrm{phot,mean}}$
is used 
for  $P_{\mathrm{eq}}$ 
(Eq. \ref{rotationsun}),
because  $P_{\mathrm{eq}}$ may be
$P_{\mathrm{phot,min}}$ or $P_{\mathrm{phot,max}}$.
% Some {\lcmethod}s
%use $P_{\mathrm{eq}}=P_{\mathrm{phot,max}}$ \citep[e.g.][]{Rei13}.
Secondly, the $k$ sign is unknown,
where $k\! >\!0$ represents solar-like and  $k \! < \!0$
anti-solar \sdr.
Thirdly, the starspot latitude range
is unknown (Eq. \ref{latrange}: $h$).
Fourthly,
the observed
$\Delta P \! = \! P_{\mathrm{phot,max}}\!-P_{\mathrm{phot,min}}$
range may underestimate the real range.

Another approach uses 
$P_{\mathrm{w}}  \pm \Delta P_{\mathrm{w}}$
(Eqs. \ref{Pw} and \ref{ePw}) to
give
the ``three sigma'' upper limit for the
$P_{\mathrm{phot}}$ changes 
\begin{eqnarray}
Z={{6 \Delta P_{\mathrm{w}}}\over{P_{\mathrm{w}}}}.
\label{zvalue}
\end{eqnarray}
\noindent
Equal weights give $P_{\mathrm{w}}=m_{\mathrm{P}}$, 
$\Delta P_{\mathrm{w}}=\sigma_{\mathrm{P}}$ and $Z= { {6 \sigma_{\mathrm{P}}}/{m_{\mathrm{P}}}}$,
where $m_{\mathrm{P}}$ and $\sigma_{\mathrm{P}}$ are 
the mean and the standard deviation
of all observed $P_{\mathrm{phot}}$ values.
The relation
\begin{eqnarray}
|k| \approx Z/h 
\label{k_three}
\end{eqnarray}
\noindent
is valid, if 
$P_{\mathrm{phot,max}}-P_{\mathrm{phot,min}} \approx 6 \Delta P_{\mathrm{w}}$
\citep[][]{Jet00}.

\citet{Hal91A} applied the \lcmethod ~to 
 277 late-type stars. 
He concluded that the stellar \sdr ~correlates 
strongly with $P_{\mathrm{phot}}\!\approx\! P_{\mathrm{rot}}$.
It decreases when $P_{\mathrm{phot}}$ decreases.
He showed that the weak
\sdr ~in rapidly rotating stars 
approached solid-body rotation.
This $P_{\mathrm{phot}}$ and
\sdr ~relation has been confirmed by 
\lcmethod ~studies of much larger 
samples \citep[e.g.][24~124 stars]{Rei13}.
We have also formulated our own {\lcmethod}s, 
and used them to measure \sdr ~
\citep[e.g.][the TSPA-method and
the \cpsmethod]{Jet99,Jet99A,Leh11,Leh16}.
These earlier findings seem to contradict the \JHLhyp,
because {\it many different} photometric rotation
periods $P_{\mathrm{phot}}$
are observed in every individual star
(\Testone).

\subsection{Sum of two different frequency 
sinusoids}
\label{explanation}

We can show that
  our general $g(t)$ light curve model
  has a physically justified connection to the starspots
of FK Com (Sect. \ref{FKsdr}).
The most simple $g(t)$ alternative,
the following sum of two sinusoids,
has been used in innumerable studies
\citep[e.g.][]{Rei13,Aig15}
\begin{eqnarray}
s(t)   & = & s_1(t)+s_2(t). \label{sinesum}     \\
s_1(t) & = & a_1\sin{(2 \pi f_1 t)} \label{sone}\\
s_2(t) & = & a_2\sin{(2 \pi f_2 t)}, \label{stwo} 
\end{eqnarray}
where
$m_0(t)\!=\!0$ (Eq. \ref{gzero}),
$s_1(t)\!=\!g_1(t)$ (Eq. \ref{gone}),
$s_2(t)\!=\!g_2(t)$ (Eq. \ref{gtwo})
and
$s(t)\!=\!g_{\mathrm{C}}(t)$  (Eq. \ref{fullmodel}).
%The amplitudes are $A_1=2a_1$ and $A_2=2a_2$.
%If $f_1 \approx f_2$, 
The $s(t)$ solutions differ for
\begin{eqnarray}
a_1 & =    & a_2 \label{fsame}    \\
a_1 & \neq & a_2. \label{fnotsame}
\end{eqnarray}

If Eq. \ref{fsame} is true, then
\begin{eqnarray}
s(t)=s_1(t)+s_2(t)= a_a(t) \sin{(2 \pi f_a t)},
\label{equalfsum}
\end{eqnarray}
where the amplitude is
\begin{eqnarray}
a_a(t)=a_1\sqrt{2+2\cos{[2\pi(f_1-f_2)t]}}.
\label{ampsame}
\end{eqnarray}
The frequency of the sum $s(t)$ remains constant
\begin{eqnarray}
f_a=(f_1+f_2)/2.
\label{freqsame}
\end{eqnarray}
\noindent
The amplitude $a_a(t)$ varies regularly between 
\begin{eqnarray}
a_{\mathrm{a,min}}  =  0, & ~~ & 
a_{\mathrm{a,max}}  =  2a_1 \label{aminima}
\end{eqnarray}
during
a lap cycle period 
\begin{eqnarray}
P_{\mathrm{lap}}=|P_1^{-1}-P_2^{-1}|^{-1}=|f_1-f_2|^{-1} = f_{\mathrm{lap}}^{-1}.
\label{lapcycle}
\end{eqnarray}
An abrupt $f_a/2$ phase shift % of $s(t)$  
occurs when $a_a(t)$ goes to zero.

\begin{table}
  \caption{One-dimensional period finding methods (Method).
$K_0$ of $m_0(t)$ (Eq. \ref{gzero}), 
$K_1$ of $g_1(t)$ (Eq. \ref{gone}) 
and reference (Formulation).
}
\label{methodstable}
\begin{center}
\addtolength{\tabcolsep}{-0.07cm} 
\begin{tabular}{lccl}
\hline
                       & \multicolumn{2}{c}{Model}  & \\
                       \cline{2-3}
                       & $m_0(t)$ & $g_1(t)$  &                     \\
 Method                & $K_0$    & $K_1$     & Formulation         \\
\hline
$g_{\mathrm{S}}(t)$-method  &  2     & 2         & This paper          \\
\lsmethod              & 0        & 1         & \citet{Lom76,Sca82} \\
\cpsmethod             & 0        & 0, 1 or 2  & \citet{Leh11}       \\
\hline
\end{tabular}
\addtolength{\tabcolsep}{+0.07cm} 
\end{center}
\end{table}

If Eq. \ref{fnotsame} is true, the sum is
\begin{eqnarray}
s(t)=s_1(t)+s_2(t)=a_b(t) \sin{[2 \pi f_2 t + \phi_b(t)] },
\label{unequalfsum}
\end{eqnarray}
where the amplitude is
\begin{eqnarray}
a_b(t)=\sqrt{a_1^2+a_2^2+2 a_1 a_2 \cos{[2 \pi (f_1-f_2)t]}}
\label{ampnotsame}
\end{eqnarray}
and the phase shift in radians is
\begin{eqnarray}
\phi_b(t)=\arctan{ 
\left[
{
{a_1 \sin{[2\pi(f_1-f_2)t]} }
\over
{a_1\cos{[2\pi(f_1-f_2)t]+a_2}}
}
\right]
}.
\end{eqnarray}
The amplitude $a_b(t)$ variation during $P_{\mathrm{lap}}$ is between
\begin{eqnarray}
a_{\mathrm{b,min}} = a_2-a_1, & ~ &
a_{\mathrm{b,max}} =  a_1+a_2.
\label{bminima}
\end{eqnarray}
The phase shift $\phi_b(t)$
induces frequency changes between 
\begin{eqnarray}
f_{\mathrm{b,max}}=f_2+ a_1(f_1-f_2)/(a_2+a_1)
\label{upperf}
\end{eqnarray}
at the $a_{\mathrm{b,max}}$ epochs, and
\begin{eqnarray}
f_{\mathrm{b,min}}=f_2- a_1(f_1-f_2)/(a_2+a_1)
\label{lowerf}
\end{eqnarray}
at the $a_{\mathrm{b,min}}$ epochs. 
The frequency $f_2$ of the stronger $s_2(t)$
signal $(a_1\!<\!a_2)$ dominates these 
regular changes 
during $P_{\mathrm{lap}}$.

The maximum $\phi_b(t)$ phase shift
to both directions is
\begin{eqnarray}
\phi_{\mathrm{b,max}}=\arcsin{(a_1/a_2)}
\label{ampratio2}
\end{eqnarray}
in radians.
For $a_1=a_2$, this shift $\phi_{\mathrm{b,max}}$
is $\pi/2$ radians, or equivalently $1/4$
in phase with $f_2$.
Its full range is
\begin{eqnarray}
\Delta \phi_{\mathrm{b}}= [\arcsin{(a_1/a_2)}]/\pi
\label{abruptphase}
\end{eqnarray}
in phase.
This gives the largest shift
$\Delta \phi_{\mathrm{b}}=0.5$ when $a_1=a_2$.
Smaller shifts occur when $a_1 \ne a_2$.

If Eq. \ref{fnotsame} is true,
the {\it observed} amplitude,
period and mimimum
of $s(t)$ vary regularly during 
$P_{\mathrm{lap}}$.

%{\color{cyan} Double check with \verb|f1f2check.py| on Jul 2nd, 2018
%\fbox{DOC1:08.02.2018}}

% 22.03.2019
% /method/progs/newsimulated.py with a1=0.050, a2=0.050 
%~/method/progs/cp kuvanow.eps /home/jetsu/method/texts/simulated1.eps
%~/method/progs/cp kuvanow.eps /home/jetsu/method/progs/valmis2/simulated1.eps
\begin{figure}
\begin{center}
\resizebox{8.0cm}{!}{\includegraphics{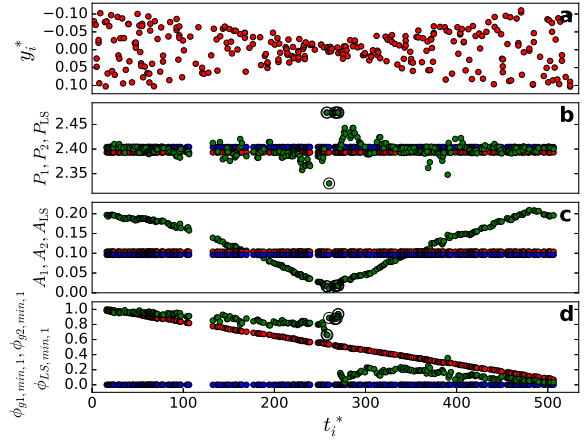}}
\end{center}
\caption{
(a) Simulated $y^*_i$ data (Eq. \ref{simulateddata}) for
$a_1=a_2=0.^{\mathrm{m}}05$ (Eqs. \ref{sone} and \ref{stwo}).
(b) Red and blue circles denote periods 
$P_1=2.^{\mathrm{d}}39321$ and $P_2=2.^{\mathrm{d}}40413$  % OK 15.04.2019
(Eqs. \ref{epheone} and \ref{ephetwo}).
Green circles 
show $P_{\mathrm{LS}}$ periods detected with \lsmethod.
Large black circles highlight cases,
where $P_{\mathrm{LS}}$
is at the end of tested period range.
Same models are highlighted in ``cd''.
(c) Red and blue circles denote
amplitudes $A_1=2a_1$ and $A_2=2a_2$ of the $g_{\mathrm{C}}(t)=s(t)$ model.
These overlapping red and blue symbols are slightly shifted apart.
Green circles show $A_{\mathrm{LS}}$ amplitudes. 
(d) Red and blue circles show 
the minima of $g_1(t)=s_1(t)$ and $g_2(t)=s_2(t)$ 
with $P_2$ in Eq. \ref{phasetwo}.
Green circles show minima of the \lsmethod ~sinusoidal models.  
}
\label{figsimulated1}
\end{figure}

% 22.03.2019 Pitaa olla vain yksi ympyra ja hyvat P&A ytiksit
% /method/progs/newsimulated.py with a1=0.025, a2=0.050 
%~/method/progs/cp kuvanow.eps /home/jetsu/method/texts/simulated2.eps
%~/method/progs/cp kuvanow.eps /home/jetsu/method/progs/valmis2/simulated2.eps
\begin{figure}
\begin{center}
\resizebox{8.0cm}{!}{\includegraphics{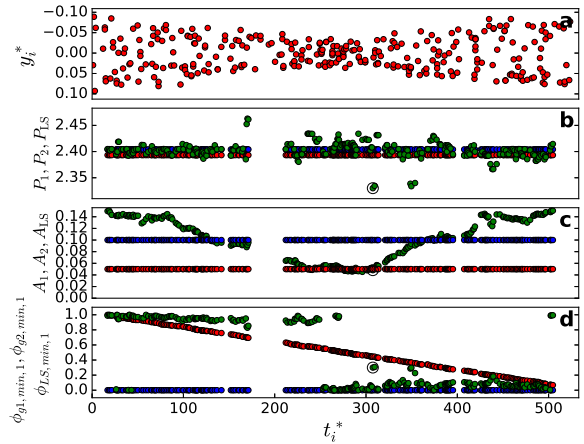}}
\end{center}
\caption{Simulated $y^*_i$ data for
$a_1=0.^{\mathrm{m}}025$ (Eq. \ref{sone}) 
and $a_2=0.^{\mathrm{m}}05$ (Eq. \ref{stwo}), 
otherwise as in Fig. \ref{figsimulated1}.} 
\label{figsimulated2}
\end{figure}

\subsection{Simulated  photometry}
\label{simulations} 

We simulate data with the $s(t)$ model 
(Eq. \ref{sinesum}) using $f_1\!=\!P_1^{-1}$ and $f_2\!=\!P_2^{-1}$,
where  
$P_1\!=\!2.^{\mathrm{d}}39321$ and
$P_2\!=\!2.^{\mathrm{d}}40413$ 
(Eqs. \ref{Kone} and \ref{Ktwo}). 
The time span is $\Delta T\!=\!527^{\mathrm{d}}$,
because $s(t)$ repeats itself
during $P_{\mathrm{lap}}$.
We use multiples of sidereal day 
$P_{\mathrm{sid}}=0.^{\mathrm{d}}9973$ 
to create $n^*=300$ time points
\begin{eqnarray}
t^*_i= i^* P_{\mathrm{sid}}+\delta t^*_i,
\nonumber
\end{eqnarray}
where $i^*$ are a random sample of integers
$1 \le i^* \le 527$. 
Each $i^*$ value is used as many times
as the random selection favours it.
The random shifts $\delta t^*_i$ 
are evenly
distributed between $-0.^{\mathrm{d}}2$ and $0.^{\mathrm{d}}2$.
The simulated data are
\begin{eqnarray}
y^*_i=s(t^*_i)+\epsilon^*_i,
\label{simulateddata}
\end{eqnarray}
\noindent
where the $\epsilon^*_i$ residuals are
drawn from a Gaussian distribution with
a zero mean and a standard deviation 
of $0.^{\mathrm{m}}008$.

From the simulated data,
we select datasets containing all $y^*_{\mathrm{i}}$
within a sliding 
$\Delta T=30^{\mathrm{d}}$ window.
We apply the \lsmethod ~to datasets having $n \ge 12$.
The tested range is $a=0.03$  
and $f_{\mathrm{mid}}=1/2.^{\mathrm{d}}4$ (Eq. \ref{fsimraja}).
\citet[][\lsmethod]{Dis16} or \citet[][\cpsmethod]{Leh16}
applied similar sliding window period
analysis to real photometry.

\subsubsection{\simuone}
 \label{simu1} 

In \simuone, we use equal amplitudes  
$a_1=a_2=0.^{\mathrm{m}}05$ (Eqs. \ref{sone} and \ref{stwo}). 
%The results are shown in Fig. \ref{figsimulated1}.
The $y^*_i$ variation first decreases to zero
and then increases back to its original
level 
(Fig. \ref{figsimulated1}a).
%This pattern is repeated during every lap cycle 
%$\Delta T = P_{\mathrm{lap}}$ (Eq. \ref{lapcycle}),
%because the interference of these two sinusoids is regular.
The simulated model periods $P_1$ and $P_2$ 
do not change (Fig. \ref{figsimulated1}b: red and blue circles).
The detected $P_{\mathrm{LS}}$
periods scatter 
when the $y^*_i$ variation approaches zero
 (Fig. \ref{figsimulated1}b: green circles).
The tested period interval is too narrow, because
some $P_{\mathrm{LS}}$ are at the ends of this interval
(Fig. \ref{figsimulated1}b: highlighted green circles).
The $P_{\mathrm{LS}}$ periods 
are mostly between 
$P_1$ and $P_2$ in Fig. \ref{figsimulated1}b,
because the 
simulated $s(t)$ model
has constant frequency,  $(f_1+f_2)/2$
(Eq. \ref{freqsame}).
The $s(t)$ model amplitudes $A_1$ and $A_2$ 
remain constant  (Fig. \ref{figsimulated1}c: red and blue circles).
The sinusoidal 
\lsmethod ~model amplitudes
$A_{\mathrm{LS}}$ vary between 
$0^{\mathrm{m}}$ and
$0.^{\mathrm{m}}2$ 
(Eq. \ref{aminima}).
The $s_2(t)$  minimum phases are stable
(Fig. \ref{figsimulated1}d: blue circles),
while those of $s_1(t)$ undergo regular linear migration 
(Fig. \ref{figsimulated1}d: red circles).
The $\phi_{\mathrm{LS,min,1}}$
minima % of the sinusoidal \lsmethod ~models 
show nearly linear changes, 
and an abrupt 0.5 phase shift 
when the $y^*_i$
variation approaches zero
(Figs. \ref{figsimulated1}d: green circles).
This predicted shift occurs in
{\it every} \simuone
~(Eq. \ref{abruptphase}: $a_1=a_2$).

\subsubsection{\simutwo }
\label{simu2} 

The amplitudes
$a_1=0.^{\mathrm{m}}025$ 
and 
$a_2=0.^{\mathrm{m}}05$
are unequal in \simutwo.
The $y^*_i$ variation decreases 
from $0.^{\mathrm{m}}15$ to $0.^{\mathrm{m}}05$,
and then increases back to  $0.^{\mathrm{m}}15$
(Fig. \ref{figsimulated2}a).
The detected $P_{\mathrm{LS}}$ remain within
the tested period range,
because the $y^*_i$ variation 
never decreases to zero 
(Figs. \ref{figsimulated2}bcd: one highlighted circle).
The $P_1$ and $P_2$ periods do not, of course, change
(Fig. \ref{figsimulated2}b: red and blue circles).
The $P_{\mathrm{LS}}$ 
changes are largest in the middle of the segment
(Fig. \ref{figsimulated2}b: green circles).
They  concentrate
on the $P_2=1/f_2$ level, because
the signal of the stronger $s_2(t)$
sinusoid $(a_1<a_2)$ dominates
(Eqs. \ref{upperf} and \ref{lowerf}).
The $A_1$ and $A_2$ amplitudes of 
the $s(t)$ model
do not change (Fig. \ref{figsimulated2}c: red and blue circles).
The $A_{\mathrm{LS}}$ amplitudes vary
regularly, as predicted by Eq. \ref{bminima}
(Fig. \ref{figsimulated2}c: green circles).
The $s_1(t)$ and $s_2(t)$ minimum changes are linear
(Fig. \ref{figsimulated2}d: red and blue circles).
The fluctuating $\phi_{\mathrm{LS,min,1}}$ minimum changes are linear 
(Fig. \ref{figsimulated2}d: green circles).
The predicted phase shift 
is only $\Delta \phi_{\mathrm{b}}=0.17$ 
(Eq. \ref{abruptphase}).

\section{Incompatibility} 
\label{incomp} 

The relations in Sect. \ref{explanation}
and the simulations in Sect. \ref{simulations}
reveal ``\Inc''.
If the \JHLhyp ~is true,
it contaminates the 
periods, the amplitudes and the minima 
determined with one-dimensional period
finding methods.
It is stronger for the real data,
where the two real curves change.

\subsection{Period-\Inc} 
\label{periodinc} 

% ~/method/progs/newsimulate.py produces 4 files
% ~/method/hammer.py produces this figure
% ~/method/$ cp hammer.eps /home/jetsu/method/valmis2/
% ~/method/progs/cp hammer.eps /home/jetsu/method/texts/
% Edit: BoundingBox: 18 330 594 612
\begin{figure}
\begin{center}
\resizebox{7.0cm}{!}{\includegraphics{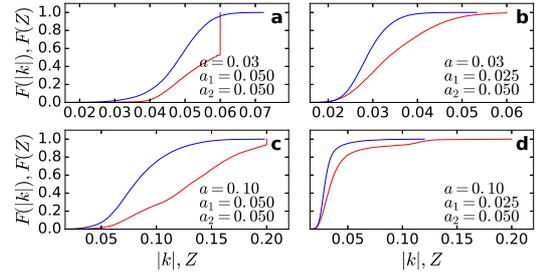}}
\end{center}
\caption{(a) \simuone.
Cumulative distribution functions 
$F(|k|)$ (red) and $F(Z)$ (blue) in 10~000 cases 
for $a=0.03$ in Eq. \ref{fsimraja}.
(b) \simutwo, otherwise as in ``a''.
(c) $a=0.10$, otherwise as in ``a''.
(d) $a=0.10$, otherwise as in ``b''.}
\label{hammer}
\end{figure}

The \lsmethod ~detects 
spurious $P_{\mathrm{LS}}$ 
period changes although 
the $g_{\mathrm{C}}(t)$-model periods $P_1$ 
and $P_2$ do not change. 
The $P_{\mathrm{LS}}$ periods in Fig. \ref{figsimulated1}b
represent only one particular 
\simuone, 
where $a=0.03$, 
$a_1=0.^{\mathrm{m}}050$ and $a_2=0.^{\mathrm{m}}050$ 
(Eqs. \ref{fsimraja}, \ref{sone} and \ref{stwo}).
We simulated 10~000 cases with this
$a$, $a_1$ and $a_2$ combination.
Each case
gave us one 
$|k|$ (Eq. \ref{k_two}) 
and $Z$ (Eq. \ref{zvalue}) estimate.
We show 
the cumulative distribution
functions $F(|k|)$ and $F(Z)$ 
in Fig. \ref{hammer}a.
The ranges are 
$0.03\le |k| \le 0.06$ and $0.02 \le Z \le 0.07$. % OK 15.04.2019
The $|k|$ values can not 
exceed $2a$,
because $P_{\mathrm{LS}}$ must be
within the tested period interval.
The steep $F(|k|)$ rise 
at $|k|=2a=0.06$
means that 
the \lsmethod ~detects $P_{\mathrm{LS}}$ periods
over the whole tested period interval
in nearly half of the cases.
The shape of $F(Z)$ resembles that of
a Gaussian cumulative distribution function
(Fig. \ref{hammer}a: blue line).

In Fig. \ref{hammer}b,
the ranges are 
$0.02\le |k| \le 0.06$ and 
$0.02 \le Z \le 0.05$ % OK 15.04.2019
for 10~000 cases of 
\simutwo ~($a=0.03$, $a_1=0.^{\mathrm{m}}025$, $a_2=0.^{\mathrm{m}}050$).
The \lsmethod ~can detect $P_{\mathrm{LS}}$
periods {\it within} the  tested period interval,
because $A_{\mathrm{LS}}$ does not decrease to zero,
unlike in \simuone.

The $a=0.10$, $a_1=0.^{\mathrm{m}}050$ and $a_2=0.^{\mathrm{m}}050$ 
\simuone ~shows results
for a longer $\pm 10$\% 
period interval (Fig. \ref{hammer}c).
A minor steep $F(|k|)$ rise still
occurs at $|k|=2a=0.20$.
The $a=0.10$, $a_1=0.^{\mathrm{m}}025$ and $a_2=0.^{\mathrm{m}}050$ 
\simutwo ~results 
are shown in Fig. \ref{hammer}d.
The \lsmethod ~still detects $P_{\mathrm{LS}}$
periods over the whole period interval,
because  $F(|k|)$ reaches $|k|=2a=0.20$.
Generally,
the $|k|$ and $Z$ ranges increase when 
the tested range increases.

Our Figs. \ref{hammer}a--d
reveal that 
$|k| \approx Z/h$ (Eq. \ref{k_three})
is a poor approximation,
because $|k|$ is computed from
{\it two} individual $P_{\mathrm{LS}}$ values, 
while $Z$ is computed 
from {\it all} $P_{\mathrm{LS}}$ values. 

% ~/method/progs/sdrsimulation.py produces this figure
\begin{figure}
\begin{center}
\resizebox{7.5cm}{!}{\includegraphics{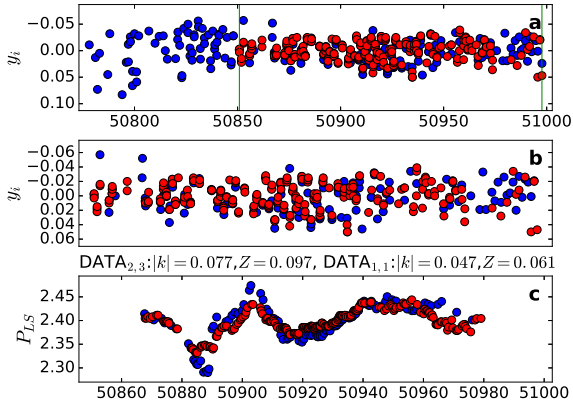}}
\end{center}
\caption{(a) Partly overlapping 
\Data{2}{3} and \Data{1}{1} 
(red and blue circles).
Fully overlapping data is between
green vertical lines.
(b) Fully overlapping data.
(c) Red and blue circles
denote $P_{\mathrm{LS}}$ periods detected 
from fully overlapping
data with 
the \lsmethod.
Results for  $|k|$ and $Z$ are 
given above this panel. }
\label{incompatibleP2}
\end{figure}

% DOUBLE CHECKED 19.06.2018
% Produced with .../progs/sdrsimulation.py
% evince nyt.eps
\begin{table}
\caption{Partly overlapping photometry
(Columns 1 and 4).
Values for $|k|$ and $Z$ in
fully overlapping photometry (other columns).}
\begin{center}
\begin{tabular}{lccclcc} 
\hline
\multicolumn{3}{c}{TEL=2}    & ~ &  \multicolumn{3}{c}{TEL=1}    \\
\cline{1-3} \cline{5-7}
Segment     &  $|k|$  & $Z$  & ~ & Segment       &   $|k|$ & $Z$   \\
\hline
\Data{2}{3}: & 0.077 & 0.097 & ~ & \Data{1}{1}:  & 0.047 & 0.061 \\
\Data{2}{4}: & 0.100 & 0.141 & ~ & \Data{1}{2}:  & 0.063 & 0.085 \\
\Data{2}{5}: & 0.022 & 0.032 & ~ & \Data{1}{3}:  & 0.012 & 0.016 \\
\Data{2}{6}: & 0.014 & 0.018 & ~ & \Data{1}{4}:  & 0.010 & 0.012 \\
\Data{2}{7}: & 0.079 & 0.068 & ~ & \Data{1}{5}:  & 0.070 & 0.096 \\
\Data{2}{8}: & 0.051 & 0.062 & ~ & \Data{1}{6}:  & 0.025 & 0.032 \\
\Data{2}{9}: & 0.088 & 0.091 & ~ & \Data{1}{7}:  & 0.048 & 0.038 \\
\hline
\end{tabular}
\end{center}
\label{compsegments}
\end{table}

Any $|k|$ or $Z$ value drawn from 
$F(|k|)$ and $F(Z)$ in Figs. \ref{hammer}a-d
is possible.
This predicts 
that two observers {\it observing} 
the {\it same} star 
during the {\it same} time interval
can get {\it different}
$P_{\mathrm{LS}}$, $|k|$ and $Z$ values with the \lsmethod.
We show the partly and fully overlapping 
\Data{2}{3} and \Data{1}{1}
photometry in Figs. \ref{incompatibleP2}ab.
We divide these data into 
subsets using a sliding
window of 30 days,
and apply \lsmethod % ~($a=0.05$) 
~to
all subsets having $n\ge12$ observations.
The $P_{\mathrm{LS}}$, $|k|$ and $Z$ results
in Fig. \ref{incompatibleP2}c
confirm the above prediction,
as well as
the respective results for other overlapping 
segment pairs (Table \ref{compsegments}).
If the \JHLhyp ~is true,
the \lsmethod ~gives different results
for the simultaneous 
photometry of the same star.
This general result  
applies to all one-dimensional
period finding methods.
This is amazing 
considering that 
the underlying
$g_{\mathrm{C}}(t)=s(t)$ model
does not change at all.

\begin{figure}
\resizebox{8.0cm}{!}{\includegraphics{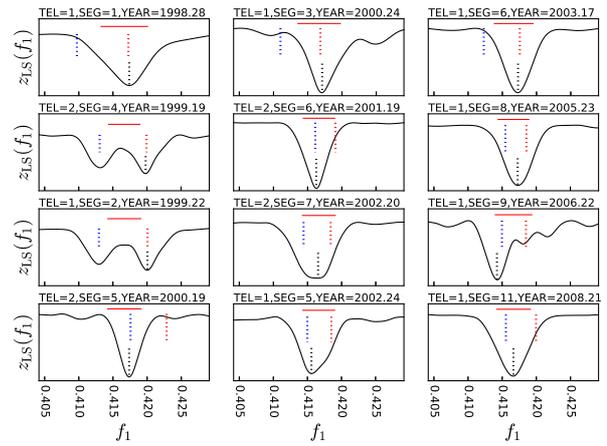}}
\caption{$z_{S}(f_1)$ 
  for segments of Table \ref{Cresults}.
  Tested $f_1$ range is $a\!=\!0.03$ 
  and $f_{\mathrm{mid}}\!=\!(2.^{\mathrm{d}}4)^{-1}$
  (Eq. \ref{fsimraja}).
Higher red and blue vertical dotted lines show 
$P_1$ and $P_2$ periods detected with
$g_{\mathrm{C}}(t)$-method. 
Lower black vertical dotted lines show
$P_1$ period detected with
$g_{\mathrm{S}}(t)$-method.
Red horizontal line is $f_{\mathrm{mid}} \!\pm\! f_0/2$ range.}
\label{segmentpeaks}
\end{figure}

The one-dimensional period finding methods can sometimes
detect signs of the two real periods. 
For overlapping \Data{2}{4} and \Data{1}{2},
the $g_{\mathrm{S}}(t)$-method succeeds in this.
It detects the $P_1$ period, which
nearly coincides with the $P_1$ period detected 
with the two-dimensional
$g_{\mathrm{C}}(t)$-method
(Fig. \ref{segmentpeaks}:
lower black and higher red dotted vertical lines).
For both segments,
the second $z_{\mathrm{S}}(f_1)$ periodogram
minimum also nearly coincides
with the $P_2$ period detected with 
$g_{\mathrm{C}}(t)$-method (Fig. \ref{segmentpeaks}: blue dotted vertical line).
In these two segments,
the distance between the 
$1/P_1$ and $1/P_2$ frequencies
exceeds $f_0$ (Fig. \ref{segmentpeaks}:
red horizontal line).
If this distance goes below $f_0$,
{\it only one} real period can be detected
(e.g. \Data{1}{3} or \Data{2}{5}),
or {\it none} at all
(e.g. \Data{1}{8} or \Data{2}{7}).
Many weaker signals are not detected 
(e.g. \Data{1}{1} or \Data{1}{6}).

% 22.03.2019
% /method/progs/segmentpeaks.py ..
% cp segmentpeaks1.eps /home/jetsu/method/texts/
% ~/method/progs$ cp segmentpeaks1.eps /home/jetsu/method/progs/valmis2/

% /home/jetsu/method/progs/peaksmerge.py
% cp peaks1.eps /home/jetsu/method/texts/
% cp peaks1.eps /home/jetsu/method/progs/valmis2/
\begin{figure}
\begin{center}
\resizebox{7.0cm}{!}{\includegraphics{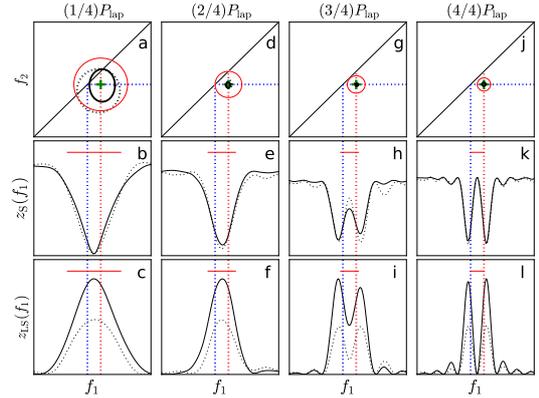}}
\end{center}
\caption{An arbitrary \simuone.
  (a)  First $(1/4)P_{\mathrm{lap}}$ quarter.
  Large green dot shows 
  $1/P_1$ and $1/P_2$.
  Blue and red dotted lines show these levels.
  Tested $f_1$ and $f_2$ ranges are $a\!=0\!.02$ 
  and $f_{\mathrm{mid}}\!=\!(2.^{\mathrm{d}}4)^{-1}$
  (Eq. \ref{fsimraja}).
  Dotted black ellipse shows accuracy for
  detected periods 
  (Eq. \ref{blackellipse}).
  Continuous black ellipse
  shows accuracy,
  if $n$ is doubled from 300 to 600.
  Red circle is $\pm f_0$ range (Eq. \ref{redcircle}).
  (b) Black dotted line is
  $z_{\mathrm{S}}(f_1)$ periodogram.
  Black continuous line is same periodogram
  for doubled sample size $(n\!=\!600)$.
  Red horizontal line is $\pm f_0/2$ range.
  Blue and red vertical lines denote
  $1/P_1$ and $1/P_2$ from ``a''.
  (c) $z_{\mathrm{LS}}(f_1)$ periodograms for same
  samples as in ``b''.
  (d-f) Two  $(2/4)P_{\mathrm{lap}}$ quarters,
  otherwise as in ``a-c''.
  (g-i) Three  $(3/4)P_{\mathrm{lap}}$ quarters,
  otherwise as in ``a-c''.
  (j-l) All  $(4/4)P_{\mathrm{lap}}$ quarters,
  otherwise as in ``a-c''.
}
\label{peaksmerge}
\end{figure}

The power spectrum methods,
like the \lsmethod, can detect two real frequencies {\it only,
if} their distance exceeds $1/\Delta T$ \citep{Lou78}.
Our $g_{\mathrm{C}}(t)$-method
does not suffer from this limitation.
We show what happens,
if $\Delta T$ remains constant,
but the sample size is
doubled from $n=300$ to $n=600$ in \simuone.
The numerical frequency
and periodogram values are ignored
in our Fig. \ref{peaksmerge},
because we are only interested in
the qualitative results. 
The results for the
first quarter $(1/4)P_{\mathrm{lap}}$ of
simulated data are shown in Figs. \ref{peaksmerge}a-c.
More accurate best periods are detected with the
$g_{\mathrm{C}}(t)$-method when the sample
size is doubled from $n=300$ to 600
(Fig. \ref{peaksmerge}a: dotted and continuous black ellipse).
\citet{Lou78} predict that the two real
periods $P_1$ and $P_2$ are not detected,
because the red $f_0/2$ circle of Eq. \ref{redcircle} 
intersects the diagonal
$f_1=f_2$ line (Fig. \ref{peaksmerge}a).
The $z_{\mathrm{S}}(f_1)$ and $z_{\mathrm{LS}}(f_1)$ 
periodograms confirm this (Figs. \ref{peaksmerge}bc).
When $\Delta T$ increases to  $(2/4)P_{\mathrm{lap}}$,
the black dotted and continuous 
ellipses of Eq. \ref{blackellipse}
shrink
(Fig. \ref{peaksmerge}d).
Since the red $f_0/2$ circle of Eq. \ref{redcircle}
still intersects the diagonal 
$f_1=f_2$ line (Fig. \ref{peaksmerge}d),
the two one-dimensional 
period finding methods can not detect the two real periods
(Figs. \ref{peaksmerge}ef).
In the $(3/4)P_{\mathrm{lap}}$ sample,
the dotted and continuous 
accuracy ellipses of Eq. \ref{blackellipse}
reduce into a single point (Fig. \ref{peaksmerge}g).
Now the $f_0/2$ circle of Eq. \ref{redcircle} does not
intersect the $f_1=f_2$ line, 
and the $z_{\mathrm{S}}(f_1)$ and $z_{\mathrm{LS}}(f_1)$ 
periodograms show signs of two periods close to, 
but not exactly at,
the real
frequencies $1/P_1$ and $1/P_2$ 
(Figs. \ref{peaksmerge}hl).
In the $(4/4)P_{\mathrm{lap}}$ sample, 
the $g_{\mathrm{C}}(t)$-method detects extremely accurate
real $P_1$ and $P_2$
values (Fig. \ref{peaksmerge}j).
The $z_{\mathrm{S}}(f_1)$ and $z_{\mathrm{LS}}(f_1)$ 
periodograms can also resolve these periods,
but not their exact values (Figs. \ref{peaksmerge}kl).
These two periodograms are ``mirror images'' of each other,
because they are sensitive to the same periodicities.
The $z_{\mathrm{LS}}(f_1)$
periodogram peak height
increases when the sample size $n$ increases,
but the resolution depends only on
the red horizontal line $f_0=1/\Delta T$
(Fig. \ref{peaksmerge}).
This line equals the resolution limit
for all one-dimensional period finding methods.
Our $g_{\mathrm{C}}(t)$-method resolution
is not limited by $\Delta T$.
It can detect two periods even in short samples, 
if the sample size $n$ and/or the accuracy $\sigma_y$ 
are sufficient.

Our Figs. \ref{segmentpeaks} and \ref{peaksmerge}
illustrate period-\Inc.
Everything
interesting happens at the lowest
$z_{\mathrm{S}}(f_1)$ valleys
and 
inside the highest
$z_{\mathrm{LS}}(f_1)$ peaks,
because
$P_{\mathrm{lap}}\!=\!527^{\mathrm{d}}$
exceeds segment $\Delta T$.
Since the discovery of starspots 
\citep{Kro47},
their light curves have been forced
to play one tune, instead of two 
(Sect. \ref{explanation}: All equations).
For example, period-\Inc ~misled
the \lsmethod ~analysis of
the photometry of 40~661 stars \citep{Rei13}.
The tested periods were between $0.^{\mathrm{d}}5$ and $45^{\mathrm{d}}$.
The resolution was $f_2=f_1-f_0$ for
$\Delta T=90^{\mathrm{d}}=1/f_0$.
For a star with $P_1=10^{\mathrm{d}}$,
the \lsmethod ~could resolve real longer 
periods $P_2 > 11.25^{\mathrm{d}}$.
Many real periods hidden within the \lsmethod ~periodogram
peaks must have gone undetected, like in
Figs. \ref{peaksmerge}cf.
If another  peak was detected in such cases,
\sdr ~was overestimated,
and even more so for 
$P>10^{\mathrm{d}}$ stars. 

\citet{Rei13A} searched for
multiple periodicities.
They determined the best 
period with the \lsmethod ~and
removed the periodic sinusoid from the data.
The \lsmethod ~analysis of the residuals 
gave the next sinusoid,
the next residuals and
their pre-whitening continued.
Our \simuone ~and \simutwo ~are
exactly {\it what} \citet{Rei13A}
tried to find in real data:
the sum of at least two sinusoids.
The \lsmethod ~can miss the real period(-s),
if the data contains {\it only two} sinusoids
(Figs. \ref{segmentpeaks} and
\ref{peaksmerge}).
Even the first stage of pre-whitening 
can fail, e.g.
a sine fit to the original
  data would eliminate both real periods in
  \Data{1}{8},
  \Data{2}{7} and
  \Data{1}{11}
  (Fig. \ref{segmentpeaks}).
For close real periods,
similar misleading pre-whitening may occur in
asteroseismology \citep[e.g.][]{Kra15,Sai18},
exoplanet detection \citep[e.g.][]{Que09,Hat13}
and other fields \citep[e.g.][]{Rey09}.

If the data 
contains two periodic signals,
the \lsmethod ~gives {\it different} $P_1$,
$|k|$ and $Z$ estimates for the {\it same}
star when it is observed with two 
different telescopes at the {\it same} time
(Fig. \ref{incompatibleP2} and Table \ref{compsegments}).
These estimates  measure the random failures of 
the \lsmethod, not SDR.
For small $|P_1\!-\!P_2|$, 
the one-dimensional period finding methods
have missed innumerable real periods
and detected short-lived starspots that never existed
\citep[e.g.][]{Rei13A,Rei15,Leh16}.
They can only detect large $|P_1\!-\!P_2|$,
and tell very little about
physical phenomena, just like
the forcing of a square through a circle.
The  correct estimate is
obtained only for the mean brightness,
just like the centres of a square and a
circle do coincide. 
The rest makes no sense,
like the periods, the amplitudes and the minima.
If the \JHLhyp ~is true,
the one-dimensional
period finding methods
fail to model the
real light curves.
{\it Many} spurious periods have been detected
for the {\it same} star (\Testone).

\subsection{Amplitude-\Inc}
\label{amplitudeinc}

The $A_{\mathrm{LS}}$ amplitudes
in Figs. \ref{figsimulated1}c and 
\ref{figsimulated2}c (green circles)
follow the $P_{\mathrm{lap}}$ cycle,
and tell nothing about the {\it real} 
amplitudes $A_1$ and $A_2$ (red and blue circles).
All one-dimensional
period finding methods
suffer from ``amplitude-\Inc''.
The period- and the amplitude-\Inc ~cause
spurious activity cycles, e.g.
the apparently quasi-periodic 
$P_{\mathrm{LS}}$ changes
in Fig. \ref{figsimulated1}b
are simply random fluctuations.

\citet{Rie84} detected a $154^{\mathrm{d}}$ cycle in the solar
$\gamma$-ray flares.
This Rieger cycle has also been detected, 
e.g. in the sunspot area \citep{Lea90,Oli98},
the solar flares \citep[][]{Bai93,Bai03}
and the Mt Wilson Sunspot Index  \citep[][]{Bal02}. 
Stellar analogues have 
been detected \citep[e.g.][]{Mas07,Lan09,Bon12,Dis17}.
The light curve amplitudes and periods were 
determined with the most widely used 
one-dimensional period finding method, the \lsmethod. 
Then quasiperiodic $P_{\mathrm{LS}}$ and $A_{\mathrm{LS}}$
short-term changes, stellar Rieger cycles,
were detected with the same method 
\citep[e.g.][]{Dis17}. 
If the \JHLhyp ~is true,
the incompatible $P_{\mathrm{LS}}$ and $A_{\mathrm{LS}}$ estimates
had nothing to do with 
the real amplitudes $(A_1,A_2)$ and the real periods $(P_1,P_2)$.
Constant brightness
does not mean that the starpots are absent, nor that
the stellar surface is fully covered with starspots.
Incompatible amplitudes have also been
used as a proxy for stellar activity \citep[e.g.][]{Jet90,Fer15},
or in an activity index which is proportional to these amplitudes
\citep[e.g.][their $A_1$ parameter]{Ark15}.

As for the long-term cycles, 
an \lsmethod ~analysis of the photometry
23601 stars revealed amplitude cycles,
but not period cycles 
\citep{Rei17}.
This supports the \JHLhyp, because long-term 
$P_{\mathrm{lap}}$ cycles can be detected
in the incompatible $A_{\mathrm{LS}}$ amplitudes, 
but only short-term random fluctuations
in the incompatible $P_{\mathrm{LS}}$ periods.
The $P_{\mathrm{lap}}$ cycle is not an activity cycle,
because it is not connected to the number,
the area or the temperature of starspots.
There should be no \Inc ~in the 
seasonal mean brightness $M$ activity cycles, 
because these estimates can be
obtained directly from the mean of the data.
\citet{Leh16} studied the activity of 21 
young solar-type stars.
The mean $M_{\mathrm{CPS}}$, 
the period $P_{\mathrm{CPS}}$,
and
the amplitude $A_{\mathrm{CPS}}$ 
estimates 
were determined with the one-dimensional \cpsmethod. 
They detected $P_{\mathrm{cyc }}$ activity cycles in 
the $M_{\mathrm{CPS}}$, $A_{\mathrm{CPS}} $, 
$M_{\mathrm{CPS}}-A_{\mathrm{CPS}}/2$ 
and $M_{\mathrm{CPS}}+A_{\mathrm{CPS}}/2$ estimates.
If the \JHLhyp ~is true, then only the $M_{\mathrm{CPS}}$ 
cycles were perhaps real,
but the $A_{\mathrm{CPS}}$, $M_{\mathrm{CPS}}-A_{\mathrm{CPS}}/2$ 
and $M_{\mathrm{CPS}}+A_{\mathrm{CPS}}/2$ 
cycles were spurious.
\citet{Leh16} compared their
$P_{\mathrm{CPS}}/P_{\mathrm{cyc}}$ ratios 
to those in \citet{Bra98} and \citet{Saa99}.
These spurious $P_{\mathrm{cyc}}$ cycles in 
$A_{\mathrm{CPS}}$, $M_{\mathrm{CPS}}-A_{\mathrm{CPS}}/2$ 
and $M_{\mathrm{CPS}}+A_{\mathrm{CPS}}/2$ 
support the results of
numerous later studies 
\citep[e.g.][]{Fab17,Bra17,Rei17,Nie17,War17A,Flo17,Lan18,Kos18,Ale18,Bra18,Ols18,War18}.
For example,
  the rejection of other similar
  spurious $P_{\mathrm{cyc}}$ cycles
may, or may not, confirm the existence
of separate $P_{\mathrm{rot}}/P_{\mathrm{cyc}}$
branches 
\citep[e.g.][]{Saa99,Boh07,Ola09,Leh16}. % Bra98, Lor05

\subsection{Minima-\Inc}
\label{minimainc}

The Kuiper method analysis 
of the 
$t_{\mathrm{LS,min,1}}$ epochs in
Fig. \ref{figsimulated1}d would not reveal
the simulated $s(t)$ model periods
$P_1$ or $P_2$,
but the migration of these
green circles in Fig. \ref{figsimulated2}d 
would yield information of the 
stronger $P_2$ signal
$(a_1 < a_2)$.
The Kuiper method can unambiguously 
detect $P_2$ when $a_1\!=\!0$ (\secondcase),
or $P_1$ when $a_2\!=\!0$ (\thirdcase),
but the light curves of
real stars give no 
information of when $a_1\!=\!0$ or $a_2\!=\!0.$
The $t_{\mathrm{LS,min,1}}$  migration alternatives 
are

\begin{description}

\item  \migrone: If $a_1 \approx a_2$ 
(Eq. \ref{fsame}),
the $t_{\mathrm{LS,min,1}}$  migration is linear 
with $(f_1+f_2)/2$ (Eq. \ref{freqsame})

\item  \migrtwo: If $a_1 < a_2$
(Eq. \ref{fnotsame}),
the $t_{\mathrm{LS,min,1}}$  migration is linear  
with $f_2$ and fluctuating
(Eqs. \ref{upperf} and \ref{lowerf})

\item  \migrthree: If $a_1 > a_2$
(Eq. \ref{fnotsame}),
the $t_{\mathrm{LS,min,1}}$  migration is linear 
with $f_1$ and fluctuating
(Eqs. \ref{upperf} and \ref{lowerf})

\end{description}

\noindent
The Kuiper method may detect
$(f_1+f_2)/2$, $f_1$ or $f_2$.
The stronger signal usually dominates 
this migration,
because it is reasonable to assume that
$a_1 \approx a_2$ occurs
less frequently than $a_1<a_2$ or $a_1>a_2$.
The Kuiper method may not
detect $f_1$, $f_2$ or $(f_1+f_2)/2$,
because the $a_1$ and $a_2$ amplitudes change
and the light curves become
a mixture of these 
three migration alternatives.
There are also other migration 
alternatives than only these three, 
because all light curves are certainly not
sums of two sinusoids.
The following minima-\Inc ~effects mislead the Kuiper 
method analysis

\begin{itemize}

\item[1.] Many different spurious migration patterns
\item[2.] Mixture of minima from all these patterns
\item[3.] Abrupt phase shifts 

\end{itemize}

\noindent 
If the \JHLhyp ~is true,
the Kuiper method can not detect an unambiguous
$P_{\mathrm{act}}$ period for FK Com (\Testtwo).
It can not separate the $P_{\mathrm{rot}}$ signal
from the $P_{\mathrm{act}}$ signal
(\Testthree),
or vice versa (\Testfour).
Abrupt phase shifts $\Delta \phi = 0.5 \equiv 180\degr$ occur
when $a_1 \approx a_2$ (\Testfive).

The $P_{\mathrm{lap}}$ cycle detection in
$P_{\mathrm{orb}}\approx P_{\mathrm{rot}}=1/f_2$  binaries
gives $ f_1 = f_{\mathrm{lap}}\pm f_2$, where
$1/f_1= P_{\mathrm{act}}$  (Eq. \ref{lapcycle}).
The correct $\pm f_2$ sign can be inferred
when $f_1$ dominates the migration.

The above migration patterns 
were named ``active longitudes'' 
\citep[e.g.][]{Zei90,Jet96}.
The starspots seemed to form 
at these long-lived longitudes, 
then detach from them, and
finally migrate with their own particular
angular velocity before decaying.
For example, \citet{Gil17} analysed
the {\it Kepler} satellite light curves
with the \lsmethod, and noted that the
light curve
amplitudes and phases vary due to rapid
starspot formation and decay. 
 %of short photometric samples 
The one-dimensional period finding methods
deliver an impression that a lot happens,
although nothing happens,
like in Figs. \ref{figsimulated1} and \ref{figsimulated2}.
The starspots appear lively 
when in fact they are not.
This failure in modelling
time intervals longer than one month 
has nothing to do with the starspot life-times.
We can model half a year
of photometry with
two long-lived starspots
(Sect. \ref{SectTel1Seg9}: \Data{1}{9}).
No active longitudes,
where the short-lived starspots form, migrate and decay,
are needed.
\citet[][]{Leh16}
detected active longitudes 
only in the strong chromospheric Ca~II H\&K emission
solar-type stars.
If no active longitudes exist, 
then the starspots in these stars
simply are more stable.

\citet{Jet91,Jet93} discovered the
\flip ~phenomenon in FK Com.
The starspots concentrated on two
active longitudes separated by 180 degrees.
An abrupt shift from one active longitude 
to another happened only three times
during a quarter of a century.
The models of non-axisymmetric stellar
magnetic fields failed to explain this
phenomenon \citep[e.g.][]{Mos95}.
Similar active longitudes were discovered
in CABSs \citep[e.g.][]{Ber98A},
as well as in the Sun \citep[e.g.][]{Jet97A}.
\citet{Mos97} modelled the 
non-axisymmetric magnetic fields of CABSs.
\cite{Mos99} demonstrated how the non-linear 
solar dynamo models could produce 
weak large-scale non-axisymmetric magnetic
fields. 
\citet{Tuo02} predicted that 
the \flip ~phenomenon 
is the stellar counterpart of 
Hale's polarity law in the bipolar sunspots,
and that observations may later confirm 
this ``active star Hale polarity rule''.
\citet{Ber03} discovered
two persistent active longitudes 
separated by $180 \degr$
from sunspot observations over 120 years.
\citet{Pel05} argued that this result was biased.
This debate continues 
\citep{Ber07,Bal07,Ver07,Tuo07,Gye14}.
Many dynamo models can
explain the  \flip ~phenomenon 
\citep[e.g.][]{Kor05,Ber06,Kor11,Col14,Pip15,Viv18}. % Man13
These models have also been used
to infer the coronal structure of
FK Com \citep{Dra08,Coh10}.

\citet{Ber98A} detected
\flip ~cycles in four CABSs.
These are not activity cycles,
because they follow the $P_{\mathrm{lap}}$ cycle period.
The
$\Delta \phi_{\mathrm{b}}$ phase shifts depend on 
the amplitudes $a_1$ and $a_2$ (Eq. \ref{abruptphase}).
The \flip ~$P_{\mathrm{lap}}$ cycles are
not so easy to detect as the $P_{\mathrm{lap}}$
cycles of incompatible amplitudes,
because a ``perfect'' $\Delta \phi_{\mathrm{b}}=0.5$ \flip ~requires
$a_1 \approx a_2$.
These cycles can be shorter than $P_{\mathrm{lap}}$.
For example, if the $g_1(t)$ and $g_2(t)$ curves are double waves,
the resulting \flip ~cycle is $P_{\mathrm{lap}}/2$.
%python /methos/progs/errorlap.py 
%python errorlap.py 
%526.8853440750929
%28.9134183947
The periods 
$P_1=2.^{\mathrm{d}}39321 \pm 0.^{\mathrm{d}}00036$ 
and 
$P_2=2.^{\mathrm{d}}40413 \pm 0.^{\mathrm{d}}00048$ 
(Eqs. \ref{Kone} and \ref{Ktwo})
give $P_{\mathrm{lap}}=527^{\mathrm{d}} \pm 29^{\mathrm{d}}$.
Thus, a perfect 
\flip ~event of FK Com
is $(a_1 \approx a_2)$, 
or is not $(a_1 \neq a_2$),
observed
after every 1.44 years.
The ``phase jumps'' and the \flip ~events of FK Com
identified by \citet{Ola06} are simply manifestations
of different $\Delta \phi_{\mathrm{b}}$ values.                        

\citet{Kor99} detected
a $6.^{\mathrm{y}}5$ \flip ~cycle 
from their Doppler images of FK Com.
Other cycles were detected by
\citet[][$3^{\mathrm{y}}$ or $6^{\mathrm{y}}$]{Kor01A},
\citet[][$6.^{\mathrm{y}}4$]{Kor02}
and
\citet[][$2.^{\mathrm{y}}6$]{Kor04}.
\citet{Kor07} detected
no cycle in their 25 Doppler images. 
FK Com can 
``fool'' its observers, because the short
$P_{\mathrm{lap}}=527^{\mathrm{d}}$ cycle misleads
comparison of Doppler images within a year.

The \JHLhyp ~neatly 
explains the \flip.
The small $|P_2-P_1|$ difference in FK Com
delayed the detection of the structure causing these events.
A quarter of a century was wasted in trying to
find this cause.
It was absurd to question
{\it who} 
discovered this phenomenon
or named it 
\citep[][Sect. 2.1]{Tuo02}.
No-one wants to take that credit now.
Nevertheless, 
we quote one
sentence from \cite{Mos97}:
{\it `The visible fields then have maxima at the
longitudes corresponding to the 
intersection of the line of centres 
with the stellar surfaces.''}
They misunderstood the cause of \flip, but
their result agreed perfectly
with the \JHLhyp, because the stationary 
starspots in all fourteen CABSs
concentrated on this intersection line
\citep[][Figs. 1-14, ``b-g'' panels]{Jet17}.

We conclude that all earlier interpretations
of the \flip ~events or the \flip ~cycles
are no longer valid
\citep[e.g.][]{Jet91,Ber98A,Mos04,Flu04,Els05,Sav08}. 
Something that never happens was successfully
modelled, although this apparently 
dramatic phenomenon is merely
interference 
(Eq. \ref{abruptphase}).
This structure predicted by 
the \JHLhyp ~may be
quite common,
because these
\flip ~events have been observed in stars of
different spectral type and different luminosity class,
including binaries \citep[e.g.][]{Ber98A,Jet17},
semi-detached and contact binaries \citep[e.g.][]{Wan15,Kou19}
and
single stars \citep[e.g.][]{Jet93,Jar08}.
Perhaps these \JHLhyp ~long-lived starspots
have not yet been detected in other stars,
because their $|P_1-P_2|$ difference is also small.
%\citep[][Contact binary]{Mit18} \\
%\citep[][Single main sequence star]{Jar08} \\
%\citep[][Contact and semi-detached binaries]{Kou19} \\
%\citep[][W Uma-binary=contact binary]{Wan15} \\
%\citep[][II Peg=K2V, EI Eri=G5 IV,Sigma Gem= K1 III]{Ber98A}
The formation, the migration and the decay
of short-lived starspots, as well as the active longitudes, 
are also spurious phenomena.
This explains the earlier contradicting 
starspot life-time estimates
\citep[e.g.][]{Hal90,Hus02,Str09,Bra14,Bas18}. % Hal94,Str94,Ber12

If the amplitudes and the minima are incompatible,
so is the light curve {\it shape}.
Studies of this shape 
tell nothing about the 
starspot distribution \citep[e.g.][]{Rei15A}.

\section{CPS-method connection}
\label{cpsexplained}

%python predictff.py 
%\begin{verbatim}
%MIN= 2.0279 PMAX= 2.7653
%13 times K=0
%PMIN= 2.017 PMAX= 2.7381
%1 times K=0
%IND=1 gives 2.39838 0.0120016 n= 136
%IND=0 or 1 gives  2.39775 0.0292693 n= 1450
%IND=0 or 1 gives (0.2 from 2.4) 2.39828 0.0105295 n= 1412
%number of removed P values 38
%number above 2.6  3
%number below 2.4 35
%13 times K=0
%1 times K=0
%13 times K=0
%53722.936 2.39313 51874.421 2.40762
%1 times K=0
%53722.936 2.39313 51874.421 2.40762
%13 times K=0
%51874.421 2.40762 53722.936 2.39313
%1 times K=0
%51874.421 2.40762 53722.936 2.39313
%TEL=1 values  760
%TEL=2 values  704
%\end{verbatim}

\begin{figure}
\resizebox{8.0cm}{!}{\includegraphics{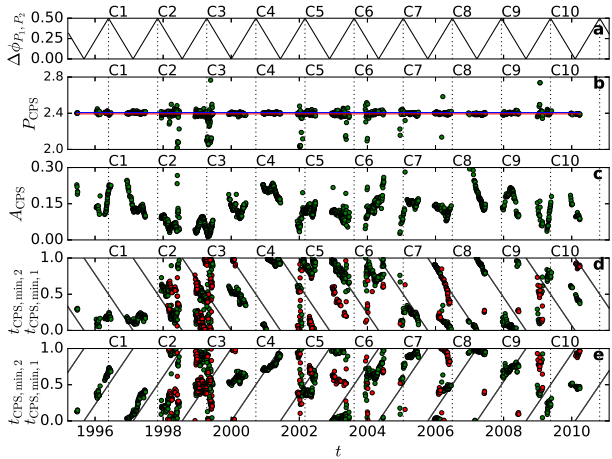}}
%\resizebox{16.5cm}{!}{\includegraphics{ff1.jpg}}
\caption{
\cpsmethod ~results (Table  \ref{electric}).
(a) Phase difference $\Delta \phi_{P_1,P_2}$ 
of Eq. \ref{phasediff} (continuous zig-zag line) and
$\Delta \phi_{P_1,P_2}=0.5$ (vertical dotted lines).
Lap cycles $P_{\mathrm{lap}}$
are from C1 to C10.
(b) Periods $P_{\mathrm{CPS}}$. 
Red and blue continuous lines
denote  $P_{\mathrm{w,1}}$ and $P_{\mathrm{w,2}}$ 
levels (Eqs. \ref{Pwone} and \ref{Pwtwo}).
(c) Amplitudes $A_{\mathrm{CPS}}$.
(d) Phases $\phi_2$ of 
Eq. \ref{ephetwo} ephemeris for
primary 
$t_{\mathrm{CPS,min,1}}$ (green circles)
and secondary $t_{\mathrm{CPS,min,2}}$ (red circles)
minima.
Tilted lines denote $\phi_1=0$ of Eq. \ref{epheone}.
(e) Same as in ``d'', except that
phases $\phi_1$ are computed from 
Eq. \ref{epheone} ephemeris.
Tilted lines denote $\phi_2=0$ of Eq. \ref{ephetwo}.}
\label{finalsolution}
\end{figure}

The \JHLhyp ~predicts 
the \cpsmethod ~results for FK Com
(Table \ref{electric}).
In Figs. \ref{finalsolution}a-e,
we connect them
to 
Eqs. \ref{epheone} and \ref{ephetwo}
ephemerides.
The zig-zag line
in Fig. \ref{finalsolution}a is the phase difference 
between the $g_1(t)$ and $g_2(t)$ minima
\begin{eqnarray}
\Delta \phi_{P_1,P_2} = (2\pi)^{-1}
\arccos{
[
\cos{\theta_1}
\cos{\theta_2}
+
\sin{\theta_1}
\sin{\theta_2}
]
}
\label{phasediff}
\end{eqnarray}
computed from
the $t_1$ zero epoch 
and the $P_1$  period of Eq. \ref{epheone},
the $t_2$ zero epoch and
the $P_2$  period of Eq. \ref{ephetwo},
and the phase angles
$\theta_1=2\pi(t-t_1)/P_1$ and 
$\theta_2=2\pi(t-t_2)/P_2$.
The $\Delta \phi_{P_1,P_2}=0.5$ epochs
are $P_{\mathrm{lap}}=527^{\mathrm{d}}$ apart
(Figs. \ref{finalsolution}a-e: vertical lines).
There are ten 
 cycles C1-C10.
The $P_{\mathrm{CPS}}$, 
$A_{\mathrm{CPS}}$,
$t_{\mathrm{CPS,min,1}}$
and 
$t_{\mathrm{CPS,min,2}}$
changes  during each 
{\it individual} $P_{\mathrm{lap}}$ cycle 
(Figs. \ref{finalsolution}b-d)
are analogous to 
the simulated changes during the {\it one}
$P_{\mathrm{lap}}$ cycle 
(Figs. \ref{figsimulated1}b-d and \ref{figsimulated2}b-d).

The red and blue lines in Fig. \ref{finalsolution}b
denote
the $P_{\mathrm{WK,1}}$ and $P_{\mathrm{WK,2}}$ period
levels (Eqs. \ref{epheone} and  \ref{ephetwo}).
\citet{Hac13} computed their weighted mean,
$P_{\mathrm{w,CPS}}\pm \Delta P_{\mathrm{w,CPS}}
=2.^{\mathrm{d}}398\pm0.^{\mathrm{d}}012$, from
independent $P_{\mathrm{CPS}}$ estimates $(n=136)$.
All $P_{\mathrm{CPS}}$ estimates 
give the same, but less accurate value,
$P_{\mathrm{w,CPS}}\pm \Delta P_{\mathrm{w,CPS}}
=2.^{\mathrm{d}}398\pm0.^{\mathrm{d}}029$  $(n=1450)$.
Removing the 38 spurious
$P_{\mathrm{CPS}} < 2.^{\mathrm{d}}2$ 
and
$P_{\mathrm{CPS}} > 2.^{\mathrm{d}}6$ values
gives  
$P_{\mathrm{w,CPS}}\pm \Delta P_{\mathrm{w,CPS}}
=2.^{\mathrm{d}}398\pm0.^{\mathrm{d}}010$  $(n=1412)$.
It differs $0.^{\mathrm{d}}001 \pm 0.^{\mathrm{d}}010$ 
from the mean 
$(P_{\mathrm{WK,1}}+P_{\mathrm{W2,2}})/2
=2.^{\mathrm{d}}39867 \pm 0.^{\mathrm{d}}00062$.
Our Eqs. \ref{freqsame}, \ref{upperf} and \ref{lowerf}
predict 
$P_{\mathrm{WK,1}} < P_{\mathrm{w,CPS}} < P_{\mathrm{WK,2}}$,
because the $P_{\mathrm{WK,1}}$ or 
the $P_{\mathrm{WK,2}}$ signal
can dominate in different segments.
This relation is
fulfilled.

 The ephemerides of Eqs. \ref{epheone} and \ref{ephetwo} 
predict that the vertical
dotted $\Delta \phi_{P_1,P_2}=0.5$ lines
coincide with  
the largest $P_{\mathrm{CPS}}$ dispersion
in Fig. \ref{finalsolution}b,
the lowest $A_{\mathrm{CPS}}$ 
in Fig. \ref{finalsolution}c,
and the abrupt phase shifts in Figs. \ref{finalsolution}de
(Eq. \ref{abruptphase}: $\Delta \phi_{\mathrm{b}}$).
As predicted by Eq. \ref{abruptphase},
the majority of secondary minima 
(Fig. \ref{finalsolution}de: red circles)
and the abrupt $\Delta \phi_{\mathrm{b}}$ shifts
occur when $A_{\mathrm{CPS}}$ are low,
like in cycles C2, C3 or C9.
Since we know that these shifts depend on the 
$a_1/a_2$ ratio, 
we can identify
``perfect'' $\Delta \phi_{\mathrm{b}}=0.5$ {\flip}s 
on the C2-C3 and C7--C8 cycle borders.
They coincide with low $A_{\mathrm{CPS}}$ 
(Eq. \ref{aminima}: $a_{\mathrm{b,min}}$ epochs),
but {\it never} with high
$A_{\mathrm{CPS}}$ 
(Eq. \ref{aminima}: $a_{\mathrm{b,max}}$ epochs).
The $P_{\mathrm{WK,2}}$ periodicity dominates 
the migration of green circles
between 1995 and 1997 (Fig. \ref{finalsolution}d).
The other periodicity, $P_{\mathrm{WK,1}}$, 
clearly dominates during C8 and C10.
This migration is regular,
when the $A_{\mathrm{CPS}}$ amplitudes are high,
like in cycles C1, C3, C4 or C8.
\citet{Jet17} detected 
similar regularities
in the photometry of fourteen CABSs.
Our Eqs. \ref{epheone} and \ref{ephetwo}
do not need to explain all details
in Figs. \ref{finalsolution}b-e,
because the real $g_1(t)$ and $g_2(t)$ light
curves of FK Com are not sinusoids,
and can change {\it between} and {\it within} segments.

The migration trends
from Fig. \ref{finalsolution}d
are reversed
in Fig. \ref{finalsolution}e
when
the phases 
are computed
from the ephemeris
of Eq. \ref{ephetwo}.
The horizontal migration 
distance scale is larger than 
the tilted migration scale.
The latter always seems more regular
than the former, e.g.
the {\it same} migration between
1995 and 1997 
seems more regular in Fig. \ref{finalsolution}e than 
in Fig. \ref{finalsolution}d. 
The human eye is fooled
by this ``corridor effect''.
\citet[][their Fig. 1]{Ber98A} utilized
this effect when they
added an arbitrary number orbital
rounds to the {\it titled} pairs of migration lines
of EI Eri,
II Peg and HR 7275.
This corridor
effect delivered the discovery of 
``permanent'' active longitudes.
If the \JHLhyp ~is true,
the minima determined with any 
one-dimensional period finding method
never follow such permanent migration
to only one direction.
There are at least 
three possible directions
(Sect. \ref{minimainc}: \migrone, \migrtwo ~and \migrthree).

% python predictff.py produces this ff1.eps
% ~/method/progs/$ cp ff1.eps /home/jetsu/method/texts/
% ~/method/progs/$ cp ff1.eps /home/jetsu/method/progs/valmis2/

\section{Spot modelling connection}
\label{sectfrontback}

Each individual starspot leaves its
  own fingerprint: its periodic signal.
  Two complementary methods may 
  retrieve this signal,
  period analysis and spot modelling.
  If the light curve contains the sum
  of such signals,
  it is possible to detect
  one or more of them.
  For example, \citet{Aig15} used
  the spot
  model of \citet{Aig12} to bury
  the periodic signals of numerous starpots
  into simulated data.
  Several research groups then tried
  to retrieve these unknown 
  periodic input signals 
  with different
  one-dimensional period analysis methods.
  They detected
  the strongest periodic input signal 
  in 90\% of cases, but 
  rarely any other
  weaker signal.
  Our approach is the same,
  except that our period finding
  method is two-dimensional.
Some spot models assume
solar law of \sdr, where
the starspot period determines its latitude,
and vice versa \citep[e.g.][his Eq. 4]{Kip12}.
\citet{How94} emphasized that
the solar \sdr ~measurements are
averages of {\it many} features over latitude.
His ``sobering reminder'' illustrated how
the latitudes of 36~708 {\it individual} sunspot groups
did not predict their rotation periods
\citep[][his Fig. 2]{How94}.
If {\it individual} sunspots do not follow the 
solar \sdr ~law, 
then why should {\it individual} starspots do so?
This limiting solar \sdr ~law assumption becomes unnecessary,
because our period finding method measures the unambiguous 
starspot periodic signals, 
longitudes and other
effects (e.g. amplitudes).
The spot models can now give the latitudes
and the other starspot parameters.

\begin{figure} 
\begin{center}
\resizebox{6.0cm}{!}{\includegraphics{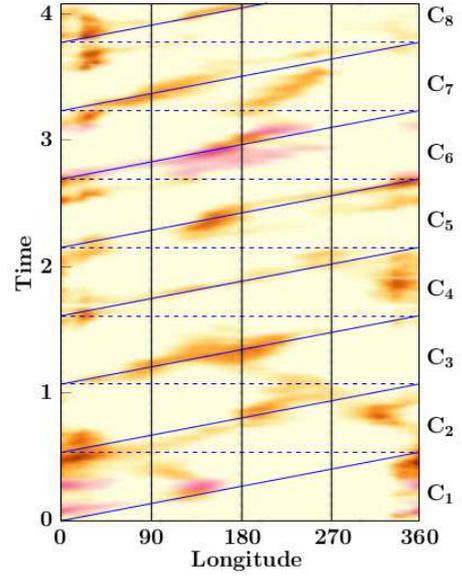}}
\end{center}
\caption{KIC 11560447 spot model
  \citep[][their Fig. 12]{Oza18}.
  Four years
  run on vertical axis.
Horizontal axis is longitude from orbital
ephemeris of Eq. \ref{ozaephe}.
Dark colour denotes 
high spot filling factor longitudes.
Dashed blue 
horizontal lines show C1-C8 cycle limits.
Continuous blue lines denote tilted
non-stationary
migration in orbital reference frame. }
\label{ozafig}
\end{figure}

\citet{Oza18}
applied their spot model to
   uninterrupted high-precision
   Kepler satellite photometry of
   KIC~11560447 (K1 IV).
   This
   offers an ideal opportunity 
   to test our \JHLhyp ~(\Testsix).
  The phases 
  were computed
   from the ephemeris
   \begin{eqnarray} 
{\mathrm{HJD~}}2454953.7303\!\pm\!0.0001 + 
 0.5276790\!\pm\!0.0000004{\mathrm{E}}
\label{ozaephe} 
\end{eqnarray}
fixed
to the primary eclipse
of this binary
\citep{Kir16}.
The stationary
starpots in the 
orbital reference frame
are at 
$0\degr$ and $180\degr$ longitudes 
(Fig. \ref{ozafig}: vertical lines), 
at the line
connecting the centres of the binary companions,
just like in fourteen CABS studied
by \citet[][their Figs. 1-14]{Jet17}.
Many tilted
non-stationary starspot migration lines
descend
from right to left,
but
none 
from left to right
(Fig. \ref{ozafig}).
The blue tilted continuous lines
denote this non-stationary
migration
during $C_1-C_8$ cycles.
Our tentative sketch relies
only on visual inspection,
but its significance during
2800 orbital cycles
can be easily confirmed later with 
nonparametric methods, like the Kuiper method.
The starspots are enhanced when the
tilted non-stationary
migration
intersects the stationary
$0\degr$  and $180\degr$ longitudes.
The spot models for  KIC 11560447
can detect 
the stationary and the
non-stationary starspots of the \JHLhyp ~(\Testsix).

% /home/jetsu/method/texts/oza.tex
% uses oza4.eps to produce
% oza.pdf with the command pdflatex oza
% gimp oza.pdf is used to produce final oxa.eps 

\citet{Rus06} showed
  that the results for the starspot distribution
  can be spurious,
  if the light curve model is a sum of lower
  order harmonics,
  like our $g(t)$ model of Eq. \ref{fullmodel}.
  This model
  fits to an infinite number of
  arbitrary starspot distributions
  \citep[][]{Rus06,Jef05,Jef09,Cow13}.
The spot modelling in Fig. \ref{ozafig}
also suffers from the ambiguity described 
by \citet{Rus06} and others,
because any starspot pattern is lumped
into one or two features 
\citep[][their Fig. 7 and 9]{Oza18}.
However, this ambiguity can not explain 
the long-lived vertical and tilted lines in 
Fig. \ref{ozafig}. Furthermore, 
  \citet{Rus06} discussed spot modelling, 
  not the detection of periodic signals.
  It is quite hard to explain,
  how arbitrary spurious spot distributions would
  consistently induce
  the same two periodic signals having 
  the same phases 
  during all four years of KIC~11560447 photometry (Fig. \ref{ozafig})
  or during all
  fourteen years of FK Com photometry
 (Fig. \ref{fkresults}ef).
  Rigorous spot modelling
  is outside the scope of our paper.
   We focus on signal detection,
  and especially on
  a new method
  for detecting two such signals simultaneously.

\section{Doppler imaging connection}
\label{SIsect}

The \dimethod
~measures \sdr ~from
the Doppler images
\citep[][]{Str09}.
The starspot periods at different latitudes
can be estimated directly
\citep[e.g.][]{Bar05,Bal16}
or the images
can be compared to the simultaneous 
light curves
\citep[e.g.][]{Ber98B,Hac04,Kor07,Col15}.
The \dimethod ~inversion is
an ``ill-posed problem'',
because an infinite
number of different solutions fit 
the observations.
Many physical parameters
must be fixed {\it before} these inversions,
e.g. inclination, rotation period
and local spectral line profiles
\citep[see][his Sect 9.2.3]{Koc16}.
The $P_{\mathrm{rot}}$ period
is usually fixed to $P_{\mathrm{phot}}$ or $P_{\mathrm{orb}}$.

There is a clear connection
  between the real light curves and the Doppler 
images of FK Com
(Sect. \ref{FKsdr}). However, FK Com is only one particular case.
Here, we show that there are
similar connections between the light curves and Doppler images
of other chromospherically active stars.

\subsection{Particular \dimethod ~results }
\label{SIparticular}

% Produced by 
% ~/method./progs/ber89.py
% /method/progs$ cp ber89.eps finalber89.eps
% cp finalber89.eps /home/jetsu/method/texts/
% cp finalber89.eps /home/jetsu/method/progs/valmis2/
% edited to %%BoundingBox: 100 400 520 620
\begin{figure} 
\begin{center}
\resizebox{6.0cm}{!}
{\includegraphics{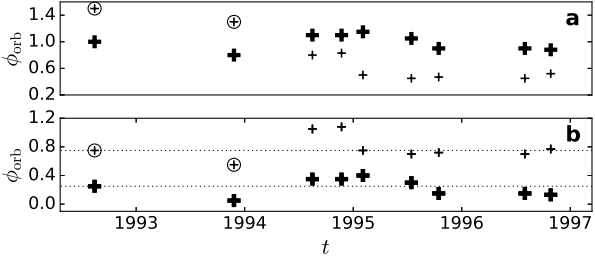}}
\end{center}
\caption{II Peg Doppler imaging.
(a) Larger (heavy crosses) and 
smaller (light crosses) starspots
with Eq. \ref{berephe} ephemeris
\citep[][]{Ber98B}.
We shift 
two highlighted crosses one
round downwards in "b".
(b) Same data with Eq. \ref{jetephe} ephemeris.
Dotted lines denote $\phi_{\mathrm{orb}}=0.25$ and 0.75.
Otherwise as in (a). }
\label{false}
\end{figure}

\citet[][]{Ber98B} published nine Doppler
images of II Peg
between 1992 and 1996.
The orbital ephemeris was
\begin{eqnarray}
{\mathrm{HJD~}}2449582.9268~+~6.724333{\mathrm{E}}.
\label{berephe}
\end{eqnarray}
\noindent
%Their zero epoch was equal to
%the  ``Ac'' epoch in
%\citet[][their Table 2]{Jet17}.
%\citet{Ber98B} 
They used $P_{\mathrm{orb}}$  
of Eq. \ref{berephe} in their inversions,
and concluded that
the larger and the smaller starspots
show two active longitudes, 
and a \flip ~event 
in the year 1994 (Fig. \ref{false}a).

\citet{Jet17} used the ephemeris
\begin{eqnarray}
{\mathrm{HJD~}}2449581.246~+~6.724333{\mathrm{E}}
\label{jetephe}
\end{eqnarray}
\noindent
for the simultaneous photometry of II Peg.
Their  $P_{\mathrm{orb}}$ was the same
as in Eq. \ref{berephe}, but
their zero epoch 
was $P_{\mathrm{orb}}/4$
earlier.
With Eq. \ref{jetephe} ephemeris,
all light curve minima of II Peg 
between 
1988 and 1997 
were close
to $\phi_{\mathrm{orb}}\!=\!0.25$
\citep[][their Fig. 27a]{Jet17}, i.e.
the stationary starspots
($P_{\mathrm{orb}}=P_{\mathrm{rot}}$)
dominated during these nine years,
which overlap the four years of the above Doppler images.

\citet[][]{Ber98B}
added one orbital
round to the first two $\phi_{\mathrm{orb}}$
of smaller starspots 
(Figs. \ref{false}ab:
two highlighted crosses).
They delivered a \flip ~event
with the corridor-effect
(see Sect. \ref{cpsexplained}: last paragraph).
We do not utilize this effect.
Our Fig. \ref{false}b shows 
the same starspots
with Eq. \ref{jetephe} ephemeris.
The stronger and the weaker starspots in the first 1992 image
are exactly at $\phi_{\mathrm{orb}}=0.25$ and
$0.75$.
All larger starspots stay close to
$\phi_{\mathrm{orb}}=0.25$,
 the smaller ones close 
to $\phi_{\mathrm{orb}}=0.75$.
The simultaneous mean light curves of II Peg confirm
that the starspots concentrated on $\phi_{\mathrm{orb}}=0.25$,
i.e. on the line connecting the centres of
the two binary companions \citep[][
Figs. 14d-e]{Jet17}.
The starspot latitudes
remained the same
in all images \citep{Ber98B}.
In this convincing {\it particular} case,
the \dimethod ~detects stationary 
starspots of II Peg
rotating with a constant period 
$P_{\mathrm{orb}}=P_{\mathrm{rot}}$
during four  years.

The long-lived starspots in 
\citet{Kor00} and \citet{Ber98B} images
rotated with a constant velocity,
as predicted by the \JHLhyp.
Such stability is common, e.g.
the longitude of the
strongest starspot ``A'' 
in V711~Tau
was stationary
%in the rotating reference frame
in 37 consecutive Doppler images
during 57 nights 
\citep[][Fig. 8]{Str00A}.

% /home/jetsu/method/progs/alfa.py produces nyt.tex _______
\begin{table}
  \caption{FK Com and thirteen CABSs in
    Jetsu et al. 2017. % \citet not allowed?
    Star,
    differential rotation
    coefficient (Eq. \ref{si_k}: $|k|$) 
    and
    angular velocity
    difference
    (Eq. \ref{si_omega}: $|\Delta \Omega|$)
    in increasing 
    $|\Delta \Omega|$ order.}
\label{falsediff}
\begin{center}
\begin{tabular}{lcc} 
\hline
Star & $|k|$ & $|\Delta \Omega|$                       \\
     &       &  ${\mathrm{rad~d}}^{-1}$                \\
\hline
              DM~UMa &
   $0.0003\pm0.0012$ &
   $0.0002\pm0.0010$ \\
              HK~Lac &
 $0.00116\pm0.00041$ &
 $0.00030\pm0.00011$ \\
           V1149~Ori &
   $0.0081\pm0.0011$ &
 $0.00096\pm0.00013$ \\
              EL~Eri &
   $0.0119\pm0.0043$ &
 $0.00156\pm0.00056$ \\
              BM~CVn &
 $0.00545\pm0.00031$ &
$0.001666\pm0.000093$ \\
              II~Peg &
 $0.00185\pm0.00010$ &
$0.001731\pm0.000098$ \\
        $\sigma$~Gem &
 $0.00550\pm0.00026$ &
$0.001767\pm0.000083$ \\
              HU~Vir &
$0.003026\pm0.000096$ &
$0.001827\pm0.000058$ \\
              XX~Tri &
 $0.00835\pm0.00042$ &
 $0.00220\pm0.00011$ \\
           V1762~Cyg &
 $0.01479\pm0.00071$ &
 $0.00327\pm0.00016$ \\
              FG~UMa &
 $0.01128\pm0.00047$ &
 $0.00334\pm0.00014$ \\
              FK~Com &
 $0.00455\pm0.00024$ &
 $0.01193\pm0.00062$ \\
              EI~Eri &
 $0.00373\pm0.00041$ &
   $0.0120\pm0.0013$ \\
            V711~Tau &
$0.019078\pm0.000069$ &
 $0.04184\pm0.00015$ \\
\hline
\end{tabular}
\end{center}
\end{table}

\subsection{General \dimethod ~results }
 \label{SIgeneral}

The {\dimethod}s ~usually measure
\sdr ~from 
\begin{eqnarray}
\Delta \Omega = 
\Omega_{\mathrm{max}}-\Omega_{\mathrm{min}}=
  {{{2 \pi} \over {P_{\mathrm{min}}}}}
  - {{{2 \pi} \over {P_{\mathrm{max}}}}}.
\label{omega}
\end{eqnarray}
\noindent
%\citep[e.g.][their Table 1]{Bar05}.
The Sun has
$\Delta \Omega \! \approx\! 0.07 {\mathrm{~rad~d^{-1}}}$,
where 
$P_{\mathrm{min}}=P_{\mathrm{eq}}\approx25^{\mathrm{d}}$
and 
$P_{\mathrm{max}}=P_{\mathrm{pole}}\approx35^{\mathrm{d}}$.
Using Eqs. \ref{rotationsun} and \ref{omega} for
$P_{\mathrm{act}}$ and $P_{\mathrm{rot}}$ of thirteen CABS 
\citep[][]{Jet17}
gives

\begin{eqnarray}
|k| & = & 
        {        
        {|P_{\mathrm{act}}-P_{\mathrm{rot}}|}
        \over
        {(P_{\mathrm{act}}+P_{\mathrm{rot}})/2}
        } 
\label{si_k} \\
|\Delta \Omega| & = & |
  {{2 \pi}\over{P_{\mathrm{act}}}}-
  {{2 \pi}\over{P_{\mathrm{rot}}}}
                  |.
\label{si_omega}
\end{eqnarray}
\noindent
We give these parameters
in 
increasing $|\Delta \Omega|$ order in Table \ref{falsediff}.
The lower and upper limits are
$|\Delta \Omega|\!=\!0.0002 {\mathrm{~rad~d^{-1}}}$
(DM~UMa) 
and
$|\Delta \Omega|\!=\!0.042{\mathrm{~rad~d^{-1}}}$
(V711~Tau).
\citet{Jet17} underestimated 
the $|P_{\mathrm{act}}-P_{\mathrm{rot}}|$ difference
when they applied the Kuiper test to {\it all}
$t_{\mathrm{CPS,min,1}}$ epochs.
Due to minima-\Inc,
some epochs were connected to the
$P_{\mathrm{rot}}$ signal, and therefore some $P_{\mathrm{act}}$
estimates were too close to $P_{\mathrm{rot}}$.
We compute $|\Delta \Omega|=0.012{\mathrm{~rad~d^{-1}}}$ of FK Com 
from the 
periods of Eqs. \ref{Kone} and \ref{Ktwo}.
Since this third largest $|\Delta \Omega|$ 
of all fourteen values in Table \ref{falsediff} 
is certainly not an underestimate,
some other $|\Delta \Omega|$ estimates may be.

Our $|\Delta \Omega| \! = \! 0.0002 {\mathrm{~rad~d^{-1}}}$ lower limit
agrees with the earlier \dimethod ~studies, but
our $|\Delta \Omega| \! = \!0.042{\mathrm{~rad~d^{-1}}}$ upper limit
is three times smaller than
in \citet[][Table 1: $0.14{\mathrm{~rad~d^{-1}}}$]{Bar05}
and ten times smaller than in
\citet[][Table 2: $0.47{\mathrm{~rad~d^{-1}}}$]{Bal16}.
For two reasons,
some earlier
$|\Delta \Omega|$ may be overestimates.
Firstly, \citet{Bar05} and \citet{Bal16} give {\it many}
$|\Delta \Omega|$ values for the {\it same} star.
This has increased
some $|\Delta \Omega|$ estimates.
Secondly, the constantly changing
$g(t)$ model (Eq. \ref{fullmodel})
has no unique longitudes.
After $P_{\mathrm{lap}}/2$,
Doppler images of the {\it same} star
can not be reliably compared
with the {\it same} period.
For FK Com,
the one year gap
between observing seasons
exceeds $P_{\mathrm{lap}}/2\!=\!264^{\mathrm{d}}$.
This ``map-\Inc'' misleads
identification of the {\it same}
structure in different images. % of the {\it same} star.
The non-stationary ($P_{\mathrm{act}}$)
and
the stationary
($P_{\mathrm{orb}}$ or $P_{\mathrm{rot}}$)
starpots get mixed.
The longitudes of
two starspots $S_1$ and $S_2$
become ambiguous
after $P_{\mathrm{lap}}/2$.
It is also uncertain, 
if $S_1$ or $S_2$ moves forwards or backwards.
This resembles the shuffling of cards.
Except for a magician,
the current order of cards,
i.e. the $S_1$ and $S_2$ longitudes,
gives no information
about the order before the shuffling.
Misidentifications may have overestimated
$|\Delta \Omega|$ 
in long \dimethod ~studies,
like in
\citet[][EI Eri: 11$^{\mathrm{y}}$]{Was09},
\citet[][II Peg: 6$^{\mathrm{y}}$]{Lin11}
or
\citet[][FK Com: 13$^{\mathrm{y}}$]{Hac13}.
% in \citet{Bar05} and \citet{Bal16}.
We underestimate some $|\Delta \Omega|$ 
(Table \ref{falsediff}),
while \citet{Bar05} or \citet{Bal16} must
have overestimated
some $|\Delta \Omega|$.

Both 
{\it particular}
and {\it general} 
evidence indicate that 
the $P_{\mathrm{act}}$ and
$P_{\mathrm{rot}}\approx P_{\mathrm{orb}}$
periods of the \JHLhyp ~have already
been detected 
with Doppler imaging (\Testseven).

\section{Ill-posed problem connection}
\label{Illposed}

Inversions of spot models, Doppler images and
non-linear models are
ill-posed problems.
There are % sophisticated 
{\it analytical} 
nonlinear $g(\bar{\beta})$ model solutions
\citep[e.g.][]{Tar82,Sni98,Kal08}.
In our {\it numerical} solution,
we divide the free parameters $\bar{\beta}$
into two groups 

\begin{description} 

\item $\bar{\beta}_{I}$ group:
Partial derivatives $\partial g / \partial \beta_i$ 
of these free parameters
{\it do not contain any} free parameters,
{\it if} the $\bar{\beta}_{II}$ group free
parameters are fixed to constant numerical values.

\item $\bar{\beta}_{II}$ group: 
Partial derivatives $\partial g / \partial \beta_i$ 
of these free parameters
{\it contain at least one} free parameter,
{\it even if} the $\bar{\beta}_{I}$ group free
parameters are fixed to constant numerical values.

\end{description}

\noindent
In our model $g(t)$ of Eq. \ref{fullmodel}, 
the coefficients $M_0 ... M_{K_0}$ and
the amplitudes $B_1 ... B_{K_1}$, $C_1 ... C_{K_1}$, 
$D_1 ... D_{K_2}$, $E_1 ... E_{K_2}$
belong to the $\bar{\beta}_{I}$ group.
The frequencies $f_1$ and $f_2$ belong
to the $\bar{\beta}_{II}$ group.
When the $\bar{\beta}_{II}$ group values are fixed,
the solutions for the second $\bar{\beta}_{I}$ group
are unambiguous.
%Similar linearization is possible for any 
%arbitrary nonlinear model.
The test statistic $z(\bar{\beta}_{II})$ can be, e.g.
$\chi^2$,
$R$ (Eq. \ref{squaredresiduals}) or
$\sqrt{R/n}$ (Eqs. \ref{twoz} or \ref{onez}).
The tested  $\bar{\beta}_{II}$ range of can be fixed
with physical or any other reasonable criteria.
The correlation of $z(\bar{\beta}_{II})$ 
for close tested $\bar{\beta}_{II}$ values
gives the suitable  
tested $\bar{\beta}_{II}$ grid density
\citep[e.g.][their Fig. 2]{Jet96A}.
Our simple recipe is
\begin{description}

\item[1.] Identify the $\bar{\beta}_{I}$ and  $\bar{\beta}_{II}$ groups.
\item[2.] Fix the tested $\bar{\beta}_{II}$ grid.
\item[3.] Compute $z(\bar{\beta}_{II})$ with a linear least squares fit.
\item[4.] Global $z(\bar{\beta}_{II})$ minimum gives
the best  $\bar{\beta}_{I}$ and  $\bar{\beta}_{II}$ values.

\end{description}

\noindent
This gives
only the {\it numerical} $\bar{\beta}$ solution. 
The bootstrap gives the errors
$\sigma_{\bar{\beta}}$ (e.g. Eq. \ref{bootsample}). 
However, this unambiguous 
$\bar{\beta} \pm \sigma_{\bar{\beta}}$ solution
consumes computation time,
because a linear least squares fit is performed at
every tested $\bar{\beta}_{II}$ grid point.
For the nested models,
the criterion of Eq. \ref{oneortwo} gives
the best model among
the different model alternatives.  

In the two-dimensional $g_{\mathrm{C}}(t)$-method, 
we utilize the symmetry $z_{\mathrm{C}}(f_1,f_2)=z_{\mathrm{C}}(f_2,f_1)$,
and test only $f_1>f_2$ pairs.
A more general three-dimensional version could test
frequency triples $f_1 > f_2 > f_3$.
If the frequencies are far from each other 
\citep[e.g.][]{Men17,Sai18},
the tested ranges could be
$l_1 < f_1 < l_2$, $l_3 < f_2 < l_4$ and $l_5 < f_3 < l_6$,
where $l_1, ..., l_6$ are the range limits.
The number of suitable $g_1(t)$, $g_2(t)$ and $g_3(t)$
functions for this three-dimensional version,
like higher harmonics, is unlimited.

The $g(t)$ model (Eq. \ref{fullmodel})
is a mathematical model, but is it
a suitable physical model?
Most models in physics are mathematical.
The solution for the free parameters 
$\bar{\beta}$
of this nonlinear model
is unambiguous for {\it any data}, not only for 
{\it photometry}.
We apply a rigorous statistical approach.
If the quality or quantity of the data
prevents the simultaneous detection
of two signals,
we accept that the complex model can not be used, 
and use the simple model.
We use the simple model $g_{\mathrm{S}}(t)$,
if ``intersect $f_1=f_2$'' or ``amplitude dispersion''
occurs.
Otherwise, we use the complex model $g_{\mathrm{C}}(t)$ 
(Sect. \ref{SectTel2Seg1}).
When these two models can be compared with the
statistical test of Eq. \ref{oneortwo},
the complex model is always better
than the simple model.
The results for FK Com are 
\begin{itemize} 
\item[1.] There are two separate period levels (Fig. \ref{fkresults}d).
\item[2.] There are two separate minimum epoch levels (Fig. \ref{fkresults}ef).
\item[3.] All these levels are connected.
\end{itemize}
\noindent
All models of one-dimensional 
period analysis methods, 
like the \lsmethod ~sinusoidal model, 
require dramatic short-term
changes in the object.
Our model would be a better mathematical
model also from the physical point of view,
because it requires no 
short-term changes.
It reveals that 
the chromospherically active stars
are not so ``lively'' objects,
as suggested by the earlier 
one-dimensional period analyses.  
It may be time to stop looking 
for something that is not there.

\section{Physical connection}
\label{Physical}

The \JHLhyp ~does not rule out 
starspot evolution.
Interferometry
has undeniably confirmed such evolution
in $\zeta$ And
\citep[][]{Roe16}.
We show that
the starspot life-times 
in FK Com are several months,
not one month
\citep[e.g.][]{Jet93}.
These life-times should
exceed the sunspot life-times,
because the magnetic fields in 
chromospherically active stars 
are stronger than in the Sun.
Slow starspot
changes in FK Com
are easier to model
than
erratic short-lived starspots 
undergoing abrupt \flip ~shifts.
The model can now be nearly stable
and have a simple geometry,
like a large long-lived magnetic loop,
where the starspots at the feet of this loop 
rotate with different angular velocities.
The radio images of the active K0~IV member of Algol
showed a similar loop \citep{Pat10}, where
the stationary 
$P_{\mathrm{orb}} \approx P_{\mathrm{rot}}$ part probably 
dominated,
because the loop was oriented towards 
the inactive B8~V member parallel
to the line connecting the centres of the two stars.

The magnetic loop is not the only suitable model.
For rapidly rotating stars, the mean-field dynamo 
solutions have long been known to become non-axisymmetric 
\citep[e.g.][]{Kra80}. The predicted
non-axisymmetric modes drift 
%in the orbital frame of the star 
due to a different azimuthal 
phase speed than the stellar rotation.
Fully non-linear global magneto convection simulations 
have verified this prediction
\citep{Col14}. This scenario would give 
rise to multiple periodicities in the magnetic activity tracers.
There are other suitable alternative models,
like two long-lived active regions \citep{Puz16}. 
The long-lived starspots
could even be vortices
resembling the Great red spot of Jupiter 
\citep[][]{Man11,Kap11},
except that the extreme chromospheric activity
of FK Com
can not be explained without 
strong magnetic fields \citep[][]{Ayr16}.

In binaries, the
$P_{\mathrm{rot}} \approx P_{\mathrm{orb}}$  relation
reveals,
if the $P_1$ or the $P_2$ period
is the stationary $P_{\mathrm{rot}}$ period.
For FK Com, this is not possible.
This single star ``acts'' like a binary.
Both $P_1$ or $P_2$ periods are still observed,
although both members of this hypothetical coalesced W~UMa binary
are no longer present.
This is not unusual for single stars.
For example, 
one period clearly dominated 
the regular {\it horizontal} 
migration in
the {\it single} star LQ Hya
in \citet[][their Fig. 4]{Leh12}.
However, another period dominated
the {\it tilted} migration during their ``SEG 14''
in the years 1998 and 1999.

\section{Uncertainties}
\label{uncertainties}

We can not prove that 
the two $P_{\mathrm{w,1}}$ and $P_{\mathrm{w,2}}$ levels 
of Eqs. \ref{Pwone} and \ref{Pwtwo} are definitely separate.
The periods after the year 2002 support this,
but admittedly some earlier
weaker signals deviate more
than $\pm 1 \sigma_{P}$ from these levels
(Fig. \ref{fkresults}d: highlighted values).
We also detect weak \sdr ~(Sect. \ref{FKsdr}). 
This would seem to support only a weaker
\JHLhyp ~
\begin{itemize}
\item[] \weakJHLtext ,

\end{itemize}
\noindent
where
the second sentence about 
$P_{\mathrm{act}}$ and $P_{\mathrm{rot}} \approx P_{\mathrm{orb}}$
is removed.
Perhaps we have ``only succeeded in''
dividing the starspots into faster and slower
rotating $S_1$ and $S_2$ groups, which are centered
at the $P_{\mathrm{w,1}}$ and $P_{\mathrm{w,2}}$ levels.
However, this weaker hypothesis can not
explain the following four results

\begin{itemize}

\item[1.] The detection of the stationary 
$P_{\mathrm{rot}} \approx P_{\mathrm{orb}}$ periods
in the light curves
of fourteen CABS 
\citep[][MLC in their Figs. 1-14]{Jet17},
or the two MLC of FK Com (Fig. \ref{figmlc}).

\item[2.] The detection of the non-stationary 
$P_{\mathrm{act}}$ periods in the light curves
of thirteen CABS 
\citep[][their Figs. 15-27]{Jet17}:
This linear parallel tilted migration 
is real, although some $P_{\mathrm{act}}$ values
may not be correct
(\Testtwo).

\item[3.]The phase coherence of the real $g_1(t)$ and
$g_2(t)$ light curve mimima (Figs. \ref{fkresults}e and f):
This indicates that the periods and the phases
of these two signals 
are connected.

\item[4.] The vertical and tilted lines in Fig. \ref{ozafig}.

\end{itemize}
\noindent
We can model FK Com light curves spanning half a year.
This explains the above mentioned vertical and tilted lines,
because even such a short six month time span already gives
a line covering
$\Delta T/P_{\mathrm{w,1}}-\Delta T/P_{\mathrm{w,2}}
=0.34$, which is about one third of $P_{\mathrm{lap}}$.
This $P_{\mathrm{lap}}$ is
one full lap around FK Com $\equiv$ $360\degr$ longitude difference.
The $P_1$ and $P_2$ correlation confirms that
both periods increase and decrease simultaneously
(Fig. \ref{FKpreltest}i).
The above four results can be explained,
if these $P_1$ and $P_2$  periods wobble at both sides of the 
$P_{\mathrm{w,1}}$ and $P_{\mathrm{w,2}}$ levels of FK Com
(Fig. \ref{FKspots}: Alternative \ref{sdrone}).
%>>> pw1=2.39321
%>>> pw2=2.40413
%>>> dt=180.
%>>> print(dt/pw1-dt/pw2)
%0.341630303488472
This does not require the removal of the second
sentence from the original \JHLhyp.
We will get very accurate $P_{\mathrm{w,1}}$ and $P_{\mathrm{w,2}}$
estimates for FK Com,
if this 
starspot pattern prevails.
If similar
``boring'' patterns 
are also found
in other chromospherically active stars,
there are striking similarities in
the geometries of magnetic fields
in
early-type and late-type stars
\citep[][Sect. 7]{Jet17}.
The similarity between the oblique rotator in early-type stars
and the magnetic loop in late-type stars is the stationary
starspot. The difference is 
the non-stationary starspot at the other
end of the magnetic loop in late-type stars.
The reason for this difference may be the interaction between
convection and magnetic fields which 
complicates things in the late-type stars, 
while the magnetic fields in the fully radiative
early-type %Ap and Bp 
stars are stable \citep[e.g.][]{Jet92,Shu18}.

\citet{Lou78} concluded that
``It is our purpose to warn those 
researching variable phenomena that 
any period-finding method may
give spurious results when applied to multiple periodicities''.
This applies also to our method,
if the data contains more than two periodic signals,
or either one of the signal amplitudes is changing.
In addition to the two stronger starspots,
there are probably  weaker starspots.
The effect of these
weaker starspots can perhaps be partly compensated
by the  $m_0(t)$ polynomial (Eq. \ref{gone}).

\section{Conclusions}
\label{Conclusions}

Our \JHLhyp ~is a total success,
  because we detect the two
  real light curves of FK Com
  with our new period analysis method.
  All parameters of these real light curves are directly
  connected to the long-lived starspots of FK Com.
These starspot parameters 
are spatially and temporally correlated
just like in the Sun,
except that dark starspots 
change the luminosity of FK Com
(Sect. \ref{FKsdr}: Eq. \ref{FKmode}).
The simultaneous Doppler images confirm weak
solar-like surface differential rotation.
Since FK Com resembles the Sun
and holds the rotation record
for single stars
\citep[][``King of spin'']{Ayr16},
its starspots provide valuable observational
and theoretical constraints for
the starspots of other slower rotating
chromospherically active stars.

We present an analogy of incompatibility,
because for seven decades
these ``creatures'' have managed to evade detection 
behind the ``veil'' of interference.
Imagine a face with a left and a right eye.
Both eyes can disappear and reappear.
At any moment, the number of eyes
may be zero, one or two.
The original stationary right eye can 
disappear and reappear only
at fixed locations.
The original non-stationary left 
eye rotates slowly around the head.
Both eyes also wobble up and down simultaneously
(Fig. \ref{FKspots}: Alternative \ref{sdrone}).
We see this head spinning.
Soon it is impossible to tell which
eye is the original left or right eye
(map-\Inc ~after $P_{\mathrm{lap}}/2$).
The only compatible pictures of this face are snapshots,
but these snapshots can not be used 
to recognize this constantly changing face.
These snapshots can capture only one side of the head,
or equivalently half of the full visible
surface of FK Com.

As for other chromospherically active stars,
this ``spinning head'' hypothesis

\begin{description}

\item[-] \JHLhyp: \JHLtext

\end{description}

\noindent
is subjected to eight \hattuuslippa ~tests.
We use the notations $P_1$ and $P_2$ for
periods of the $g(t)$ model (Eq. \ref{fullmodel}).
They fulfill $P_1<P_2$.
Depending on the object,
these $P_1$ and $P_2$ periods can refer to the
$P_{\mathrm{act}}$, $P_{\mathrm{rot}}$ or $P_{\mathrm{orb}}$
periods.
The {\it binaries} may have 
$P_1=P_{\mathrm{act}}$ and $P_2=P_{\mathrm{rot}} \approx P_{\mathrm{orb}}$,
or
$P_1=P_{\mathrm{rot}} \approx P_{\mathrm{orb}}$ 
and $P_2=P_{\mathrm{act}}$.
The {\it single} stars may have
$P_1=P_{\mathrm{act}}$ and $P_2=P_{\mathrm{rot}}$,
or
$P_1=P_{\mathrm{rot}} $ and $P_2=P_{\mathrm{act}}$.
For the {\it binaries},
the relation $P_{\mathrm{rot}} \approx P_{\mathrm{orb}}$ can be used
to check, if $P_1$ or $P_2$ represents the stationary
$P_{\mathrm{rot}}$ period.
This is not
possible for the {\it single} stars, like FK Com.

Here are our eight tests and our results for these tests:

\begin{description}

\item[-]  \Testonetext
\item[] \result{
One-dimensional period
finding methods
``detect'' {\it many different spurious} 
periods for the {\it same}
star, although the 
{\it real} $P_1$ and $P_2$ periods
remain the {\it same} (Sect. \ref{periodinc}).}

\item[-] \Testtwotext

\item[] \result{The Kuiper method 
analysis of the incompatible
epochs did
not give an unambiguous 
$P_{\mathrm{act}}$ period (Sect. \ref{minimainc}).}

\item[-]  \Testthreetext

\item[] \result{This method 
does not detect only the 
$P_{\mathrm{act}}$ period.
The detected period 
can be the $[(f_1+f_2)/2]^{-1}$, $f_2^{-1}$ or $f_1^{-1}$
period of $g(t)$ model,
or none of these,
because the amplitudes of the real light
curves change
(Sect. \ref{minimainc})}

\item[-] \Testfourtext

\item[] \result{
\citet{Jet17} detected 
the long-term mean
light curves following
the $P_{\mathrm{rot}}\approx P_{\mathrm{orb}}$ period.
Due to minima-\Inc, their
$P_{\mathrm{act}}$ estimates
failed (Sect. \ref{minimainc}).}

\item[-] \Testfivetext

\item \result{{\it Interference} 
causes these events.
The {\it abrupt} shift is $\Delta \phi_{\mathrm{b}}=
0.5\equiv 180\degr$,
if the real light curves
are equal amplitude sinusoids 
with constant periods $P_1$ and $P_2$
(Eq. \ref{abruptphase}).}

\item[-] \Testsixtext

\item \result{\citet{Oza18} have detected 
the 
stationary $P_{\mathrm{rot}} \approx P_{\mathrm{orb}}$ period
and
the
non-stationary $P_{\mathrm{act}}$ period 
in the spot models for 
%high-precision Kepler satellite
light curves of KIC~11560447 (Fig. \ref{ozafig}).}

\item[-] \Testseventext 

\item[] \result{{\it Particular} and {\it general} evidence 
indicate that these periods
have been detected in the Doppler images
(Sect. \ref{SIsect}).}

\item[-] \Testeighttext

\item[] \result{The one-dimensional period finding methods
detect {\it spurious} periods from the light curves.
The Doppler images can detect {\it real} periods.}

\end{description}

The \JHLhyp ~neatly explains many 
phenomena that have been detected 
earlier from the light curves
with one-dimensional period finding methods

\begin{description}
\item[-] Spurious {\it observed} rapid light curve changes
\item[-] Spurious short starspot life-times
\item[-] Spurious rapid photometric rotation period changes
\item[-] Spurious active longitudes
\item[-] Spurious starspot migration
\item[-] Spurious cycles in periods and amplitudes 
\item[-] Spurious  \flip ~events and \flip ~cycles
\item[-] Long-term mean light curves
\item[-] \lcmethod ~and \dimethod ~discrepancies for \sdr

\end{description}
\noindent
While the above list of phenomena is by no means
complete (e.g. Sect. \ref{minimainc}: light curve {\it shape}),
it does contain at least the most obvious ones.
It took a quarter of a century to find 
the correct result for the \Testfive:
``Farewell \flip.''

\section*{Acknowledgements}
We thank the anonymous referee,
Thomas  Hackman, Ilana Hiilesmaa,
Emre {I{\c s}{\i}k}, 
Maarit K\"a\-py\-l\"a, Kauko Palmu and Igor Savanov
for their comments of this manuscript. 
This work has made use of NASA's 
Astrophysics Data System (ADS) services.

\bibliographystyle{mnras}
\bibliography{jetsufk}

\begin{thebibliography}{}
\makeatletter
\relax
\def\mn@urlcharsother{\let\do\@makeother \do\$\do\&\do\#\do\^\do\_\do\%\do\~}
\def\mn@doi{\begingroup\mn@urlcharsother \@ifnextchar [ {\mn@doi@}
  {\mn@doi@[]}}
\def\mn@doi@[#1]#2{\def\@tempa{#1}\ifx\@tempa\@empty \href
  {http://dx.doi.org/#2} {doi:#2}\else \href {http://dx.doi.org/#2} {#1}\fi
  \endgroup}
\def\mn@eprint#1#2{\mn@eprint@#1:#2::\@nil}
\def\mn@eprint@arXiv#1{\href {http://arxiv.org/abs/#1} {{\tt arXiv:#1}}}
\def\mn@eprint@dblp#1{\href {http://dblp.uni-trier.de/rec/bibtex/#1.xml}
  {dblp:#1}}
\def\mn@eprint@#1:#2:#3:#4\@nil{\def\@tempa {#1}\def\@tempb {#2}\def\@tempc
  {#3}\ifx \@tempc \@empty \let \@tempc \@tempb \let \@tempb \@tempa \fi \ifx
  \@tempb \@empty \def\@tempb {arXiv}\fi \@ifundefined
  {mn@eprint@\@tempb}{\@tempb:\@tempc}{\expandafter \expandafter \csname
  mn@eprint@\@tempb\endcsname \expandafter{\@tempc}}}

\bibitem[\protect\citeauthoryear{{Aigrain}, {Pont}  \& {Zucker}}{{Aigrain}
  et~al.}{2012}]{Aig12}
{Aigrain} S.,  {Pont} F.,   {Zucker} S.,  2012, \mn@doi [\mnras]
  {10.1111/j.1365-2966.2011.19960.x}, \href
  {http://adsabs.harvard.edu/abs/2012MNRAS.419.3147A} {419, 3147}

\bibitem[\protect\citeauthoryear{{Aigrain} et~al.,}{{Aigrain}
  et~al.}{2015}]{Aig15}
{Aigrain} S.,  et~al., 2015, \mn@doi [\mnras] {10.1093/mnras/stv853}, \href
  {http://adsabs.harvard.edu/abs/2015MNRAS.450.3211A} {450, 3211}

\bibitem[\protect\citeauthoryear{{Alekseev} \& {Kozhevnikova}}{{Alekseev} \&
  {Kozhevnikova}}{2018}]{Ale18}
{Alekseev} I.~Y.,  {Kozhevnikova} A.~V.,  2018, \mn@doi [Astronomy Reports]
  {10.1134/S1063772918050013}, \href
  {http://cdsads.u-strasbg.fr/abs/2018ARep...62..396A} {62, 396}

\bibitem[\protect\citeauthoryear{{Arkhypov}, {Khodachenko}, {Lammer},
  {G{\"u}del}, {L{\"u}ftinger}  \& {Johnstone}}{{Arkhypov}
  et~al.}{2015}]{Ark15}
{Arkhypov} O.~V.,  {Khodachenko} M.~L.,  {Lammer} H.,  {G{\"u}del} M.,
  {L{\"u}ftinger} T.,   {Johnstone} C.~P.,  2015, \mn@doi [\apj]
  {10.1088/0004-637X/807/1/109}, \href
  {http://adsabs.harvard.edu/abs/2015ApJ...807..109A} {807, 109}

\bibitem[\protect\citeauthoryear{{Ayres} et~al.,}{{Ayres} et~al.}{2016}]{Ayr16}
{Ayres} T.~R.,  et~al., 2016, \mn@doi [\apjs] {10.3847/0067-0049/223/1/5},
  \href {http://cdsads.u-strasbg.fr/abs/2016ApJS..223....5A} {223, 5}

\bibitem[\protect\citeauthoryear{{Bai}}{{Bai}}{2003}]{Bai03}
{Bai} T.,  2003, \mn@doi [\apj] {10.1086/375295}, \href
  {http://cdsads.u-strasbg.fr/abs/2003ApJ...591..406B} {591, 406}

\bibitem[\protect\citeauthoryear{{Bai} \& {Sturrock}}{{Bai} \&
  {Sturrock}}{1993}]{Bai93}
{Bai} T.,  {Sturrock} P.~A.,  1993, \mn@doi [\apj] {10.1086/172680}, \href
  {http://cdsads.u-strasbg.fr/abs/1993ApJ...409..476B} {409, 476}

\bibitem[\protect\citeauthoryear{{Ballester}, {Oliver}  \&
  {Carbonell}}{{Ballester} et~al.}{2002}]{Bal02}
{Ballester} J.~L.,  {Oliver} R.,   {Carbonell} M.,  2002, \mn@doi [\apj]
  {10.1086/338075}, \href {http://adsabs.harvard.edu/abs/2002ApJ...566..505B}
  {566, 505}

\bibitem[\protect\citeauthoryear{{Balona} \& {Abedigamba}}{{Balona} \&
  {Abedigamba}}{2016}]{Bal16}
{Balona} L.~A.,  {Abedigamba} O.~P.,  2016, \mn@doi [\mnras]
  {10.1093/mnras/stw1443}, \href
  {http://cdsads.u-strasbg.fr/abs/2016MNRAS.461..497B} {461, 497}

\bibitem[\protect\citeauthoryear{{Balthasar}}{{Balthasar}}{2007}]{Bal07}
{Balthasar} H.,  2007, \mn@doi [\aap] {10.1051/0004-6361:20077475}, \href
  {http://cdsads.u-strasbg.fr/abs/2007A%26A...471..281B} {471, 281}

\bibitem[\protect\citeauthoryear{{Barnes}, {Collier Cameron}, {Donati},
  {James}, {Marsden}  \& {Petit}}{{Barnes} et~al.}{2005}]{Bar05}
{Barnes} J.~R.,  {Collier Cameron} A.,  {Donati} J.-F.,  {James} D.~J.,
  {Marsden} S.~C.,   {Petit} P.,  2005, \mn@doi [\mnras]
  {10.1111/j.1745-3933.2005.08587.x}, \href
  {http://cdsads.u-strasbg.fr/abs/2005MNRAS.357L...1B} {357, L1}

\bibitem[\protect\citeauthoryear{{Basri} \& {Nguyen}}{{Basri} \&
  {Nguyen}}{2018}]{Bas18}
{Basri} G.,  {Nguyen} H.~T.,  2018, \mn@doi [\apj] {10.3847/1538-4357/aad3b6},
  \href {http://adsabs.harvard.edu/abs/2018ApJ...863..190B} {863, 190}

\bibitem[\protect\citeauthoryear{{Berdyugina}}{{Berdyugina}}{2007}]{Ber07}
{Berdyugina} S.~V.,  2007, \memsai, \href
  {http://cdsads.u-strasbg.fr/abs/2007MmSAI..78..242B} {78, 242}

\bibitem[\protect\citeauthoryear{{Berdyugina} \& {Tuominen}}{{Berdyugina} \&
  {Tuominen}}{1998}]{Ber98A}
{Berdyugina} S.~V.,  {Tuominen} I.,  1998, \aap, \href
  {http://cdsads.u-strasbg.fr/abs/1998A%26A...336L..25B} {336, L25}

\bibitem[\protect\citeauthoryear{{Berdyugina} \& {Usoskin}}{{Berdyugina} \&
  {Usoskin}}{2003}]{Ber03}
{Berdyugina} S.~V.,  {Usoskin} I.~G.,  2003, \mn@doi [\aap]
  {10.1051/0004-6361:20030748}, \href
  {http://cdsads.u-strasbg.fr/abs/2003A%26A...405.1121B} {405, 1121}

\bibitem[\protect\citeauthoryear{{Berdyugina}, {Berdyugin}, {Ilyin}  \&
  {Tuominen}}{{Berdyugina} et~al.}{1998}]{Ber98B}
{Berdyugina} S.~V.,  {Berdyugin} A.~V.,  {Ilyin} I.,   {Tuominen} I.,  1998,
  \aap, \href {http://cdsads.u-strasbg.fr/abs/1998A%26A...340..437B} {340, 437}

\bibitem[\protect\citeauthoryear{{Berdyugina}, {Moss}, {Sokoloff}  \&
  {Usoskin}}{{Berdyugina} et~al.}{2006}]{Ber06}
{Berdyugina} S.~V.,  {Moss} D.,  {Sokoloff} D.,   {Usoskin} I.~G.,  2006,
  \mn@doi [\aap] {10.1051/0004-6361:20053454}, \href
  {http://cdsads.u-strasbg.fr/abs/2006A%26A...445..703B} {445, 703}

\bibitem[\protect\citeauthoryear{{B{\"o}hm-Vitense}}{{B{\"o}hm-Vitense}}{2007}]{Boh07}
{B{\"o}hm-Vitense} E.,  2007, \mn@doi [\apj] {10.1086/510482}, \href
  {http://cdsads.u-strasbg.fr/abs/2007ApJ...657..486B} {657, 486}

\bibitem[\protect\citeauthoryear{{Bonomo} \& {Lanza}}{{Bonomo} \&
  {Lanza}}{2012}]{Bon12}
{Bonomo} A.~S.,  {Lanza} A.~F.,  2012, \mn@doi [\aap]
  {10.1051/0004-6361/201219999}, \href
  {http://adsabs.harvard.edu/abs/2012A%26A...547A..37B} {547, A37}

\bibitem[\protect\citeauthoryear{{Bopp} \& {Stencel}}{{Bopp} \&
  {Stencel}}{1981}]{Bop81}
{Bopp} B.~W.,  {Stencel} R.~E.,  1981, \mn@doi [\apjl] {10.1086/183606}, \href
  {http://cdsads.u-strasbg.fr/abs/1981ApJ...247L.131B} {247, L131}

\bibitem[\protect\citeauthoryear{{Bradshaw} \& {Hartigan}}{{Bradshaw} \&
  {Hartigan}}{2014}]{Bra14}
{Bradshaw} S.~J.,  {Hartigan} P.,  2014, \mn@doi [\apj]
  {10.1088/0004-637X/795/1/79}, \href
  {http://adsabs.harvard.edu/abs/2014ApJ...795...79B} {795, 79}

\bibitem[\protect\citeauthoryear{{Brandenburg}}{{Brandenburg}}{2018}]{Bra18}
{Brandenburg} A.,  2018, \mn@doi [Journal of Plasma Physics]
  {10.1017/S0022377818000806}, \href
  {http://adsabs.harvard.edu/abs/2018JPlPh..84d7304B} {84, 735840404}

\bibitem[\protect\citeauthoryear{{Brandenburg}, {Saar}  \&
  {Turpin}}{{Brandenburg} et~al.}{1998}]{Bra98}
{Brandenburg} A.,  {Saar} S.~H.,   {Turpin} C.~R.,  1998, \mn@doi [\apjl]
  {10.1086/311297}, \href {http://cdsads.u-strasbg.fr/abs/1998ApJ...498L..51B}
  {498, L51}

\bibitem[\protect\citeauthoryear{{Brandenburg}, {Mathur}  \&
  {Metcalfe}}{{Brandenburg} et~al.}{2017}]{Bra17}
{Brandenburg} A.,  {Mathur} S.,   {Metcalfe} T.~S.,  2017, \mn@doi [\apj]
  {10.3847/1538-4357/aa7cfa}, \href
  {http://cdsads.u-strasbg.fr/abs/2017ApJ...845...79B} {845, 79}

\bibitem[\protect\citeauthoryear{{Chugainov}}{{Chugainov}}{1966}]{Chu66}
{Chugainov} P.~F.,  1966, Information Bulletin on Variable Stars, \href
  {http://cdsads.u-strasbg.fr/abs/1966IBVS..172....1C} {172}

\bibitem[\protect\citeauthoryear{{Cohen}, {Drake}, {Kashyap}, {Korhonen},
  {Elstner}  \& {Gombosi}}{{Cohen} et~al.}{2010}]{Coh10}
{Cohen} O.,  {Drake} J.~J.,  {Kashyap} V.~L.,  {Korhonen} H.,  {Elstner} D.,
  {Gombosi} T.~I.,  2010, \mn@doi [\apj] {10.1088/0004-637X/719/1/299}, \href
  {http://cdsads.u-strasbg.fr/abs/2010ApJ...719..299C} {719, 299}

\bibitem[\protect\citeauthoryear{{Cole}, {K{\"a}pyl{\"a}}, {Mantere}  \&
  {Brandenburg}}{{Cole} et~al.}{2014}]{Col14}
{Cole} E.,  {K{\"a}pyl{\"a}} P.~J.,  {Mantere} M.~J.,   {Brandenburg} A.,
  2014, \mn@doi [\apjl] {10.1088/2041-8205/780/2/L22}, \href
  {http://cdsads.u-strasbg.fr/abs/2014ApJ...780L..22C} {780, L22}

\bibitem[\protect\citeauthoryear{{Cole}, {Hackman}, {K{\"a}pyl{\"a}}, {Ilyin},
  {Kochukhov}  \& {Piskunov}}{{Cole} et~al.}{2015}]{Col15}
{Cole} E.~M.,  {Hackman} T.,  {K{\"a}pyl{\"a}} M.~J.,  {Ilyin} I.,  {Kochukhov}
  O.,   {Piskunov} N.,  2015, \mn@doi [\aap] {10.1051/0004-6361/201425440},
  \href {http://cdsads.u-strasbg.fr/abs/2015A%26A...581A..69C} {581, A69}

\bibitem[\protect\citeauthoryear{{Collier Cameron}}{{Collier
  Cameron}}{2007}]{Col07}
{Collier Cameron} A.,  2007, \mn@doi [Astronomische Nachrichten]
  {10.1002/asna.200710880}, \href
  {http://cdsads.u-strasbg.fr/abs/2007AN....328.1030C} {328, 1030}

\bibitem[\protect\citeauthoryear{{Cowan}, {Fuentes}  \& {Haggard}}{{Cowan}
  et~al.}{2013}]{Cow13}
{Cowan} N.~B.,  {Fuentes} P.~A.,   {Haggard} H.~M.,  2013, \mn@doi [\mnras]
  {10.1093/mnras/stt1191}, \href
  {http://adsabs.harvard.edu/abs/2013MNRAS.434.2465C} {434, 2465}

\bibitem[\protect\citeauthoryear{{Distefano}, {Lanzafame}, {Lanza}, {Messina}
  \& {Spada}}{{Distefano} et~al.}{2016}]{Dis16}
{Distefano} E.,  {Lanzafame} A.~C.,  {Lanza} A.~F.,  {Messina} S.,   {Spada}
  F.,  2016, \mn@doi [\aap] {10.1051/0004-6361/201527698}, \href
  {http://cdsads.u-strasbg.fr/abs/2016A%26A...591A..43D} {591, A43}

\bibitem[\protect\citeauthoryear{{Distefano}, {Lanzafame}, {Lanza}, {Messina}
  \& {Spada}}{{Distefano} et~al.}{2017}]{Dis17}
{Distefano} E.,  {Lanzafame} A.~C.,  {Lanza} A.~F.,  {Messina} S.,   {Spada}
  F.,  2017, \mn@doi [\aap] {10.1051/0004-6361/201730967}, \href
  {http://cdsads.u-strasbg.fr/abs/2017A%26A...606A..58D} {606, A58}

\bibitem[\protect\citeauthoryear{{Drake}, {Chung}, {Kashyap}, {Korhonen}, {Van
  Ballegooijen}  \& {Elstner}}{{Drake} et~al.}{2008}]{Dra08}
{Drake} J.~J.,  {Chung} S.~M.,  {Kashyap} V.,  {Korhonen} H.,  {Van
  Ballegooijen} A.,   {Elstner} D.,  2008, \mn@doi [\apj] {10.1086/587443},
  \href {http://cdsads.u-strasbg.fr/abs/2008ApJ...679.1522D} {679, 1522}

\bibitem[\protect\citeauthoryear{Draper \& Smith}{Draper \&
  Smith}{1998}]{Dra98}
Draper N.~R.,  Smith H.,  1998, Applied Regression Analysis.
John Wiley {\&} Sons, Inc., \mn@doi{10.1002/9781118625590}

\bibitem[\protect\citeauthoryear{{Efron} \& {Tibshirani}}{{Efron} \&
  {Tibshirani}}{1986}]{Efr86}
{Efron} B.,  {Tibshirani} R.,  1986, Statistical Science, 1, 54

\bibitem[\protect\citeauthoryear{{Eggen} \& {Iben}}{{Eggen} \&
  {Iben}}{1989}]{Egg89}
{Eggen} O.~J.,  {Iben} Jr. I.,  1989, \mn@doi [\aj] {10.1086/114993}, \href
  {http://cdsads.u-strasbg.fr/abs/1989AJ.....97..431E} {97, 431}

\bibitem[\protect\citeauthoryear{{Elstner} \& {Korhonen}}{{Elstner} \&
  {Korhonen}}{2005}]{Els05}
{Elstner} D.,  {Korhonen} H.,  2005, \mn@doi [Astronomische Nachrichten]
  {10.1002/asna.200410389}, \href
  {http://cdsads.u-strasbg.fr/abs/2005AN....326..278E} {326, 278}

\bibitem[\protect\citeauthoryear{{Fabbian} et~al.,}{{Fabbian}
  et~al.}{2017}]{Fab17}
{Fabbian} D.,  et~al., 2017, \mn@doi [Astronomische Nachrichten]
  {10.1002/asna.201713403}, \href
  {http://cdsads.u-strasbg.fr/abs/2017AN....338..753F} {338, 753}

\bibitem[\protect\citeauthoryear{{Ferreira Lopes}, {Le{\~a}o}, {de Freitas},
  {Canto Martins}, {Catelan}  \& {De Medeiros}}{{Ferreira Lopes}
  et~al.}{2015}]{Fer15}
{Ferreira Lopes} C.~E.,  {Le{\~a}o} I.~C.,  {de Freitas} D.~B.,  {Canto
  Martins} B.~L.,  {Catelan} M.,   {De Medeiros} J.~R.,  2015, \mn@doi [\aap]
  {10.1051/0004-6361/201424900}, \href
  {http://cdsads.u-strasbg.fr/abs/2015A%26A...583A.134F} {583, A134}

\bibitem[\protect\citeauthoryear{{Flores Soriano} \& {Strassmeier}}{{Flores
  Soriano} \& {Strassmeier}}{2017}]{Flo17}
{Flores Soriano} M.,  {Strassmeier} K.~G.,  2017, \mn@doi [\aap]
  {10.1051/0004-6361/201629338}, \href
  {http://cdsads.u-strasbg.fr/abs/2017A%26A...597A.101F} {597, A101}

\bibitem[\protect\citeauthoryear{{Fluri} \& {Berdyugina}}{{Fluri} \&
  {Berdyugina}}{2004}]{Flu04}
{Fluri} D.~M.,  {Berdyugina} S.~V.,  2004, \mn@doi [\solphys]
  {10.1007/s11207-005-4147-y}, \href
  {http://cdsads.u-strasbg.fr/abs/2004SoPh..224..153F} {224, 153}

\bibitem[\protect\citeauthoryear{{Giles}, {Collier Cameron}  \&
  {Haywood}}{{Giles} et~al.}{2017}]{Gil17}
{Giles} H.~A.~C.,  {Collier Cameron} A.,   {Haywood} R.~D.,  2017, \mn@doi
  [\mnras] {10.1093/mnras/stx1931}, \href
  {http://cdsads.u-strasbg.fr/abs/2017MNRAS.472.1618G} {472, 1618}

\bibitem[\protect\citeauthoryear{{Gyenge}, {Baranyi}  \&
  {Ludm{\'a}ny}}{{Gyenge} et~al.}{2014}]{Gye14}
{Gyenge} N.,  {Baranyi} T.,   {Ludm{\'a}ny} A.,  2014, \mn@doi [\solphys]
  {10.1007/s11207-013-0424-3}, \href
  {http://cdsads.u-strasbg.fr/abs/2014SoPh..289..579G} {289, 579}

\bibitem[\protect\citeauthoryear{{Hackman}}{{Hackman}}{2004}]{Hac04}
{Hackman} T.,  2004, PhD thesis, University of Helsinki, Finland

\bibitem[\protect\citeauthoryear{{Hackman} et~al.,}{{Hackman}
  et~al.}{2013}]{Hac13}
{Hackman} T.,  et~al., 2013, \mn@doi [\aap] {10.1051/0004-6361/201220690},
  \href {http://cdsads.u-strasbg.fr/abs/2013A%26A...553A..40H} {553, A40}

\bibitem[\protect\citeauthoryear{{Hale}}{{Hale}}{1908}]{Hal08}
{Hale} G.~E.,  1908, \mn@doi [\apj] {10.1086/141602}, \href
  {http://adsabs.harvard.edu/abs/1908ApJ....28..315H} {28, 315}

\bibitem[\protect\citeauthoryear{{Hall}}{{Hall}}{1991}]{Hal91A}
{Hall} D.~S.,  1991, in {Tuominen} I.,  {Moss} D.,   {R{\"u}diger} G.,  eds,
  Lecture Notes in Physics, Berlin Springer Verlag Vol. 380, IAU Colloq. 130:
  The Sun and Cool Stars. Activity, Magnetism, Dynamos. p.~353,
  \mn@doi{10.1007/3-540-53955-7_156}

\bibitem[\protect\citeauthoryear{{Hall} \& {Busby}}{{Hall} \&
  {Busby}}{1990}]{Hal90}
{Hall} D.~S.,  {Busby} M.~R.,  1990, in NATO Advanced Science Institutes (ASI)
  Series C. p.~377

\bibitem[\protect\citeauthoryear{{Hatzes}}{{Hatzes}}{2013}]{Hat13}
{Hatzes} A.~P.,  2013, \mn@doi [\apj] {10.1088/0004-637X/770/2/133}, \href
  {http://cdsads.u-strasbg.fr/abs/2013ApJ...770..133H} {770, 133}

\bibitem[\protect\citeauthoryear{{Henry}}{{Henry}}{1995}]{Hen95A}
{Henry} G.~W.,  1995, in {Henry} G.~W.,  {Eaton} J.~A.,  eds,  Astronomical
  Society of the Pacific Conference Series Vol. 79, Robotic Telescopes. Current
  Capabilities, Present Developments, and Future Prospects for Automated
  Astronomy. p.~44

\bibitem[\protect\citeauthoryear{{Howard}}{{Howard}}{1994}]{How94}
{Howard} R.~F.,  1994, in {Balasubramaniam} K.~S.,  {Simon} G.~W.,  eds,
  Astronomical Society of the Pacific Conference Series Vol. 68, Solar Active
  Region Evolution: Comparing Models with Observations. p.~1

\bibitem[\protect\citeauthoryear{{Howell}, {Mason}, {Boyd}, {Smith}  \&
  {Gelino}}{{Howell} et~al.}{2016}]{How16}
{Howell} S.~B.,  {Mason} E.,  {Boyd} P.,  {Smith} K.~L.,   {Gelino} D.~M.,
  2016, \mn@doi [\apj] {10.3847/0004-637X/831/1/27}, \href
  {http://cdsads.u-strasbg.fr/abs/2016ApJ...831...27H} {831, 27}

\bibitem[\protect\citeauthoryear{{Hussain}}{{Hussain}}{2002}]{Hus02}
{Hussain} G.~A.~J.,  2002, \mn@doi [Astronomische Nachrichten]
  {10.1002/1521-3994(200208)323:3/4<349::AID-ASNA349>3.0.CO;2-E}, \href
  {http://cdsads.u-strasbg.fr/abs/2002AN....323..349H} {323, 349}

\bibitem[\protect\citeauthoryear{{J{\"a}rvinen}, {Korhonen}, {Berdyugina},
  {Ilyin}, {Strassmeier}, {Weber}, {Savanov}  \& {Tuominen}}{{J{\"a}rvinen}
  et~al.}{2008}]{Jar08}
{J{\"a}rvinen} S.~P.,  {Korhonen} H.,  {Berdyugina} S.~V.,  {Ilyin} I.,
  {Strassmeier} K.~G.,  {Weber} M.,  {Savanov} I.,   {Tuominen} I.,  2008,
  \mn@doi [\aap] {10.1051/0004-6361:200809837}, \href
  {http://adsabs.harvard.edu/abs/2008A%26A...488.1047J} {488, 1047}

\bibitem[\protect\citeauthoryear{{Jeffers}}{{Jeffers}}{2005}]{Jef05}
{Jeffers} S.~V.,  2005, \mn@doi [\mnras] {10.1111/j.1365-2966.2005.08951.x},
  \href {http://adsabs.harvard.edu/abs/2005MNRAS.359..729J} {359, 729}

\bibitem[\protect\citeauthoryear{{Jeffers} \& {Keller}}{{Jeffers} \&
  {Keller}}{2009}]{Jef09}
{Jeffers} S.~V.,  {Keller} C.~U.,  2009, in {Stempels} E.,  ed.,  American
  Institute of Physics Conference Series Vol. 1094, 15th Cambridge Workshop on
  Cool Stars, Stellar Systems, and the Sun. pp 664--667,
  \mn@doi{10.1063/1.3099201}

\bibitem[\protect\citeauthoryear{{Jetsu}}{{Jetsu}}{1996}]{Jet96}
{Jetsu} L.,  1996, \aap, \href
  {http://cdsads.u-strasbg.fr/abs/1996A%26A...314..153J} {314, 153}

\bibitem[\protect\citeauthoryear{{Jetsu} \& {Pelt}}{{Jetsu} \&
  {Pelt}}{1996}]{Jet96A}
{Jetsu} L.,  {Pelt} J.,  1996, \aaps, \href
  {http://cdsads.u-strasbg.fr/abs/1996A%26AS..118..587J} {118, 587}

\bibitem[\protect\citeauthoryear{{Jetsu} \& {Pelt}}{{Jetsu} \&
  {Pelt}}{1999}]{Jet99}
{Jetsu} L.,  {Pelt} J.,  1999, \mn@doi [\aaps] {10.1051/aas:1999411}, \href
  {http://cdsads.u-strasbg.fr/abs/1999A%26AS..139..629J} {139, 629}

\bibitem[\protect\citeauthoryear{{Jetsu} \& {Porceddu}}{{Jetsu} \&
  {Porceddu}}{2015}]{Jet15}
{Jetsu} L.,  {Porceddu} S.,  2015, \mn@doi [PLoS ONE]
  {10.1371/journal.pone.0144140}, \href
  {http://cdsads.u-strasbg.fr/abs/2015PLoSO..1044140J} {10(12), e0144140}

\bibitem[\protect\citeauthoryear{{Jetsu}, {Huovelin}, {Tuominen}, {Vilhu},
  {Bopp}  \& {Piirola}}{{Jetsu} et~al.}{1990}]{Jet90}
{Jetsu} L.,  {Huovelin} J.,  {Tuominen} I.,  {Vilhu} O.,  {Bopp} B.~.~W.,
  {Piirola} V.,  1990, \aap, \href
  {http://adsabs.harvard.edu/abs/1990A%26A...236..423J} {236, 423}

\bibitem[\protect\citeauthoryear{{Jetsu}, {Pelt}, {Tuominen}  \&
  {Nations}}{{Jetsu} et~al.}{1991}]{Jet91}
{Jetsu} L.,  {Pelt} J.,  {Tuominen} I.,   {Nations} H.,  1991, in {Tuominen}
  I.,  {Moss} D.,   {R{\"u}diger} G.,  eds,  Lecture Notes in Physics, Berlin
  Springer Verlag Vol. 380, IAU Colloq. 130: The Sun and Cool Stars. Activity,
  Magnetism, Dynamos. p.~381, \mn@doi{10.1007/3-540-53955-7_161}

\bibitem[\protect\citeauthoryear{{Jetsu}, {Kokko}  \& {Tuominen}}{{Jetsu}
  et~al.}{1992}]{Jet92}
{Jetsu} L.,  {Kokko} M.,   {Tuominen} I.,  1992, \aap, \href
  {http://adsabs.harvard.edu/abs/1992A%26A...265..547J} {265, 547}

\bibitem[\protect\citeauthoryear{{Jetsu}, {Pelt}  \& {Tuominen}}{{Jetsu}
  et~al.}{1993}]{Jet93}
{Jetsu} L.,  {Pelt} J.,   {Tuominen} I.,  1993, \aap, \href
  {http://cdsads.u-strasbg.fr/abs/1993A%26A...278..449J} {278, 449}

\bibitem[\protect\citeauthoryear{{Jetsu}, {Pohjolainen}, {Pelt}  \&
  {Tuominen}}{{Jetsu} et~al.}{1997}]{Jet97A}
{Jetsu} L.,  {Pohjolainen} S.,  {Pelt} J.,   {Tuominen} I.,  1997, \aap, \href
  {http://cdsads.u-strasbg.fr/abs/1997A%26A...318..293J} {318, 293}

\bibitem[\protect\citeauthoryear{{Jetsu}, {Pelt}  \& {Tuominen}}{{Jetsu}
  et~al.}{1999}]{Jet99A}
{Jetsu} L.,  {Pelt} J.,   {Tuominen} I.,  1999, \aap, \href
  {http://cdsads.u-strasbg.fr/abs/1999A%26A...351..212J} {351, 212}

\bibitem[\protect\citeauthoryear{{Jetsu}, {Hackman}, {Hall}, {Henry}, {Kokko}
  \& {You}}{{Jetsu} et~al.}{2000}]{Jet00}
{Jetsu} L.,  {Hackman} T.,  {Hall} D.~S.,  {Henry} G.~W.,  {Kokko} M.,   {You}
  J.,  2000, \aap, \href {http://cdsads.u-strasbg.fr/abs/2000A%26A...362..223J}
  {362, 223}

\bibitem[\protect\citeauthoryear{{Jetsu}, {Porceddu}, {Lyytinen}, {Kajatkari},
  {Lehtinen}, {Markkanen}  \& {Toivari-Viitala}}{{Jetsu} et~al.}{2013}]{Jet13}
{Jetsu} L.,  {Porceddu} S.,  {Lyytinen} J.,  {Kajatkari} P.,  {Lehtinen} J.,
  {Markkanen} T.,   {Toivari-Viitala} J.,  2013, \mn@doi [\apj]
  {10.1088/0004-637X/773/1/1}, \href
  {http://cdsads.u-strasbg.fr/abs/2013ApJ...773....1J} {773, 1}

\bibitem[\protect\citeauthoryear{{Jetsu}, {Henry}  \& {Lehtinen}}{{Jetsu}
  et~al.}{2017}]{Jet17}
{Jetsu} L.,  {Henry} G.~W.,   {Lehtinen} J.,  2017, \mn@doi [\apj]
  {10.3847/1538-4357/aa65cb}, \href
  {http://cdsads.u-strasbg.fr/abs/2017ApJ...838..122J} {838, 122}

\bibitem[\protect\citeauthoryear{{K{\H o}v{\'a}ri}, {Ol{\'a}h}, {Kriskovics},
  {Vida}, {Forg{\'a}cs-Dajka}  \& {Strassmeier}}{{K{\H o}v{\'a}ri}
  et~al.}{2017}]{Kov17}
{K{\H o}v{\'a}ri} Z.,  {Ol{\'a}h} K.,  {Kriskovics} L.,  {Vida} K.,
  {Forg{\'a}cs-Dajka} E.,   {Strassmeier} K.~G.,  2017, \mn@doi [Astronomische
  Nachrichten] {10.1002/asna.201713400}, \href
  {http://adsabs.harvard.edu/abs/2017AN....338..903K} {338, 903}

\bibitem[\protect\citeauthoryear{Kaltenbacher, Neubauer  \&
  Scherzer}{Kaltenbacher et~al.}{2008}]{Kal08}
Kaltenbacher B.,  Neubauer A.,   Scherzer O.,  2008, Iterative Regularization
  Methods for Nonlinear Ill-Posed Problems.
Radon Series on Computational and Applied Mathematics, De Gruyter

\bibitem[\protect\citeauthoryear{{K{\"a}pyl{\"a}}, {Mantere}  \&
  {Hackman}}{{K{\"a}pyl{\"a}} et~al.}{2011}]{Kap11}
{K{\"a}pyl{\"a}} P.~J.,  {Mantere} M.~J.,   {Hackman} T.,  2011, \mn@doi [\apj]
  {10.1088/0004-637X/742/1/34}, \href
  {http://cdsads.u-strasbg.fr/abs/2011ApJ...742...34K} {742, 34}

\bibitem[\protect\citeauthoryear{{Kipping}}{{Kipping}}{2012}]{Kip12}
{Kipping} D.~M.,  2012, \mn@doi [\mnras] {10.1111/j.1365-2966.2012.22124.x},
  \href {http://adsabs.harvard.edu/abs/2012MNRAS.427.2487K} {427, 2487}

\bibitem[\protect\citeauthoryear{{Kirk} et~al.,}{{Kirk} et~al.}{2016}]{Kir16}
{Kirk} B.,  et~al., 2016, \mn@doi [\aj] {10.3847/0004-6256/151/3/68}, \href
  {http://adsabs.harvard.edu/abs/2016AJ....151...68K} {151, 68}

\bibitem[\protect\citeauthoryear{{Kochukhov}}{{Kochukhov}}{2016}]{Koc16}
{Kochukhov} O.,  2016, in {Rozelot} J.-P.,  {Neiner} C.,  eds,  Lecture Notes
  in Physics, Berlin Springer Verlag Vol. 914, Lecture Notes in Physics, Berlin
  Springer Verlag. p.~177, \mn@doi{10.1007/978-3-319-24151-7_9}

\bibitem[\protect\citeauthoryear{{Korhonen} \& {Elstner}}{{Korhonen} \&
  {Elstner}}{2005}]{Kor05}
{Korhonen} H.,  {Elstner} D.,  2005, \mn@doi [\aap]
  {10.1051/0004-6361:20053562}, \href
  {http://cdsads.u-strasbg.fr/abs/2005A%26A...440.1161K} {440, 1161}

\bibitem[\protect\citeauthoryear{{Korhonen} \& {Elstner}}{{Korhonen} \&
  {Elstner}}{2011}]{Kor11}
{Korhonen} H.,  {Elstner} D.,  2011, \mn@doi [\aap]
  {10.1051/0004-6361/201117016}, \href
  {http://cdsads.u-strasbg.fr/abs/2011A%26A...532A.106K} {532, A106}

\bibitem[\protect\citeauthoryear{{Korhonen}, {Berdyugina}, {Hackman},
  {Duemmler}, {Ilyin}  \& {Tuominen}}{{Korhonen} et~al.}{1999}]{Kor99}
{Korhonen} H.,  {Berdyugina} S.~V.,  {Hackman} T.,  {Duemmler} R.,  {Ilyin}
  I.~V.,   {Tuominen} I.,  1999, \aap, \href
  {http://cdsads.u-strasbg.fr/abs/1999A%26A...346..101K} {346, 101}

\bibitem[\protect\citeauthoryear{{Korhonen}, {Berdyugina}, {Hackman},
  {Strassmeier}  \& {Tuominen}}{{Korhonen} et~al.}{2000}]{Kor00}
{Korhonen} H.,  {Berdyugina} S.~V.,  {Hackman} T.,  {Strassmeier} K.~G.,
  {Tuominen} I.,  2000, \aap, \href
  {http://cdsads.u-strasbg.fr/abs/2000A%26A...360.1067K} {360, 1067}

\bibitem[\protect\citeauthoryear{{Korhonen} et~al.,}{{Korhonen}
  et~al.}{2001}]{Kor01A}
{Korhonen} H.,  et~al., 2001, \mn@doi [\aap] {10.1051/0004-6361:20010804},
  \href {http://cdsads.u-strasbg.fr/abs/2001A%26A...374.1049K} {374, 1049}

\bibitem[\protect\citeauthoryear{{Korhonen}, {Berdyugina}  \&
  {Tuominen}}{{Korhonen} et~al.}{2002}]{Kor02}
{Korhonen} H.,  {Berdyugina} S.~V.,   {Tuominen} I.,  2002, \mn@doi [\aap]
  {10.1051/0004-6361:20020674}, \href
  {http://cdsads.u-strasbg.fr/abs/2002A%26A...390..179K} {390, 179}

\bibitem[\protect\citeauthoryear{{Korhonen}, {Berdyugina}  \&
  {Tuominen}}{{Korhonen} et~al.}{2004}]{Kor04}
{Korhonen} H.,  {Berdyugina} S.~V.,   {Tuominen} I.,  2004, \mn@doi
  [Astronomische Nachrichten] {10.1002/asna.200310240}, \href
  {http://cdsads.u-strasbg.fr/abs/2004AN....325..402K} {325, 402}

\bibitem[\protect\citeauthoryear{{Korhonen}, {Berdyugina}, {Hackman}, {Ilyin},
  {Strassmeier}  \& {Tuominen}}{{Korhonen} et~al.}{2007}]{Kor07}
{Korhonen} H.,  {Berdyugina} S.~V.,  {Hackman} T.,  {Ilyin} I.~V.,
  {Strassmeier} K.~G.,   {Tuominen} I.,  2007, \mn@doi [\aap]
  {10.1051/0004-6361:20041806}, \href
  {http://cdsads.u-strasbg.fr/abs/2007A%26A...476..881K} {476, 881}

\bibitem[\protect\citeauthoryear{{K{\'o}sp{\'a}l}, {{\'A}brah{\'a}m}, {Zsidi},
  {Vida}, {Szab{\'o}}, {Mo{\'o}r}  \& {P{\'a}l}}{{K{\'o}sp{\'a}l}
  et~al.}{2018}]{Kos18}
{K{\'o}sp{\'a}l} {\'A}.,  {{\'A}brah{\'a}m} P.,  {Zsidi} G.,  {Vida} K.,
  {Szab{\'o}} R.,  {Mo{\'o}r} A.,   {P{\'a}l} A.,  2018, \mn@doi [\apj]
  {10.3847/1538-4357/aacafa}, \href
  {http://adsabs.harvard.edu/abs/2018ApJ...862...44K} {862, 44}

\bibitem[\protect\citeauthoryear{{Kouzuma}}{{Kouzuma}}{2019}]{Kou19}
{Kouzuma} S.,  2019, \mn@doi [\pasj] {10.1093/pasj/psy140}, \href
  {http://adsabs.harvard.edu/abs/2019PASJ...71...21K} {71, 21}

\bibitem[\protect\citeauthoryear{{Kraus} et~al.,}{{Kraus} et~al.}{2015}]{Kra15}
{Kraus} M.,  et~al., 2015, \mn@doi [\aap] {10.1051/0004-6361/201425383}, \href
  {http://cdsads.u-strasbg.fr/abs/2015A%26A...581A..75K} {581, A75}

\bibitem[\protect\citeauthoryear{{Krause} \& {Raedler}}{{Krause} \&
  {Raedler}}{1980}]{Kra80}
{Krause} F.,  {Raedler} K.-H.,  1980, {Mean-field magnetohydrodynamics and
  dynamo theory}.
~

\bibitem[\protect\citeauthoryear{{Kron}}{{Kron}}{1947}]{Kro47}
{Kron} G.~E.,  1947, \mn@doi [\pasp] {10.1086/125964}, \href
  {http://adsabs.harvard.edu/abs/1947PASP...59..261K} {59, 261}

\bibitem[\protect\citeauthoryear{{Lanza} et~al.,}{{Lanza} et~al.}{2009}]{Lan09}
{Lanza} A.~F.,  et~al., 2009, \mn@doi [\aap] {10.1051/0004-6361:200810591},
  \href {http://adsabs.harvard.edu/abs/2009A%26A...493..193L} {493, 193}

\bibitem[\protect\citeauthoryear{{Lanza}, {Das Chagas}  \& {De
  Medeiros}}{{Lanza} et~al.}{2014}]{Lan14}
{Lanza} A.~F.,  {Das Chagas} M.~L.,   {De Medeiros} J.~R.,  2014, \mn@doi
  [\aap] {10.1051/0004-6361/201323172}, \href
  {http://adsabs.harvard.edu/abs/2014A%26A...564A..50L} {564, A50}

\bibitem[\protect\citeauthoryear{{Lanzafame} et~al.,}{{Lanzafame}
  et~al.}{2018}]{Lan18}
{Lanzafame} A.~C.,  et~al., 2018, \mn@doi [\aap] {10.1051/0004-6361/201833334},
  \href {http://adsabs.harvard.edu/abs/2018A%26A...616A..16L} {616, A16}

\bibitem[\protect\citeauthoryear{{Lean}}{{Lean}}{1990}]{Lea90}
{Lean} J.,  1990, \mn@doi [\apj] {10.1086/169378}, \href
  {http://adsabs.harvard.edu/abs/1990ApJ...363..718L} {363, 718}

\bibitem[\protect\citeauthoryear{{Lehtinen}, {Jetsu}, {Hackman}, {Kajatkari}
  \& {Henry}}{{Lehtinen} et~al.}{2011}]{Leh11}
{Lehtinen} J.,  {Jetsu} L.,  {Hackman} T.,  {Kajatkari} P.,   {Henry} G.~W.,
  2011, \mn@doi [\aap] {10.1051/0004-6361/201015454}, \href
  {http://adsabs.harvard.edu/abs/2011A} {527, A136}

\bibitem[\protect\citeauthoryear{{Lehtinen}, {Jetsu}, {Hackman}, {Kajatkari}
  \& {Henry}}{{Lehtinen} et~al.}{2012}]{Leh12}
{Lehtinen} J.,  {Jetsu} L.,  {Hackman} T.,  {Kajatkari} P.,   {Henry} G.~W.,
  2012, \mn@doi [\aap] {10.1051/0004-6361/201219185}, \href
  {http://cdsads.u-strasbg.fr/abs/2012A%26A...542A..38L} {542, A38}

\bibitem[\protect\citeauthoryear{{Lehtinen}, {Jetsu}, {Hackman}, {Kajatkari}
  \& {Henry}}{{Lehtinen} et~al.}{2016}]{Leh16}
{Lehtinen} J.,  {Jetsu} L.,  {Hackman} T.,  {Kajatkari} P.,   {Henry} G.~W.,
  2016, \mn@doi [\aap] {10.1051/0004-6361/201527420}, \href
  {http://cdsads.u-strasbg.fr/abs/2016A%26A...588A..38L} {588, A38}

\bibitem[\protect\citeauthoryear{{Lindborg}, {Korpi}, {Hackman}, {Tuominen},
  {Ilyin}  \& {Piskunov}}{{Lindborg} et~al.}{2011}]{Lin11}
{Lindborg} M.,  {Korpi} M.~J.,  {Hackman} T.,  {Tuominen} I.,  {Ilyin} I.,
  {Piskunov} N.,  2011, \mn@doi [\aap] {10.1051/0004-6361/201015203}, \href
  {http://adsabs.harvard.edu/abs/2011A%26A...526A..44L} {526, A44}

\bibitem[\protect\citeauthoryear{{Lomb}}{{Lomb}}{1976}]{Lom76}
{Lomb} N.~R.,  1976, \mn@doi [\apss] {10.1007/BF00648343}, \href
  {http://cdsads.u-strasbg.fr/abs/1976Ap%26SS..39..447L} {39, 447}

\bibitem[\protect\citeauthoryear{{Loumos} \& {Deeming}}{{Loumos} \&
  {Deeming}}{1978}]{Lou78}
{Loumos} G.~L.,  {Deeming} T.~J.,  1978, \mn@doi [\apss] {10.1007/BF01879560},
  \href {http://cdsads.u-strasbg.fr/abs/1978Ap%26SS..56..285L} {56, 285}

\bibitem[\protect\citeauthoryear{{Mantere}, {K{\"a}pyl{\"a}}  \&
  {Hackman}}{{Mantere} et~al.}{2011}]{Man11}
{Mantere} M.~J.,  {K{\"a}pyl{\"a}} P.~J.,   {Hackman} T.,  2011, \mn@doi
  [Astronomische Nachrichten] {10.1002/asna.201111620}, \href
  {http://cdsads.u-strasbg.fr/abs/2011AN....332..876M} {332, 876}

\bibitem[\protect\citeauthoryear{{Massi}}{{Massi}}{2007}]{Mas07}
{Massi} M.,  2007, \memsai, \href
  {http://cdsads.u-strasbg.fr/abs/2007MmSAI..78..247M} {78, 247}

\bibitem[\protect\citeauthoryear{{Mennickent}}{{Mennickent}}{2017}]{Men17}
{Mennickent} R.~E.,  2017, \mn@doi [Serbian Astronomical Journal]
  {10.2298/SAJ1794001M}, \href
  {http://adsabs.harvard.edu/abs/2017SerAJ.194....1M} {194, 1}

\bibitem[\protect\citeauthoryear{{Moss}}{{Moss}}{1999}]{Mos99}
{Moss} D.,  1999, \mn@doi [\mnras] {10.1046/j.1365-8711.1999.02510.x}, \href
  {http://cdsads.u-strasbg.fr/abs/1999MNRAS.306..300M} {306, 300}

\bibitem[\protect\citeauthoryear{{Moss}}{{Moss}}{2004}]{Mos04}
{Moss} D.,  2004, \mn@doi [\mnras] {10.1111/j.1365-2966.2004.08125.x}, \href
  {http://cdsads.u-strasbg.fr/abs/2004MNRAS.352L..17M} {352, L17}

\bibitem[\protect\citeauthoryear{{Moss} \& {Tuominen}}{{Moss} \&
  {Tuominen}}{1997}]{Mos97}
{Moss} D.,  {Tuominen} I.,  1997, \aap, \href
  {http://cdsads.u-strasbg.fr/abs/1997A%26A...321..151M} {321, 151}

\bibitem[\protect\citeauthoryear{{Moss}, {Barker}, {Brandenburg}  \&
  {Tuominen}}{{Moss} et~al.}{1995}]{Mos95}
{Moss} D.,  {Barker} D.~M.,  {Brandenburg} A.,   {Tuominen} I.,  1995, \aap,
  \href {http://cdsads.u-strasbg.fr/abs/1995A%26A...294..155M} {294, 155}

\bibitem[\protect\citeauthoryear{{Nielsen}, {Schunker}, {Gizon}, {Schou}  \&
  {Ball}}{{Nielsen} et~al.}{2017}]{Nie17}
{Nielsen} M.~B.,  {Schunker} H.,  {Gizon} L.,  {Schou} J.,   {Ball} W.~H.,
  2017, \mn@doi [\aap] {10.1051/0004-6361/201730896}, \href
  {http://cdsads.u-strasbg.fr/abs/2017A%26A...603A...6N} {603, A6}

\bibitem[\protect\citeauthoryear{{Ol{\'a}h}, {Korhonen}, {K{\H o}v{\'a}ri},
  {Forg{\'a}cs-Dajka}  \& {Strassmeier}}{{Ol{\'a}h} et~al.}{2006}]{Ola06}
{Ol{\'a}h} K.,  {Korhonen} H.,  {K{\H o}v{\'a}ri} Z.,  {Forg{\'a}cs-Dajka} E.,
   {Strassmeier} K.~G.,  2006, \mn@doi [\aap] {10.1051/0004-6361:20054539},
  \href {http://adsabs.harvard.edu/abs/2006A%26A...452..303O} {452, 303}

\bibitem[\protect\citeauthoryear{{Ol{\'a}h} et~al.,}{{Ol{\'a}h}
  et~al.}{2009}]{Ola09}
{Ol{\'a}h} K.,  et~al., 2009, \mn@doi [\aap] {10.1051/0004-6361/200811304},
  \href {http://cdsads.u-strasbg.fr/abs/2009A%26A...501..703O} {501, 703}

\bibitem[\protect\citeauthoryear{{Oliver}, {Ballester}  \& {Baudin}}{{Oliver}
  et~al.}{1998}]{Oli98}
{Oliver} R.,  {Ballester} J.~L.,   {Baudin} F.,  1998, \mn@doi [\nat]
  {10.1038/29012}, \href {http://adsabs.harvard.edu/abs/1998Natur.394..552O}
  {394, 552}

\bibitem[\protect\citeauthoryear{{Olspert}, {Lehtinen}, {K{\"a}pyl{\"a}},
  {Pelt}  \& {Grigorievskiy}}{{Olspert} et~al.}{2018}]{Ols18}
{Olspert} N.,  {Lehtinen} J.~J.,  {K{\"a}pyl{\"a}} M.~J.,  {Pelt} J.,
  {Grigorievskiy} A.,  2018, \mn@doi [\aap] {10.1051/0004-6361/201732525},
  \href {http://adsabs.harvard.edu/abs/2018A%26A...619A...6O} {619, A6}

\bibitem[\protect\citeauthoryear{{{\"O}zavc{\i}}, {{\c S}enavc{\i}}, {I{\c
  s}{\i}k}, {Hussain}, {O'Neal}, {Y{\i}lmaz}  \& {Selam}}{{{\"O}zavc{\i}}
  et~al.}{2018}]{Oza18}
{{\"O}zavc{\i}} I.,  {{\c S}enavc{\i}} H.~V.,  {I{\c s}{\i}k} E.,  {Hussain}
  G.~A.~J.,  {O'Neal} D.,  {Y{\i}lmaz} M.,   {Selam} S.~O.,  2018, \mn@doi
  [\mnras] {10.1093/mnras/stx3053}, \href
  {http://cdsads.u-strasbg.fr/abs/2018MNRAS.474.5534O} {474, 5534}

\bibitem[\protect\citeauthoryear{{Panov} \& {Dimitrov}}{{Panov} \&
  {Dimitrov}}{2007}]{Pan07}
{Panov} K.,  {Dimitrov} D.,  2007, \mn@doi [\aap] {10.1051/0004-6361:20065596},
  \href {http://cdsads.u-strasbg.fr/abs/2007A%26A...467..229P} {467, 229}

\bibitem[\protect\citeauthoryear{{Pelt}, {Tuominen}  \& {Brooke}}{{Pelt}
  et~al.}{2005}]{Pel05}
{Pelt} J.,  {Tuominen} I.,   {Brooke} J.,  2005, \mn@doi [\aap]
  {10.1051/0004-6361:20041357}, \href
  {http://cdsads.u-strasbg.fr/abs/2005A%26A...429.1093P} {429, 1093}

\bibitem[\protect\citeauthoryear{{Peterson}, {Mutel}, {G{\"u}del}  \&
  {Goss}}{{Peterson} et~al.}{2010}]{Pat10}
{Peterson} W.~M.,  {Mutel} R.~L.,  {G{\"u}del} M.,   {Goss} W.~M.,  2010,
  \mn@doi [\nat] {10.1038/nature08643}, \href
  {http://cdsads.u-strasbg.fr/abs/2010Natur.463..207P} {463, 207}

\bibitem[\protect\citeauthoryear{{Petit} et~al.,}{{Petit} et~al.}{2004}]{Pet04}
{Petit} P.,  et~al., 2004, \mn@doi [\mnras] {10.1111/j.1365-2966.2004.07827.x},
  \href {http://cdsads.u-strasbg.fr/abs/2004MNRAS.351..826P} {351, 826}

\bibitem[\protect\citeauthoryear{{Pipin} \& {Kosovichev}}{{Pipin} \&
  {Kosovichev}}{2015}]{Pip15}
{Pipin} V.~V.,  {Kosovichev} A.~G.,  2015, \mn@doi [\apj]
  {10.1088/0004-637X/813/2/134}, \href
  {https://ui.adsabs.harvard.edu/#abs/2015ApJ...813..134P} {813, 134}

\bibitem[\protect\citeauthoryear{Porceddu, Jetsu, Markkanen  \&
  Toivari-Viitala}{Porceddu et~al.}{2008}]{Por08}
Porceddu S.,  Jetsu L.,  Markkanen T.,   Toivari-Viitala J.,  2008, \mn@doi
  [Cambridge Archaeological Journal] {10.1017/s0959774308000395}, 18, 327

\bibitem[\protect\citeauthoryear{Porceddu, Jetsu, Markkanen, Lyytinen,
  Kajatkari, Lehtinen  \& Toivari-Viitala}{Porceddu et~al.}{2018}]{Por18}
Porceddu S.,  Jetsu L.,  Markkanen T.,  Lyytinen J.,  Kajatkari P.,  Lehtinen
  J.,   Toivari-Viitala J.,  2018, Open Astronomy, 27, 232

\bibitem[\protect\citeauthoryear{{Puzin}, {Savanov}  \& {Dmitrienko}}{{Puzin}
  et~al.}{2014}]{Puz14}
{Puzin} V.~B.,  {Savanov} I.~S.,   {Dmitrienko} E.~S.,  2014, \mn@doi
  [Astronomy Reports] {10.1134/S1063772914070075}, \href
  {http://cdsads.u-strasbg.fr/abs/2014ARep...58..471P} {58, 471}

\bibitem[\protect\citeauthoryear{{Puzin}, {Savanov}, {Dmitrienko}, {Romanyuk},
  {Semenko}, {Yakunin}  \& {Burdanov}}{{Puzin} et~al.}{2016}]{Puz16}
{Puzin} V.~B.,  {Savanov} I.~S.,  {Dmitrienko} E.~S.,  {Romanyuk} I.~I.,
  {Semenko} E.~A.,  {Yakunin} I.~A.,   {Burdanov} A.~Y.,  2016, \mn@doi
  [Astrophysical Bulletin] {10.1134/S1990341316020061}, \href
  {http://cdsads.u-strasbg.fr/abs/2016AstBu..71..189P} {71, 189}

\bibitem[\protect\citeauthoryear{{Puzin}, {Savanov}  \& {Dmitrienko}}{{Puzin}
  et~al.}{2017}]{Puz17}
{Puzin} V.~B.,  {Savanov} I.~S.,   {Dmitrienko} E.~S.,  2017, \mn@doi
  [Astronomy Reports] {10.1134/S1063772917070071}, \href
  {http://cdsads.u-strasbg.fr/abs/2017ARep...61..693P} {61, 693}

\bibitem[\protect\citeauthoryear{{Queloz} et~al.,}{{Queloz}
  et~al.}{2009}]{Que09}
{Queloz} D.,  et~al., 2009, \mn@doi [\aap] {10.1051/0004-6361/200913096}, \href
  {http://cdsads.u-strasbg.fr/abs/2009A%26A...506..303Q} {506, 303}

\bibitem[\protect\citeauthoryear{{Radick} \& {Foukal}}{{Radick} \&
  {Foukal}}{1994}]{Rad94}
{Radick} R.~R.,  {Foukal} P.,  1994, \mn@doi [Science]
  {10.1126/science.266.5187.1072}, \href
  {https://ui.adsabs.harvard.edu/abs/1994Sci...266.1072R} {266, 1072}

\bibitem[\protect\citeauthoryear{{Radick}, {Lockwood}, {Henry}, {Hall}  \&
  {Pevtsov}}{{Radick} et~al.}{2018}]{Rad18}
{Radick} R.~R.,  {Lockwood} G.~W.,  {Henry} G.~W.,  {Hall} J.~C.,   {Pevtsov}
  A.~A.,  2018, \mn@doi [\apj] {10.3847/1538-4357/aaaae3}, \href
  {http://cdsads.u-strasbg.fr/abs/2018ApJ...855...75R} {855, 75}

\bibitem[\protect\citeauthoryear{{Reinhold} \& {Arlt}}{{Reinhold} \&
  {Arlt}}{2015}]{Rei15A}
{Reinhold} T.,  {Arlt} R.,  2015, \mn@doi [\aap] {10.1051/0004-6361/201425337},
  \href {http://adsabs.harvard.edu/abs/2015A%26A...576A..15R} {576, A15}

\bibitem[\protect\citeauthoryear{{Reinhold} \& {Gizon}}{{Reinhold} \&
  {Gizon}}{2015}]{Rei15}
{Reinhold} T.,  {Gizon} L.,  2015, \mn@doi [\aap]
  {10.1051/0004-6361/201526216}, \href
  {http://cdsads.u-strasbg.fr/abs/2015A%26A...583A..65R} {583, A65}

\bibitem[\protect\citeauthoryear{{Reinhold} \& {Reiners}}{{Reinhold} \&
  {Reiners}}{2013}]{Rei13A}
{Reinhold} T.,  {Reiners} A.,  2013, \mn@doi [\aap]
  {10.1051/0004-6361/201321161}, \href
  {http://cdsads.u-strasbg.fr/abs/2013A%26A...557A..11R} {557, A11}

\bibitem[\protect\citeauthoryear{{Reinhold}, {Reiners}  \& {Basri}}{{Reinhold}
  et~al.}{2013}]{Rei13}
{Reinhold} T.,  {Reiners} A.,   {Basri} G.,  2013, \mn@doi [\aap]
  {10.1051/0004-6361/201321970}, \href
  {http://cdsads.u-strasbg.fr/abs/2013A%26A...560A...4R} {560, A4}

\bibitem[\protect\citeauthoryear{{Reinhold}, {Cameron}  \& {Gizon}}{{Reinhold}
  et~al.}{2017}]{Rei17}
{Reinhold} T.,  {Cameron} R.~H.,   {Gizon} L.,  2017, \mn@doi [\aap]
  {10.1051/0004-6361/201730599}, \href
  {http://cdsads.u-strasbg.fr/abs/2017A%26A...603A..52R} {603, A52}

\bibitem[\protect\citeauthoryear{{Reyniers}, {Degroote}, {Bodewits}, {Cuypers}
  \& {Waelkens}}{{Reyniers} et~al.}{2009}]{Rey09}
{Reyniers} M.,  {Degroote} P.,  {Bodewits} D.,  {Cuypers} J.,   {Waelkens} C.,
  2009, \mn@doi [\aap] {10.1051/0004-6361:20079225}, \href
  {http://cdsads.u-strasbg.fr/abs/2009A%26A...494..379R} {494, 379}

\bibitem[\protect\citeauthoryear{{Rieger}, {Kanbach}, {Reppin}, {Share},
  {Forrest}  \& {Chupp}}{{Rieger} et~al.}{1984}]{Rie84}
{Rieger} E.,  {Kanbach} G.,  {Reppin} C.,  {Share} G.~H.,  {Forrest} D.~J.,
  {Chupp} E.~L.,  1984, \mn@doi [\nat] {10.1038/312623a0}, \href
  {http://cdsads.u-strasbg.fr/abs/1984Natur.312..623R} {312, 623}

\bibitem[\protect\citeauthoryear{{Roettenbacher} et~al.,}{{Roettenbacher}
  et~al.}{2016}]{Roe16}
{Roettenbacher} R.~M.,  et~al., 2016, \mn@doi [\nat] {10.1038/nature17444},
  \href {http://cdsads.u-strasbg.fr/abs/2016Natur.533..217R} {533, 217}

\bibitem[\protect\citeauthoryear{{Russell}}{{Russell}}{1906}]{Rus06}
{Russell} H.~N.,  1906, \mn@doi [\apj] {10.1086/141361}, \href
  {http://adsabs.harvard.edu/abs/1906ApJ....24....1R} {24, 1}

\bibitem[\protect\citeauthoryear{{Saar} \& {Brandenburg}}{{Saar} \&
  {Brandenburg}}{1999}]{Saa99}
{Saar} S.~H.,  {Brandenburg} A.,  1999, \mn@doi [\apj] {10.1086/307794}, \href
  {http://cdsads.u-strasbg.fr/abs/1999ApJ...524..295S} {524, 295}

\bibitem[\protect\citeauthoryear{{Saio}, {Bedding}, {Kurtz}, {Murphy},
  {Antoci}, {Shibahashi}, {Li}  \& {Takata}}{{Saio} et~al.}{2018}]{Sai18}
{Saio} H.,  {Bedding} T.~R.,  {Kurtz} D.~W.,  {Murphy} S.~J.,  {Antoci} V.,
  {Shibahashi} H.,  {Li} G.,   {Takata} M.,  2018, \mn@doi [\mnras]
  {10.1093/mnras/sty784}, \href
  {http://cdsads.u-strasbg.fr/abs/2018MNRAS.477.2183S} {477, 2183}

\bibitem[\protect\citeauthoryear{{Savanov} \& {Strassmeier}}{{Savanov} \&
  {Strassmeier}}{2008}]{Sav08}
{Savanov} I.~S.,  {Strassmeier} K.~G.,  2008, \mn@doi [Astronomische
  Nachrichten] {10.1002/asna.200710963}, \href
  {http://cdsads.u-strasbg.fr/abs/2008AN....329..364S} {329, 364}

\bibitem[\protect\citeauthoryear{{Scargle}}{{Scargle}}{1982}]{Sca82}
{Scargle} J.~D.,  1982, \mn@doi [\apj] {10.1086/160554}, \href
  {http://cdsads.u-strasbg.fr/abs/1982ApJ...263..835S} {263, 835}

\bibitem[\protect\citeauthoryear{{Schwabe}}{{Schwabe}}{1844}]{Sch44}
{Schwabe} M.,  1844, Astronomische Nachrichten, \href
  {http://cdsads.u-strasbg.fr/abs/1844AN.....21..233S} {21, 233}

\bibitem[\protect\citeauthoryear{{Shultz} et~al.,}{{Shultz}
  et~al.}{2018}]{Shu18}
{Shultz} M.~E.,  et~al., 2018, \mn@doi [\mnras] {10.1093/mnras/sty103}, \href
  {http://cdsads.u-strasbg.fr/abs/2018MNRAS.475.5144S} {475, 5144}

\bibitem[\protect\citeauthoryear{{Snieder}}{{Snieder}}{1998}]{Sni98}
{Snieder} R.,  1998, Inverse Problems, 14, 387

\bibitem[\protect\citeauthoryear{{Strassmeier}}{{Strassmeier}}{2009}]{Str09}
{Strassmeier} K.~G.,  2009, \mn@doi [\aapr] {10.1007/s00159-009-0020-6}, \href
  {http://adsabs.harvard.edu/abs/2009A%26ARv..17..251S} {17, 251}

\bibitem[\protect\citeauthoryear{{Strassmeier} \& {Bartus}}{{Strassmeier} \&
  {Bartus}}{2000}]{Str00A}
{Strassmeier} K.~G.,  {Bartus} J.,  2000, \aap, \href
  {http://cdsads.u-strasbg.fr/abs/2000A%26A...354..537S} {354, 537}

\bibitem[\protect\citeauthoryear{{Strassmeier}, {Boyd}, {Epand}  \&
  {Granzer}}{{Strassmeier} et~al.}{1997}]{Str97}
{Strassmeier} K.~G.,  {Boyd} L.~J.,  {Epand} D.~H.,   {Granzer} T.,  1997,
  \mn@doi [\pasp] {10.1086/133934}, \href
  {http://cdsads.u-strasbg.fr/abs/1997PASP..109..697S} {109, 697}

\bibitem[\protect\citeauthoryear{{Tarantola} \& {Valette}}{{Tarantola} \&
  {Valette}}{1982}]{Tar82}
{Tarantola} A.,  {Valette} B.,  1982, \mn@doi [Reviews of Geophysics and Space
  Physics] {10.1029/RG020i002p00219}, \href
  {http://adsabs.harvard.edu/abs/1982RvGSP..20..219T} {20, 219}

\bibitem[\protect\citeauthoryear{{Tuominen}, {Berdyugina}  \&
  {Korpi}}{{Tuominen} et~al.}{2002}]{Tuo02}
{Tuominen} I.,  {Berdyugina} S.~V.,   {Korpi} M.~J.,  2002, \mn@doi
  [Astronomische Nachrichten]
  {10.1002/1521-3994(200208)323:3/4<367::AID-ASNA367>3.0.CO;2-E}, \href
  {http://cdsads.u-strasbg.fr/abs/2002AN....323..367T} {323, 367}

\bibitem[\protect\citeauthoryear{{Tuominen}, {Pelt}, {Brooke}, {Korpi}  \&
  {Ostwriter}}{{Tuominen} et~al.}{2007}]{Tuo07}
{Tuominen} I.,  {Pelt} J.,  {Brooke} J.,  {Korpi} M.,   {Ostwriter} G.~H.,
  2007, \mn@doi [Astronomische Nachrichten] {10.1002/asna.200710845}, \href
  {http://cdsads.u-strasbg.fr/abs/2007AN....328.1020T} {328, 1020}

\bibitem[\protect\citeauthoryear{{Van Helden}}{{Van Helden}}{1996}]{van96}
{Van Helden} A.,  1996, Proc. Am. Phil. Soc., 140, 358

\bibitem[\protect\citeauthoryear{{Vernova}, {Tyasto}  \& {Baranov}}{{Vernova}
  et~al.}{2007}]{Ver07}
{Vernova} E.~S.,  {Tyasto} M.~I.,   {Baranov} D.~G.,  2007, \mn@doi [\solphys]
  {10.1007/s11207-007-9018-2}, \href
  {http://cdsads.u-strasbg.fr/abs/2007SoPh..245..177V} {245, 177}

\bibitem[\protect\citeauthoryear{{Vida}, {Korhonen}, {Ilyin}, {Ol{\'a}h},
  {Andersen}  \& {Hackman}}{{Vida} et~al.}{2015}]{Vid15}
{Vida} K.,  {Korhonen} H.,  {Ilyin} I.~V.,  {Ol{\'a}h} K.,  {Andersen} M.~I.,
  {Hackman} T.,  2015, \mn@doi [\aap] {10.1051/0004-6361/201526066}, \href
  {http://cdsads.u-strasbg.fr/abs/2015A%26A...580A..64V} {580, A64}

\bibitem[\protect\citeauthoryear{{Viviani}, {Warnecke}, {K{\"a}pyl{\"a}},
  {K{\"a}pyl{\"a}}, {Olspert}, {Cole-Kodikara}, {Lehtinen}  \&
  {Brandenburg}}{{Viviani} et~al.}{2018}]{Viv18}
{Viviani} M.,  {Warnecke} J.,  {K{\"a}pyl{\"a}} M.~J.,  {K{\"a}pyl{\"a}} P.~J.,
   {Olspert} N.,  {Cole-Kodikara} E.~M.,  {Lehtinen} J.~J.,   {Brandenburg} A.,
   2018, \mn@doi [\aap] {10.1051/0004-6361/201732191}, \href
  {http://adsabs.harvard.edu/abs/2018A%26A...616A.160V} {616, A160}

\bibitem[\protect\citeauthoryear{{Wang}, {Zhang}, {Deng}, {Luo}, {Luo}  \&
  {Zhang}}{{Wang} et~al.}{2015}]{Wan15}
{Wang} K.,  {Zhang} X.,  {Deng} L.,  {Luo} C.,  {Luo} Y.,   {Zhang} J.,  2015,
  \mn@doi [\apj] {10.1088/0004-637X/805/1/22}, \href
  {http://cdsads.u-strasbg.fr/abs/2015ApJ...805...22W} {805, 22}

\bibitem[\protect\citeauthoryear{{Wargelin}, {Saar}, {Pojma{\'n}ski}, {Drake}
  \& {Kashyap}}{{Wargelin} et~al.}{2017}]{War17A}
{Wargelin} B.~J.,  {Saar} S.~H.,  {Pojma{\'n}ski} G.,  {Drake} J.~J.,
  {Kashyap} V.~L.,  2017, \mn@doi [\mnras] {10.1093/mnras/stw2570}, \href
  {http://cdsads.u-strasbg.fr/abs/2017MNRAS.464.3281W} {464, 3281}

\bibitem[\protect\citeauthoryear{{Warnecke}}{{Warnecke}}{2018}]{War18}
{Warnecke} J.,  2018, \mn@doi [\aap] {10.1051/0004-6361/201732413}, \href
  {http://adsabs.harvard.edu/abs/2018A%26A...616A..72W} {616, A72}

\bibitem[\protect\citeauthoryear{{Washuettl}, {Strassmeier}  \&
  {Weber}}{{Washuettl} et~al.}{2009}]{Was09}
{Washuettl} A.,  {Strassmeier} K.~G.,   {Weber} M.,  2009, \mn@doi
  [Astronomische Nachrichten] {10.1002/asna.200811136}, \href
  {http://adsabs.harvard.edu/abs/2009AN....330..366W} {330, 366}

\bibitem[\protect\citeauthoryear{{Webbink}}{{Webbink}}{1976}]{Web76}
{Webbink} R.~F.,  1976, \mn@doi [\apj] {10.1086/154781}, \href
  {http://cdsads.u-strasbg.fr/abs/1976ApJ...209..829W} {209, 829}

\bibitem[\protect\citeauthoryear{{Welty} \& {Ramsey}}{{Welty} \&
  {Ramsey}}{1994}]{Wel94}
{Welty} A.~D.,  {Ramsey} L.~W.,  1994, \mn@doi [\apj] {10.1086/174864}, \href
  {http://cdsads.u-strasbg.fr/abs/1994ApJ...435..848W} {435, 848}

\bibitem[\protect\citeauthoryear{{Willson} \& {Hudson}}{{Willson} \&
  {Hudson}}{1991}]{Wil91}
{Willson} R.~C.,  {Hudson} H.~S.,  1991, \mn@doi [\nat] {10.1038/351042a0},
  \href {http://cdsads.u-strasbg.fr/abs/1991Natur.351...42W} {351, 42}

\bibitem[\protect\citeauthoryear{{Zeilik}, {Cox}, {Ledlow}, {Rhodes}, {Heckert}
   \& {Budding}}{{Zeilik} et~al.}{1990}]{Zei90}
{Zeilik} M.,  {Cox} D.~A.,  {Ledlow} M.~J.,  {Rhodes} M.,  {Heckert} P.~A.,
  {Budding} E.,  1990, \mn@doi [\apj] {10.1086/169373}, \href
  {http://adsabs.harvard.edu/abs/1990ApJ...363..647Z} {363, 647}

\makeatother
\end{thebibliography}

\appendix

\section{Supplementary material}

\begin{table*}
\caption{Abbreviations. }gypt
\begin{center}
\begin{tabular}{ll}
\hline
Abbreviation &  \\
\hline
CABS       & Chromospherically Active Binary Star     (Sect. \ref{Intro}) \\
MLC        & Mean Light Curve                         (Sect. \ref{Intro}) \\
\JHLhyp    & Jetsu, Henry, Lehtinen hypothesis        (Sect. \ref{Intro}) \\
\sdr       & Surface Differential Rotation            (Sect. \ref{Intro}) \\
\dimethod  & Doppler Images measure \sdr             ~(Sect. \ref{Intro}) \\
\lcmethod  & Light Curves measure \sdr               ~(Sect. \ref{Intro}) \\
\cpsmethod & Continuous Period Search method          (Sect. \ref{Intro}) \\
\lsmethod  & Lomb-Scargle method                      (Sect. \ref{lcmethodsect}) \\
\simuone   & Simulated $s(t)$ data having $a_1=a_2=0.^{\mathrm{m}}05$                  (Sect. \ref{simu1}) \\
\simutwo   & Simulated $s(t)$ data having $0.^{\mathrm{m}}025=a_1<a_2=0.^{\mathrm{m}}05$  (Sect. \ref{simu2}) \\
\hline
\end{tabular}
\end{center}
\label{abbre}
\end{table*}

\begin{table}
\caption{Photometry of FK Com. 
Heliocentric Julian Date (HJD),
magnitude $(V)$,
segment (SEG) and
telescope (TEL).
We show only the two first lines of all 3807 lines.
}
\begin{center}
\begin{tabular}{cccc}
\hline
HJD           & $V$    & SEG &  TEL \\
\hline
2450850.8754 &  8.198  & 1   &  1 \\
2450850.8815 &  8.197  & 1   &  1 \\
\hline
\end{tabular}
\end{center}
\label{fkdata}
\end{table}

% /home/jetsu/pinkmlc/newprogs/finaldata.py produces segementinfo.tex
\begin{table}
\caption{Standard Johnson $V$ photometry of FK Com. 
Telescope (TEL), 
segments (SEG),
first and last observing time ($t_1$ and $t_n$),
time span and number
of observations ($\Delta T$ and $n$)}
\begin{center}
\begin{tabular}{cccccc}
\hline
TEL              &
SEG              & $t_1$          & $t_n$        & $\Delta T$ & $n$ \\
                 &
                 & HJD            & HJD          &  d & \\
\hline
   1 &    1 & 2450850.875 & 2450997.687 &    146.8 &    221 \\ 
   1 &    2 & 2451153.047 & 2451360.673 &    207.6 &    378 \\ 
   1 &    3 & 2451534.046 & 2451694.803 &    160.8 &     67 \\ 
   1 &    4 & 2451883.044 & 2452078.743 &    195.7 &     88 \\ 
   1 &    5 & 2452248.044 & 2452461.698 &    213.7 &    220 \\ 
   1 &    6 & 2452614.048 & 2452790.765 &    176.7 &    170 \\ 
   1 &    7 & 2452978.048 & 2453144.694 &    166.6 &    127 \\ 
   1 &    8 & 2453343.046 & 2453565.681 &    222.6 &    192 \\ 
   1 &    9 & 2453721.051 & 2453907.661 &    186.6 &    189 \\ 
   1 &   10 & 2454075.045 & 2454285.686 &    210.6 &    181 \\ 
   1 &   11 & 2454440.049 & 2454643.730 &    203.7 &    187 \\ 
   1 &   12 & 2454807.040 & 2455004.744 &    197.7 &    129 \\ 
   1 &   13 & 2455172.044 & 2455297.831 &    125.8 &    111 \\ 
   2 &    1 & 2450085.055 & 2450265.674 &    180.6 &    111 \\ 
   2 &    2 & 2450412.038 & 2450636.678 &    224.6 &    209 \\ 
   2 &    3 & 2450778.042 & 2450997.706 &    219.7 &    205 \\ 
   2 &    4 & 2451144.039 & 2451362.728 &    218.7 &    179 \\ 
   2 &    5 & 2451508.043 & 2451712.687 &    204.6 &    157 \\ 
   2 &    6 & 2451873.042 & 2452089.727 &    216.7 &    193 \\ 
   2 &    7 & 2452242.033 & 2452448.752 &    206.7 &    188 \\ 
   2 &    8 & 2452613.009 & 2452826.679 &    213.7 &    159 \\ 
   2 &    9 & 2452972.029 & 2453194.682 &    222.7 &    154 \\ 
\hline
\end{tabular}
\end{center}
\label{tabledata}
\end{table}

% Produced with ~/fk/newprogs/predictff.py
\begin{table*}
\caption{CPS-method results from \citet{Hac13}.
The first column gives the telescope (TEL).
The remaining columns give the same subset parameters as
described in the Appendix of 
\citet[][their Table A.1]{Leh11}:
first observing time $(t_1)$,
last observing time $(t_n)$,
mean observing time $(\tau)$,
statistically independent estimates (IND: 1=Yes, 0=No),
number of observations $(n)$,
period $(P_{\mathrm{CPS}} \pm \sigma_{P_{\mathrm{CPS}}})$,
peak to peak light curve amplitude 
$(A_{\mathrm{CPS}} \pm \sigma_{A_{\mathrm{CPS}}})$,
epochs of primary and secondary minima 
$(t_{\mathrm{CPS,min,1}} \pm \sigma_{\mathrm{t_{CPS,min,1}}} ,
t_{\mathrm{CPS,min,2}} \pm \sigma_{\mathrm{t_{CPS,min,2}}})$.
The dummy value ``-1.000'' denotes the
cases where no estimate was obtained.
The units of Heliocentric Julian Days are HJD-2~400~000. 
We show only the two first lines of this table.
The full table contains
1464 lines (TEL=1: 760 lines) and (TEL=2: 704 lines).}
\begin{center}
\addtolength{\tabcolsep}{-0.09cm} 
\begin{tabular}{cccccccccccccc}
\hline
TEL & $t_1$ & $t_n$ & $\tau$ & IND & $n$ & 
$P_{\mathrm{CPS}}$ & $\sigma_{P_{\mathrm{CPS}}}$ &
$A_{\mathrm{CPS}}$ & $\sigma_{A_{\mathrm{CPS}}}$ &
$t_{\mathrm{CPS,min,1}}$ & $\sigma_{\mathrm{CPS,t_{min,1}}}$ &
$t_{\mathrm{CPS,min,2}}$ & $\sigma_{\mathrm{CPS,t_{min,2}}}$ \\
    & HJD & HJD & HJD &  &  & d & d & mag & mag 
    & HJD & d 
    & HJD & d \\
\hline
  1 &   50850.875 &   50874.887 &   50863.555 &  1 &  29 & 2.4129 & 0.0036 &  0.059 &  0.006 &   50851.594 &   0.032 &      -1.000 &  -1.000 \\ 
  1 &   50851.871 &   50875.883 &   50865.793 &  0 &  28 & 2.4154 & 0.0029 &  0.059 &  0.004 &   50854.043 &   0.034 &      -1.000 &  -1.000 \\ 
\hline
\end{tabular}
\addtolength{\tabcolsep}{+0.09cm} 
\end{center}
\label{electric}
\end{table*}

% DOC 28.05.2019A &  DOC 28.05.2019B: Contain all relevant info, including double-check
% ..method/progs/figlatex.py produces latexcomplex.tex
% cp latexcomplex.tex /home/jetsu/method/Revision2/
% Now I can rerun figlatex.py ...
\begin{table*}
  \caption{Complex model segments.
    First three columns give
    telescope (TEL), segment (SEG) and mean observing time (YEAR).
    Next two columns give mean residuals
    (Eq. \ref{meanresiduals}: $|\epsilon|_{\mathrm{C}}$)
    and $z_{\mathrm{C}}(f_1,f_2)$ periodogram global minimum
    $(z_{\mathrm{C,min}})$.
    Mean of observations $(M)$ and eight complex $g_{\mathrm{C}}(t)$
    model parameters
    $(P_1,A_1,t_{\mathrm{g1,min,1}},t_{\mathrm{g1,min,2}},
      P_2,A_2,t_{\mathrm{g2,min,1}},t_{\mathrm{g2,min,2}})$
      are given in next columns.
      Critical level for
      rejecting $H_0$ of Sect. \ref{sectallsegments}
      is given in last column (Eq. \ref{oneortwo}: $Q_{\mathrm{F}}$). }
\label{Cresults}
\addtolength{\tabcolsep}{-0.16cm} 
\begin{tabular}{cccccccccccccccc}
\hline
      &      &      &                        &                  &     &
\multicolumn{4}{c}{$g_1(t)$}                                          & &
\multicolumn{4}{c}{$g_2(t)$}                                          &  \\
\cline{7-10} \cline{12-15} 
  TEL & SEG  & YEAR & $|\epsilon|_{\mathrm{C}}$ & $z_{\mathrm{C,min}}$ & $M$ &
$P_1$ & $A_1$ & $t_{\mathrm{g1,min,1}}$ &  $t_{\mathrm{g1,min,2}}$ & &
$P_2$ & $A_2$ & $t_{\mathrm{g2,min,1}}$ &  $t_{\mathrm{g2,min,2}}$ & $Q_F$ \\
      &      & $[\mathrm{y}]$ & $[\mathrm{mag}]$ & $[\mathrm{mag}]$ & $[\mathrm{mag}]$ &
      $[\mathrm{d}]$  & $[\mathrm{mag}]$ &  $[\mathrm{HJD}]$ &  $[\mathrm{HJD}]$ & &
      $[\mathrm{d}]$  & $[\mathrm{mag}]$ &  $[\mathrm{HJD}]$ &  $[\mathrm{HJD}]$ & \\                                                                                     
\hline
 1 &  1 & 1998.28 &  0.008 &  0.011 &     8.2032 &     2.3966 &     0.043 &    50851.0072 &    50851.5944 & ~ &     2.4407 &     0.014 &    50851.9761 &         -     & 7e-08\\ 
   &    &        &        &         &$\pm 0.0012$&$\pm 0.0026$&$\pm 0.003$&$\pm    0.0802$&$\pm    0.1139$& ~ &$\pm 0.0085$&$\pm 0.003$&$\pm    0.2978$&$             $&       \\
 2 &  4 & 1999.19 &  0.011 &  0.014 &     8.1792 &     2.3813 &     0.042 &    51145.4801 &    51144.5276 & ~ &     2.4209 &     0.032 &    51146.4361 &         -     & 5e-15\\ 
   &    &        &        &         &$\pm 0.0019$&$\pm 0.0018$&$\pm 0.004$&$\pm    0.0802$&$\pm    0.0828$& ~ &$\pm 0.0029$&$\pm 0.004$&$\pm    0.1405$&$             $&       \\
 1 &  2 & 1999.22 &  0.009 &  0.011 &     8.1747 &     2.3805 &     0.040 &    51155.0321 &       -       & ~ &     2.4216 &     0.032 &    51153.4754 &         -     & $<$ 1e-16\\ 
   &    &        &        &         &$\pm 0.0011$&$\pm 0.0012$&$\pm 0.002$&$\pm    0.0598$&$             $& ~ &$\pm 0.0017$&$\pm 0.002$&$\pm    0.0957$&$             $&       \\
 2 &  5 & 2000.19 &  0.011 &  0.015 &     8.2623 &     2.3649 &     0.029 &    51510.2775 &       -       & ~ &     2.3948 &     0.125 &    51508.1169 &         -     & 2e-09\\ 
   &    &        &        &         &$\pm 0.0038$&$\pm 0.0052$&$\pm 0.007$&$\pm    0.2344$&$             $& ~ &$\pm 0.0012$&$\pm 0.007$&$\pm    0.0516$&$             $&       \\
 1 &  3 & 2000.24 &  0.009 &  0.011 &     8.2576 &     2.3990 &     0.128 &    51534.3431 &       -       & ~ &     2.4330 &     0.030 &    51534.5784 &    51536.2426 &  0.00025\\ 
   &    &        &        &         &$\pm 0.0054$&$\pm 0.0016$&$\pm 0.008$&$\pm    0.0608$&$             $& ~ &$\pm 0.0057$&$\pm 0.011$&$\pm    0.1999$&$\pm    0.2979$&       \\
 2 &  6 & 2001.19 &  0.011 &  0.014 &     8.3019 &     2.3862 &     0.039 &    51874.9892 &    51873.7961 & ~ &     2.4032 &     0.189 &    51874.4216 &         -     & $<$ 1e-16\\ 
   &    &        &        &         &$\pm 0.0050$&$\pm 0.0016$&$\pm 0.006$&$\pm    0.0766$&$\pm    0.0798$& ~ &$\pm 0.0009$&$\pm 0.007$&$\pm    0.0397$&$             $&       \\
 2 &  7 & 2002.20 &  0.012 &  0.016 &     8.2998 &     2.3901 &     0.069 &    52242.5921 &    52243.4023 & ~ &     2.4130 &     0.071 &    52243.8981 &         -     & $<$ 1e-16\\ 
   &    &        &        &         &$\pm 0.0033$&$\pm 0.0019$&$\pm 0.008$&$\pm    0.0783$&$\pm    0.1209$& ~ &$\pm 0.0026$&$\pm 0.007$&$\pm    0.1572$&$             $&       \\
 1 &  5 & 2002.24 &  0.012 &  0.016 &     8.2903 &     2.3894 &     0.056 &    52248.2426 &    52249.8340 & ~ &     2.4097 &     0.085 &    52248.7552 &         -     & $<$ 1e-16\\ 
   &    &        &        &         &$\pm 0.0034$&$\pm 0.0017$&$\pm 0.006$&$\pm    0.1092$&$\pm    0.0811$& ~ &$\pm 0.0016$&$\pm 0.005$&$\pm    0.0854$&$             $&       \\
 1 &  6 & 2003.17 &  0.009 &  0.011 &     8.2394 &     2.3949 &     0.103 &    52614.8074 &       -       & ~ &     2.4255 &     0.018 &    52616.3039 &         -     & 4e-06\\ 
   &    &        &        &         &$\pm 0.0030$&$\pm 0.0011$&$\pm 0.005$&$\pm    0.0360$&$             $& ~ &$\pm 0.0074$&$\pm 0.004$&$\pm    0.2626$&$             $&       \\
 1 &  8 & 2005.23 &  0.011 &  0.015 &     8.2483 &     2.3891 &     0.092 &    53344.8189 &       -       & ~ &     2.4069 &     0.070 &    53343.6720 &         -     & $<$ 1e-16\\ 
   &    &        &        &         &$\pm 0.0034$&$\pm 0.0016$&$\pm 0.010$&$\pm    0.0983$&$             $& ~ &$\pm 0.0028$&$\pm 0.010$&$\pm    0.1422$&$             $&       \\
 1 &  9 & 2006.22 &  0.008 &  0.010 &     8.2432 &     2.3895 &     0.085 &    53722.9408 &    53721.8201 & ~ &     2.4098 &     0.106 &    53723.1062 &         -     & $<$ 1e-16\\ 
   &    &        &        &         &$\pm 0.0034$&$\pm 0.0005$&$\pm 0.005$&$\pm    0.0222$&$\pm    0.0258$& ~ &$\pm 0.0011$&$\pm 0.004$&$\pm    0.0475$&$             $&       \\
 1 & 11 & 2008.21 &  0.012 &  0.019 &     8.3074 &     2.3812 &     0.078 &    54440.8657 &       -       & ~ &     2.4065 &     0.136 &    54442.0416 &         -     & $<$ 1e-16\\ 
   &    &        &        &         &$\pm 0.0045$&$\pm 0.0025$&$\pm 0.011$&$\pm    0.1238$&$             $& ~ &$\pm 0.0016$&$\pm 0.010$&$\pm    0.0769$&$             $&       \\
\hline 
\end{tabular}
\addtolength{\tabcolsep}{+0.16cm} 
\end{table*}

% DOC 28.05.2019A &  DOC 28.05.2019B: Contain all relevant info, including double-check
% ../method/progs/figlatex.py produces latexsimple.tex
% cp latexsimple.tex /home/jetsu/method/Revision2/
% Now I can rerun figlatex.py ...
\begin{table*}
  \caption{Simple model segments.
    First two columns indicate, if black ellipse 
of Eq. \ref{blackellipse}
    ``intersects $f_1=f_2$'' or ``amplitude dispersion'' occurs.
    Telescope (TEL), segment (SEG) and mean observing time (YEAR)
    are given in next three columns.
    Mean of observations $(M)$ and four simple $g_{\mathrm{C}}(t)$
    model parameters
    $(P_1,A_1,t_{\mathrm{g1,min,1}},t_{\mathrm{g1,min,2}})$
    are given in next five columns.
    Last column indicates connection
    to $P_{\mathrm{w,1}}$ or $P_{\mathrm{w,2}}$ 
    ($P_{\mathrm{w}}$).}
\label{Sresults}
\addtolength{\tabcolsep}{-0.05cm} 
\begin{tabular}{ccccccccccccc}
\hline
  intersects & amplitude & & & & & & & & & & &  \\
  $f_1=f_2$  & dispersion   & TEL & SEG  & YEAR 
  & $|\epsilon|_{\mathrm{S}}$ & $z_{\mathrm{S,min}}$ & $M$ &
$P_1$ & $A_1$ & $t_{\mathrm{g1,min,1}}$ &  $t_{\mathrm{g1,min,2}}$ & 
$P_{\mathrm{w}}$ \\
\hline
                               Yes & Yes &  2 &  1 & 1996.25 &  0.016 &  0.021 &     8.2060 &     2.4063 &     0.149 &    50086.0730 &       -       &$P_{\mathrm{w,2}}$\\
                                            &     &    &    &         &        &        &$\pm 0.0055$&$\pm 0.0013$&$\pm 0.005$&$\pm    0.0521$&$             $&         \\
                               Yes & Yes &  2 &  2 & 1997.23 &  0.014 &  0.018 &     8.2337 &     2.4066 &     0.140 &    50412.9504 &       -       &$P_{\mathrm{w,2}}$\\
                                            &     &    &    &         &        &        &$\pm 0.0037$&$\pm 0.0010$&$\pm 0.004$&$\pm    0.0546$&$             $&         \\
                               Yes & Yes &  2 &  3 & 1998.21 &  0.013 &  0.017 &     8.2036 &     2.3949 &     0.052 &    50779.7092 &       -       &$P_{\mathrm{w,1}}$\\
                                            &     &    &    &         &        &        &$\pm 0.0018$&$\pm 0.0027$&$\pm 0.003$&$\pm    0.1360$&$             $&         \\
                               Yes & Yes &  1 &  4 & 2001.26 &  0.013 &  0.016 &     8.2983 &     2.4022 &     0.200 &    51884.0363 &       -       &$P_{\mathrm{w,2}}$\\
                                            &     &    &    &         &        &        &$\pm 0.0076$&$\pm 0.0007$&$\pm 0.006$&$\pm    0.0440$&$             $&         \\
                               Yes & Yes &  2 &  8 & 2003.21 &  0.015 &  0.019 &     8.2428 &     2.3978 &     0.093 &    52614.6877 &       -       &$P_{\mathrm{w,1}}$\\
                                            &     &    &    &         &        &        &$\pm 0.0032$&$\pm 0.0016$&$\pm 0.005$&$\pm    0.0768$&$             $&         \\
                                No & Yes &  1 &  7 & 2004.13 &  0.014 &  0.018 &     8.1844 &     2.4023 &     0.121 &    52979.7627 &       -       &$P_{\mathrm{w,2}}$\\
                                            &     &    &    &         &        &        &$\pm 0.0043$&$\pm 0.0015$&$\pm 0.004$&$\pm    0.0710$&$             $&         \\
                               Yes & Yes &  2 &  9 & 2004.23 &  0.025 &  0.036 &     8.1928 &     2.4065 &     0.137 &    52972.4620 &       -       &$P_{\mathrm{w,2}}$\\
                                            &     &    &    &         &        &        &$\pm 0.0050$&$\pm 0.0019$&$\pm 0.008$&$\pm    0.1065$&$             $&         \\
                                No & Yes &  1 & 10 & 2007.26 &  0.021 &  0.026 &     8.2548 &     2.3937 &     0.188 &    54076.9722 &       -       &$P_{\mathrm{w,1}}$\\
                                            &     &    &    &         &        &        &$\pm 0.0053$&$\pm 0.0008$&$\pm 0.006$&$\pm    0.0400$&$             $&         \\
                                        Yes & Yes &  1 & 12 & 2009.23 &  0.021 &  0.025 &     8.2816 &      -     &     -     &       -       &       -       &    -    \\
                                            &     &    &    &         &        &        &$\pm 0.0032$&$          $&$         $&$             $&$             $&         \\
                               Yes & Yes &  1 & 13 & 2010.12 &  0.014 &  0.017 &     8.2916 &     2.3896 &     0.115 &    55174.4076 &    55173.0240 &$P_{\mathrm{w,1}}$\\
                                            &     &    &    &         &        &        &$\pm 0.0037$&$\pm 0.0017$&$\pm 0.005$&$\pm    0.0871$&$\pm    0.0632$&         \\
\hline
\end{tabular}
\addtolength{\tabcolsep}{+0.05cm} 
\end{table*}

% Don't change these lines
\bsp    % typesetting comment
\label{lastpage}
\end{document}